\pdfoutput=1
\documentclass[12pt,]{article}
\usepackage{times}
\usepackage{amsmath,amssymb,amsthm}
\usepackage{graphicx}
\usepackage{xspace}
\usepackage{rotating} % to rotate tables
\usepackage{booktabs}
\usepackage{float}
\usepackage{subfig}
\usepackage{setspace}
\usepackage{mdwlist}
\usepackage{paralist}
\usepackage{tabularx}	
\usepackage{tabulary}	
\usepackage{multirow}
\usepackage{bigstrut}
\usepackage{wrapfig}
\usepackage[margin=1.5in]{geometry}

\newcommand{\uec}{\ensuremath{D}\xspace}
\newcommand{\distincts}{\ensuremath{d}\xspace}
\newcommand{\sample}{\ensuremath{q}\xspace}
\newcommand{\NoverD}{\ensuremath{\frac{N}{A}}\xspace}

\newcommand{\Z}{\ensuremath{\theta}\xspace}
\newcommand{\popsize}{\ensuremath{N}\xspace}
\newcommand{\sampsize}{\ensuremath{n}\xspace}

\newcommand{\jkone}{\ensuremath{\hat{D}_{\mathrm{uj1}}}\xspace}
\newcommand{\jktwo}{\ensuremath{\hat{D}_{\mathrm{uj2}}}\xspace}
\newcommand{\jktwos}{\ensuremath{\hat{D}_{\mathrm{sj2}}}\xspace}
\newcommand{\jktwoa}{\ensuremath{\hat{D}_{\mathrm{uj2a}}}\xspace}
\newcommand{\sh}{\ensuremath{\hat{D}_{\mathrm{Sh}}}\xspace}
\newcommand{\shtwo}{\ensuremath{\hat{D}_{\mathrm{Sh2}}}\xspace}
\newcommand{\shthree}{\ensuremath{\hat{D}_{\mathrm{Sh3}}}\xspace}
\newcommand{\clone}{\ensuremath{\hat{D}_{\mathrm{CL1}}}\xspace}
\newcommand{\cltwo}{\ensuremath{\hat{D}_{\mathrm{CL2}}}\xspace}
\newcommand{\gee}{\ensuremath{\hat{D}_{\mathrm{GEE}}}\xspace}
\newcommand{\ade}{\ensuremath{\hat{D}_{\mathrm{AE}}}\xspace}

\newcommand{\cvar}{\ensuremath{\gamma^2}\xspace}
\newcommand{\estcvar}[1]{\ensuremath{\hat{\gamma}^2(#1)}\xspace}

\newcommand{\est}[1][]{\ensuremath{\hat{D}_{\mathrm{#1}}}\xspace}
\newcommand{\sampcov}{\ensuremath{\hat{C}}\xspace}
\newcommand{\tildecvar}{\ensuremath{\tilde{\cvar}} \xspace}
\newcommand{\hatcvar}{\ensuremath{\hat{\cvar}} \xspace}

\usepackage[small,compact]{titlesec} 
\usepackage[usenames]{color}
\usepackage{colortbl}
\usepackage{multicol}
\usepackage{floatpag}
\newcommand{\captionfonts}{\footnotesize}

\makeatletter  % Allow the use of @ in command names
\long\def\@makecaption#1#2{%
  \vskip\abovecaptionskip
  \sbox\@tempboxa{{\captionfonts #1: #2}}%
  \ifdim \wd\@tempboxa >\hsize
    {\captionfonts #1: #2\par}
  \else
    \hbox to\hsize{\hfil\box\@tempboxa\hfil}%
  \fi
  \vskip\belowcaptionskip}
\makeatother   % Cancel the effect of \makeatletter

\title{Extensive Large-Scale Study of Error in Samping-Based Distinct Value
	Estimators \\ for Databases\footnote{This is the full-length
		version of a shorter published paper, and includes
		supplementary material for the published paper. Please cite as
		``Vinay Deolalikar and Hernan Laffitte: Extensive Large-Scale
		Study of Error in Samping-Based Distinct Value Estimators
		for Databases, IEEE Big Data Conference, Washington DC,
	December 2016."}
}

\author{Vinay Deolalikar\footnote{contact author,
	{\tt	deolalikar.academic@gmail.com}, work done  at HP Labs, Palo Alto.}  \hspace{0.1in} Hernan
Laffitte\footnote{{\tt hernan.laffitte@hp.com}}}

\date{}

\begin{document}

\maketitle

\begin{abstract}
The problem of distinct value estimation has many applications.
Being a critical component of query optimizers in databases, it also has high
commercial impact. Many
distinct value estimators have been proposed, using various
statistical approaches. 
However, characterizing the errors incurred by these 
estimators is an open problem: existing analytical approaches are not
powerful enough,  and
extensive empirical studies at large scale do not exist. We conduct an extensive large-scale
empirical study of 11
distinct value estimators from four different approaches to the problem
over families of Zipfian distributions whose parameters model real-world
applications. Our study is the first that \emph{scales to the size of a
billion-rows} that today's large commercial databases have to operate in. This allows us to characterize the error that is
encountered in real-world
applications of distinct value estimation. By mining the generated data,
we show that estimator error depends on a key latent
parameter --- the average uniform class size --- that has not been studied
previously. This parameter also allows
us to unearth error patterns that were previously unknown. Importantly, ours
is the first approach that provides a framework for
 \emph{visualizing the error patterns} in distinct value
estimation, facilitating discussion of this problem in
enterprise settings.  Our  characterization of errors can be used
for several problems in  distinct value estimation, such as the design of
hybrid estimators.  This work aims at the practitioner and the
researcher alike, and addresses questions frequently asked by both
audiences. 
\end{abstract}

\section{Introduction}

% what is the problem
Consider the following  problem: estimate the number of distinct values (or
classes)  in a population by statistically analyzing its sample. 
% why is it interesting
This is the problem of \emph{distinct value estimation}, and it 
arises in a surprisingly large variety of applications, where it is used to
estimate a number of interest: in census studies,  the number of individuals
in lists having duplications \cite{HS98};
in economics, the number of investors from samples
of share registers of companies \cite{Mos49};
in ecology, the number of species from some sampling scheme \cite{SvB84};
in numismatics,  the number of coins produced by a mint when a selection of
such coins has been found \cite{CL92}.

However, the application that has arguably the most commercial impact arises
in databases\footnote{
	The high-end large data warehouse market for
	2022 is estimated to be \$22B \cite{TADW16}.
	}. 
Large data warehouses rely on complex query
optimizers to formulate their query plans.  A query optimizer needs an
estimate of the number of distinct values in its attributes in order to
formulate its query plan \cite{HNSS95}. On attributes that are not indexed,
single pass (or scan) estimators  provide fairly accurate estimates of
distinct values while using a small memory footprint \cite{FM85, Beyer09}.
However, databases have witnessed explosive growth over the past decade;
today's large commercial data warehouses have billions of rows.
Therefore, accessing all the data  in this manner is seldom possible.  This
makes sampling based estimators of distinct values the only practical solution
for distinct value estimation. All the large scale commercial databases that
we are aware of use sampling based estimators of distinct values as part of
their query optimizer.

A great range of sampling based estimators for distinct values have been proposed in both the
statistics and database literature, see \cite{BF93} for a survey. Distinct
value estimation is a hard problem, and estimators generally incur
(significant) errors in performing the estimates.  
Therefore, users with some knowledge about the datasets they will encounter in
applications, are interested in the following questions.
\begin{itemize*}
\item[Q1] From this plethora of distinct value estimators, which estimator will
give the least error on their dataset which is of the size of a typical
database --- (millions to a billion rows)? 
\item[Q2] For a choice of estimator, can
one characterize the errors that will be incurred on such a large dataset?  
\item[Q3] How high should the sampling fraction be in order to keep error within a
tolerable margin? Conversely, given a sampling fraction and a desired error
margin, what choice of estimator will restrict errors to that margin?
\item[Q4] Do the errors occur in patterns that can be effectively visualized? 
\end{itemize*}

These problems have serious practical ramifications in database
design: 
a poor estimate of distinct values can result in a considerably more expensive
query plan \cite{HNSS95}. Furthermore, a query estimator may have more tolerance
for a certain region of bias, as opposed to other regions (depending on
the region where the query plan changes from a good one to an inefficient one).
  Unfortunately, due to the difficulty of
the problem, and the lack of an adequate characterization of errors,
commercial systems often suffer from 
unexpected poor estimates that seem to occur ``at random". These cause 
the query optimizer to formulate highly inefficient query plans, resulting in
intolerable delays to the system user, especially when the database has in excess of 
hundred million rows. Simply put, our state of knowledge about error in
distinct value estimators is untenable due to the large sizes of today's databases.
  
One might think that a way out would be to increase the sampling
fraction. However, the cost of even 1\% sampling in a
commercial database with  billions of rows, is significant, and a larger
than 2\% sample	 is often simply not possible.
Also, because data is stored in blocks, generating a 10\% random sample is
sometimes as expensive as scanning all the data. 
Therefore, there is great practical interest in the least sample size that
can deliver a reasonable estimate of distinct values. 

Finally, visual descriptions of errors, in addition to facilitating one's own
understanding,  permit effective communication to
non-specialists, which is an important aspect of working with statistical technologies in an enterprise setting.

Unfortunately, there do not yet exist analytical techniques that
characterize error satisfactorily (or answer any of (Q1)-(Q4)) for any distinct value estimator in
literature (except, to a degree, for the \gee, see \ref{sec:relatedwork}).
Instead, there is a powerful result in the ``opposite direction" in \cite{CMN98,CCMN00}
that says that every estimator will give a large error on some dataset. This
suggests that new analytical approaches may have to be developed for specific
datasets/distributions. 

In the absence of analytical approaches to characterizing error, we must turn
to empirical characterizations performed through a large-scale extensive
study over distributions that represent real-world applications. However, here, 
the situation is best described by \cite{HNSS95}: 
\begin{quote} \begin{small}
``Unfortunately, analysis of distinct value estimators is non-trivial, and few analytic results are available. To make matters worse, there has been no extensive empirical testing or comparison of estimators either in the database or statistical literature...the testing that has been done in the statistical literature has generally involved small sets of data in which the frequencies of different attribute values are fairly uniform; real data is seldom so well-behaved."
\end{small}
\end{quote}

The impetus for our work arose from an effort, during the year 2007, to design a distinct value estimator for a large
commercial HP database product that had, among its customers, a Fortune-10
company. 
At that time, we did not find an extensive, large-scale, comparative study on distinct value
estimators that would allow us to answer questions (Q1)-(Q4) reliably for critical database applications. 
Several studies in literature showed different estimators to be the best over
the datasets in the purview of the particular study. These are valuable data
points, but they are almost never comparable, and it is not clear how far the
results generalize (see $\S$~\ref{sec:relatedWork:empirical}). This
is the gap that the current paper aims to fill. 

This brings us to our approach. Our study characterizes error in distinct value
estimators, and provides answers to (Q1)-(Q4). Our approach can be described
in the following steps.
\begin{enumerate*}
\item Conduct an extensive empirical study of the 
relative performance of various estimators on a well-chosen \emph{parameter
space}.
\item Mine the generated data for \emph{stable patterns} to the relative performance of estimators on
datasets,
\item Find parameters that organize these stable patterns. 
\item Present these stable patterns in a manner that is \emph{easy to
visualize} and
communicate among practitioners. 
\item Characterize error through bias, ratio-error, and RMSE, for each 
region of the parameter space that is delineated by stable patterns of
behavior. 
\end{enumerate*}

% why is it hard

%\subsection{Our Motivation and Approach}
%	The situation is In light of our past
%	experience with the difficulties of designing distinct value estimators, we framed the following  two broad
%	questions to guide our present study. Both questions are posed at the large
%	scales of today's commercial databases. 
%	\begin{enumerate*}
%	\item Which estimators do well on which datasets?
%	\item What properties of datasets allow certain estimators to do well on them,
%	and not others? 
%	\end{enumerate*}
%	
 
%We stress on characterization,
%as opposed to just an evaluation of errors. A characterization should keep
%track of the relative sign and magnitude of biases in the various estimators
%as the parameters are varied in a controlled manner. The hope is that such a
%characterization will yield \emph{stable patterns} of relative behavior of the
%estimators. To cite a toy example where a characterization can be far more
%useful than a error evaluation, consider the case where two estimators have
%errors in a region of the parameter space, but that these errors result from
%positive, and negative biases, respectively, of approximately equal magnitude.
%Then, we would simply be able to combine them and obtain an accurate estimator
%for this region. Note that a basic error evaluation (using RMSE) would not
%provide this richer information. 

\subsection{Our Contributions}

Our extensive empirical study allowed us to construct 
 a detailed
large-scale characterization of the relative behavior of important families of
distinct value estimators. Our study is the first that scales to the
billion-row size that today's large commercial databases operate on. Our characterization describes both inter- and
intra-family behavior over a parameter space that models the variation of
real-world data.  We identify stable patterns of behavior that allow us to
``make sense" of the huge amount of data generated in our empirical study.
We identify a critical latent parameter --- the average uniform class size ---  that estimators
are sensitive to, and that allows us to find patterns in our results. This
variable has not yet been studied in literature.

%Let $N$ be the size of a Zipfian  population, $D$ be the size of the alphabet over which Zipfian data of Zipfian parameter \Z and  size $N$ is drawn. 
Our study allows us to answer the following types of questions. 
\begin{enumerate*}
\item What are scale effects on each estimator?
\item What are skewness effects on each estimator?
\item How does the actual distinct value count affect an estimator?
\item What is the ``best" estimator under various definitions?
\item What sampling percentage is adequate
 for various error requirements?
 \end{enumerate*}

Finally, our study allows us to comparatively evaluate, at large-scale, 
the four different approaches to distinct value estimation that are within our
purview. 

\noindent{\bf Limitations and Scope of our Study.}
Our study is
most useful for distributions that approximate a power law, and with
population parameters that resemble those of our study.  However, both these
are intended to reflect real-world problems.

\section{Related Work}
\label{sec:relatedwork}

% general intuition, stated by Goodman, theorem by Chaudhuri
%	The difficulty of distinct values estimation is intuitively clear: we are
%	asked to estimate classes that we do not see in the sample. The precise reason
%	is that ``there is nearly always a good chance that there ar a very large
%	number of extremely rare species" \cite[quote by I. J. Good]{BF93}. A formal
%	theorem based on this idea is provided in \cite{CMN98}, where it is shown that
%	any estimator that draws $r$ samples at random from a population of size $N$
%	must incur an error of $O(\sqrt{N/r})$ on some distribution. 
%	\cite{CCMN00} extends the result to any (possibly adaptive) sampling scheme.
%	
Over the years, a growing number of methods have been used to
construct estimators for the distinct values problem \cite{BF93}. 
As pointed out by \cite{HS98}, there are few papers that focus on the case where the population size is known.
This is the case that models applications to databases. 
First, we briefly survey results on \emph{unsuitability} of several proposed
estimators for large scale applications. These include the earliest
distinct value estimators from statistics that were used for database
applications \cite{HOT88,ODT91}.  
\begin{itemize*}
\item The estimators \est[Good1] and \est[Good2] from \cite{Goo49} are unbiased, but have extremely high variance, and are numerically
unstable at small sample sizes \cite{HOT88,NS90,HNSS95,BF93}. 
\item The estimator 
\est[Chao] from \cite{ODT91,Cha84} underestimates, except where $f_2 = 0$, where
it blows up \cite{HNSS95}. 
\item The jacknife estimator \est[CJ] \cite{ODT91} based on \cite{BO78,BO79}
is derived from  inapplicable assumptions \cite{HNSS95}.
\item The estimator  
\est[Sichel]
% \cite{Sic86a,Sic86b,Sic92}
\cite{Sic92} is unsuitable since the  two-parameter GIGP does not fit
several datasets \cite{HNSS95}. 
\item The method of moments estimators 
\est[MM0], \est[MM1], and \est[MM2] \cite{HNSS95} yield poor estimates when $\cvar > 1$
\cite{HNSS95}.
\item \est[Boot] \cite{SvB84} has the property that $\est[Boot] < 2\distincts$, therefore poor
performance is likely at large \uec and small \sample \cite{HNSS95}.
\end{itemize*}
  
\subsection{The Estimators Under Study}

The 11 estimators that are selected for our study come four
different approaches to distinct value estimation, and have been proposed as
being suitable for the scale of database applications.  We set notation before
we commence their description.

Consider a population of size \popsize that has $D$ classes of sizes $N_1,\ldots,N_D$ so
that $\popsize = \sum^D_{j=1} N_j$. We denote by $F_i$ the number of classes
of size $i$ in the population, so that $D = \sum^N_{i=1} F_i$.
Samples of size $n =  \sample \popsize$ may
be drawn from the population. Here \sample is the sampling fraction. The
resulting sample has $d$ classes (therefore $d \leq D$). We denote by $f_i$
the number of classes in the sample that occur $i$ times, so that $d =
\sum^n_{i=1} f_i$ and $\sum^n_{i=1} if_i = n$.    
We wish to form an estimate \est of $D$ by analyzing the sample, and using the
known value $N$.  

For convenience, we place the notation defined earlier, as well as
that to be defined shortly, in Table~\ref{tbl:notation}.

\begin{table}
\footnotesize
\caption{Notation}
\centering
\begin{tabular}{cll}
\toprule
Symbol & Meaning & Notes\\
\midrule
\popsize & size of population & very large (millions to billion)\\
\uec & number of classes in population & quantity to be estimated   \\
$N_j$ & size of $j^{\mathrm{th}}$ class in population & $\sum N_j =
\popsize$\\ 
$\bar{N}$ & average class size in population & $\bar{N} =
\popsize/\uec$ \\
\cvar & squared coefficient of variation of class sizes & $\cvar =
{(1/D) (\sum_{j=1}^D (N_j - \bar{N})^2)}/{(\bar{N})}^2$ \\
$F_i$ & number of classes of size $i$ in the population & $\uec =
\sum F_i$, $\popsize = \sum i F_i$ \\
\sampsize & size of sample drawn from population & preferably small \\
\sample & sampling fraction  & $\sample = \sampsize / \popsize$ \\
$n_j$ &  size of $j^{\mathrm{th}}$ class in sample & may be zero   \\ 
$d$ & distinct classes in sample & number of non-zero $n_j$  \\
$f_i$ & number of classes of size $i$ in the sample & $d = \sum
f_i$, $\sampsize = \sum i f_i$ \\
\est & estimator for \uec & subscripts indicate different estimators \\
\Z & skewness parameter for a Zipfian population & defined in
$\S\ref{sec:large scale study}$ \\
$A$ & size of alphabet for Zipfian population & defined in  $\S\ref{sec:large scale study}$\\
\NoverD & average uniform class size & defined in  $\S\ref{sec:large scale study}$\\
\bottomrule
\end{tabular}
\label{tbl:notation}
\end{table}

%\begin{tabulary}{6.5cm}{ccc}
%\text{Estimator} & \text{Ref} & \text{Comments} \\
%First order Schlosser \cite{Sch81}& $ \sh = \distincts + f_1 \frac{\sum_{i=1}^{n} (1-q)^i f_i } { \sum_{i=1}{n} i q
%(1-q)^{i-1} f_i}. $ &   $ \frac{E[f_i]}{E[f_1]} = \frac{F_i}{F_1}.$
%\end{tabulary}
%
%\begin{table}
%\centering	
%$
%\begin{array}{l}
%\shtwo = 
%d + f_1 \left( \frac{q (1 + q)^{\tilde{N} -1 } }
%{(1 + q)^{\tilde{N} - 1}}
%\right)
%\left( \frac{ \sum_{i=1}^{n} (1 - q)^i f_i } {
%\sum_{i=1}^{n} i q (1 - q)^{i-1} f_i} \right)  \\
%\shthree =
%d + f_1 \left(
%\frac{\sum_{i=1}^n   i q^2 (1 -q^2)^{i-1} f_i }
%{ \sum_{i=1}^n ( 1 - q)^i ((1 + q)^i  - 1) f_i }
%\right)
%\left( \frac{ \sum_{i=1}^{n} (1 - q)^i f_i } {
%\sum_{i=1}^{n} i q (1 - q)^{i-1} f_i} \right)  \\
%\jkone =  \left( 1 - \frac{(1-q) f_1}{n} \right)^{-1} d_n  \\
%\jktwo =  \left( 1 - \frac{(1-q) f_1}{n} \right)^{-1} \left( d -
%\frac{f_1(1-q) \ln(1-q) \estcvar \jkone }{q} \right) \\
%\jktwos = (1 - (1 - q)^{\frac{N}{\jkone}})^{-1} ( d - (1-q)^{\frac{N}{\jkone}}
%\ln(1-q) \estcvar \jkone   ) 
%\end{array}
%$
%\end{table}
%

Now we come to the estimators in our study.
We will find it convenient to describe the remaining estimators with the model
\begin{equation} \label{eq:model}
\est = d + f_0,
\end{equation}
where $f_0$ is the number of classes that are not represented in the sample.
The first estimator is proposed by \cite{Sch81} who consider the task of building estimators for dictionaries. 
The estimators constructed assume that
the population size is large, the sampling fraction is non-negligible, and
that the proportions of classes in the sample reflects that in the population,
namely
\(
\frac{E[f_i]}{E[f_1]} = \frac{F_i}{F_1}. 
\) 
Under these assumptions, Schlosser derived the estimator
\[
\sh = \distincts + f_1 \frac{\sum_{i=1}^{n} (1-q)^i f_i } { \sum_{i=1}^{n} i q
(1-q)^{i-1} f_i}.
\]
Note that the estimate of classes not represented in the sample is a 
function of $f_1$.  
This estimator was constructed for language dictionaries, and word
usage is known to be Zipfian with a high skewness parameter, and the
assumptions are indeed valid in the application under study. 

The next family of estimators is from \cite{HS98}, and  
uses the bias reducing jackknifing technique of \cite{GS72}. It is of the form
\begin{equation} \label{eq:model:jackknife}
\est = d + K \frac{f_1}{n}.
\end{equation}
Namely, the jackknife estimate of $f_0$ is $K f_1/n$. 
Using first and second order methods, \cite{HS98} derive different
values of the parameter $K$, resulting in the estimators \jkone and \jktwo,
respectively. The second order estimator \jktwo requires estimation of the
squared coefficient of variation
\(
\cvar = {(1/D) (\sum_{j=1}^D (N_j - \bar{N})^2)}/{\bar{N}^2}, 
\)
where $\bar{N} = N/D$.  
A method of moments estimate based on \est is used as follows
\begin{equation}
\estcvar{\est} = \max \left( 0, \frac{\est}{n^2} \sum_{i=1}^n i(i-1)f_i +
\frac{\est}{N} - 1 \right).
\end{equation}

 They then construct an estimator that ``smooths"
the second order estimator, resulting in \jktwos. 
These estimators are given below.
\begin{align}
\jkone = & \left( 1 - \frac{(1-q) f_1}{n} \right)^{-1} d, \\
\jktwo = & \left( 1 - \frac{(1-q) f_1}{n} \right)^{-1} \left( d -
\frac{f_1(1-q) \ln(1-q) \estcvar{\jkone}}{q} \right), \\
\jktwos =& (1 - (1 - q)^{\tilde{N}})^{-1} ( d - (1-q)^{\tilde{N}}
\ln(1-q)\popsize \estcvar{\jkone}), 
\end{align}
where $\tilde{N}$ is an estimate of the average class size, and is set to
$N/\jkone$. 
Finally, they use a ``stabilizing" technique from \cite{CMY93} to construct the
estimator \jktwoa as follows. Fix $c >1$ and remove all classes whose
frequency in the sample exceeds $c$. Then compute \jktwo of the reduced
sample, and increment it by the number of the previously removed classes,
giving \jktwoa. In our experiments, we used $c=50$.   
   
\cite{HS98} then observe that the estimator \sh also conforms to the model
\eqref{eq:model} with parameter 
\[
K = K_{\mathrm{Sh}} = n \frac{\sum_{i=1}^n (1-q)^i f_i }{ \sum_{i=1}^n i q (1-q)^{i -1}
f_i }.
\]

Replacing $K_{\mathrm{Sh}}$ with alternative expressions, they obtain the following two
estimators.
\begin{align} 
\shtwo = &
d + f_1 \left( \frac{q (1 + q)^{\tilde{N} -1 } }
{(1 + q)^{\tilde{N} }- 1}
\right)
\left( \frac{ \sum_{i=1}^{n} (1 - q)^i f_i } {
\sum_{i=1}^{n} i q (1 - q)^{i-1} f_i} \right), \label{eq:shtwo}\\
\shthree = & 
d + f_1 \left(
\frac{\sum_{i=1}^n   i q^2 (1 -q^2)^{i-1} f_i }
{ \sum_{i=1}^n ( 1 - q)^i ((1 + q)^i  - 1) f_i }
\right)
\left( \frac{ \sum_{i=1}^{n} (1 - q)^i f_i } {
\sum_{i=1}^{n} i q (1 - q)^{i-1} f_i} \right).\end{align}

We encountered floating point errors when evaluating \shtwo for large $N$, and so
approximated the term in the first parentheses in \eqref{eq:shtwo} by
$q/(1+q)$ in those cases. 
 
%The jacknife approach of \cite{HS98}, which was begun in \cite{HNSS95},
%differs from previous jackknife estimators for distinct values developed in
%\cite{BO78, BO79,HF83,SvB84} in that these previous estimators used the
%following form of distinct values
%\[ 
%E[d_n] = \uec + \sum_{k=1}{\infty} \frac{a_k}{n^k}. 
%\] 
%This equation implies that $d_n$ has a bias of $O(n^{-1})$ as an estimator of
%$D$. An $m^{th}$ order jackknife can be obtained by correcting for biases
%up to order $O(n^{-m + 1})$. However, this approach to estimation is not
%suitable for the application at hand since $E[d_n]$ is not of the form above. 

In another stream of research \cite{CMN98, CCMN00} reason that classes that occur
frequently in the sample represent a single class in the population.
However, classes that occur infrequently could represent multiple classes in
the population.  By estimating how many classes each such infrequent class
represents, they
construct two estimators: \gee\cite{CMN98} and \ade\cite{CCMN00},  a heuristic
estimator that 
adapts to the distribution skew. 
\begin{align}
\gee =& \sqrt{\frac{N}{n}} f_1 + \sum_{j=2}^{n} f_j,\\
\ade =& d + K f_1, \text{ where }\\
K =& \frac{\sum_{i=3}^n  e^{-i} f_i + m e^{(f_1 + 2f_2)/m}          }
{\sum_{i=3}^n i e^{-i}f_i + (f_1 + 2f_2) e^{-(f_1 + 2f_2)/m} }.
\end{align}
 To solve for $m$ above, we use $m - f_1 - f_2 = K f_1$. Notice that \gee
conforms to \eqref{eq:model} with $f_0 = (\sqrt{\frac{N}{n}} - 1)f_1$.  

%Furthermore, \cite{CMN98} also establish a powerful
%negative result in sampling based distinct value estimation. Namely, they show
%that the ratio error of distinct value estimation, defined as
%\[
%error ( \est{} ) = 
%\begin{cases}
%\est/\uec & \text{ if $\est \geq \uec$ }\\
%\uec/\est{} & \text{ if $\uec \geq \est{}$ }. 
%\end{cases}
%\]
%in the case of sampling $r$ rows out of $n$, must satisfy
%
%\[
%error( \est{}) \geq \sqrt{\frac{n \ln 1/\gamma}{r} },
%\]
%for any $\gamma > e^{-r}$. 
%In \cite{CCMN00}, the lower bound is computed to  $\sqrt{\frac{(n-r) \ln
%1/\gamma}{2r} }$. Further, it is shown that the \gee estimator is optimal with
%respect to this bound.

The next approach to distinct value estimation evaluated in our study is
through the notion of ``sample
coverage". Sample coverage $C$ is defined as the fraction of classes in the
population that appears in the sample. This approach was proposed by Turing \cite{Goo53}, who suggested the following estimator of sample coverage.
\[
\sampcov = 1 - f_1/n.
\]
Therefore, an estimate of the distinct values would be
\begin{equation} \label{eq:samplecoverage:uniform}
\est[1] = d/\sampcov.
\end{equation}
\cite{DR80} show that the above estimate is quite efficient relative to the
MLE of the same quantity.  
Further development of the sample coverage method includes
\cite{GT56,Rob68,Est82,Est86}. The estimators we use are
constructed in 
\cite{CL92}, and are ``second order" in the sense that they require an estimate
of the coefficient of variation of the class distribution to correct the
estimate \eqref{eq:samplecoverage:uniform} that holds for uniformly
distributed classes.  Two estimators  of \cvar are constructed.
\begin{align}
 \tildecvar = & \max \left[ \frac{\est[1] \sum i (i - 1) f_i}{n^2- n - 1}, 0
\right],\\
\hatcvar = & \max
\left[ \tildecvar \left( \frac{ 1 + n ( 1 - \sampcov) \sum i (i - 1) f_i}{n(n - 1)\sampcov} \right), 0 \right].
\end{align}

The above estimates of \cvar result in the following two estimators of \uec. 
\begin{align}
\clone = & \frac{\distincts}{\sampcov} + \frac{n( 1 - \sampcov)}{\sampcov}
\tildecvar,\\
\cltwo = & \frac{\distincts}{\sampcov} + \frac{n( 1 - \sampcov)}{\sampcov}
\hatcvar.
\end{align}
These estimators are for a multinomial sampling regime (sampling with
replacement) for an infinite population.  However, given the large populations
that we consider in our work, the adjustment for sampling without replacement
from finite populations is negligible, and therefore conceivably these
estimators could be used for the distinct value problem in databases as well.

The aforementioned works provide strong theoretical contributions, and give us
a range of estimators, from different approaches to the distinct value
problem.

\subsection{Deficiencies in Empirical Studies}
\label{sec:relatedWork:empirical}
We note the following types of deficiencies in previous studies. 
\begin{asparaenum}
\item The scale of the study is far too small and does not reflect today's commercial databases
that handle billions of rows.
\item The studies involve too few datasets for stable patterns to emerge,
so no generalized conclusions can be drawn.
\item The variation in parameters is over a significantly smaller range than
occurs in real-world applications.
\item Most studies report the RMSE, not the individual biases of
each estimator.
\item Most studies do not consider sampling percentages lower than 1\%,
(most assume sampling percentages above 5\%). In real world
commercial data warehouses,  even 1\% sampling is 
expensive, and one would like to make do with less. 
\item Comparisons are made between members of one or two approaches to
distinct value estimation only. 
\item Individual estimators are compared to hybrid estimators.
\item Differing notions of skewness, that are not proxies for each
other, are used. 
\end{asparaenum}

\section{Methodology}
\label{sec:methodology}
\label{sec:large scale study}

%The literature on distinct value estimation is vast. Even if we restrict to
%the subproblem that is immediately applicable to databases, where the size of
%the population is known, there are over a dozen papers. Each paper that
%proposes an estimator of distinct values does indeed
%test the estimator they propose on some datasets. However, there is no
%standardized method to evaluate these estimators. 
%
%\subsection{Our Approach to Characterization}
%
%Our approach is motivated by the following three questions.
%\begin{enumerate}
%\item What features of an estimator make it perform well in certain datasets?
%\item What features of a dataset make it suited to a certain estimator, and
%not to others?
%\item Are there stable patterns to the relative performance of estimators on
%datasets? 
%\end{enumerate}
%
%Clearly, these questions lend themselves both to theoretical, and empirical
%approaches. Our approach in this paper is largely empirical; in forthcoming
%work, we will use this empirical characterization to inform theoretical work.
%
%In order to systematize our approach,  
%we wish to draw general conclusions by testing the estimators on a carefully
%controlled parameter space.  The idea is to vary the parameters of the
%underlying population, and the sampling regime, in a stepwise manner, across a
%wide domain of values, so that stable
%patterns will emerge.  We want enough variation so that we can generalize our
%results, but not too much variation that will prevent any conclusions from
%being drawn.  

\subsection{Desiderata of Study}

Our empirical study was designed with the following high-level desiderata.
\begin{asparaenum}
\item We wish to standardize our benchmarking setup so that in the future,
different estimators can be systematically compared.
\item We want to understand the behavior of estimators as an
absolute function of parameters that are entirely in our control.
\item The parameter space should model real-world data. 
\end{asparaenum}

Characteristics of real-world data are numerous, and often interacting. We can never be certain which characteristic
of the data is causing or influencing the performance of the estimator. In this way, the
performance of the estimator becomes \emph{relative}, and not absolute as a
function of a chosen set of parameters.  

This points us to the use of artificially generated data from a well-chosen
parameter space, as the means to our desired characterization. 
The potential hazard in doing so is that the artificial data may not represent any real-world
application. So while desiderata (1) and (2) are met, we might fail on (3). 

Recall that the family of
Zipfian distributions $Z_{A,\Z}$ parametrized by their skewness parameter \Z, and the
size of the alphabet $A$, have probability masses $P_{A,\Z}(i)$ satisfying 
\(
P_{A,\Z}(i) \propto \frac{1}{i^{\Z}}.
\)
Normalizing so that we get a distribution, we obtain the probability mass
\[
P_{A,\Z}(i) = \left(\sum_{i=0}^{D} \frac{1}{i^{\Z}}\right)^{-1} \frac{1}{i^{\Z}}.
\]

It is by now accepted
that several naturally occurring distributions of importance are power laws,
see \cite{Mit04} for diverse examples. Equally importantly, several distributions approximate
power laws once their outliers are removed.  Several quantities stored in our
commercial database systems also follow power law distributions after similar
processing. Therefore, a study on a Zipfian parameter space is
important in and of itself. Secondly, the Zipfian distribution allows us to
vary just the right parameters --- namely skewness and the size of the unique
alphabet --- that are most germane to the performance of distinct value
estimators. 

In light of the discussion above, we choose to characterize our estimators
over a Zipfian parameter space, by varying the parameters of the Zipfian
population over a wide range of values that reflect real-world applications.
In this way, we get the ``best of both worlds" --- we are able to vary
parameters, understand estimator behavior as an absolute function of these
parameters, \emph{and} model real-world applications.

\subsection{Datasets and Protocol}

In order to obtain a fine grained characterization, we use the design of
(population size, alphabet size) pairs  as shown in
Table~\ref{tbl:pop-regimes}. Each such pair will be called a \emph{regime}.
Therefore, there are 20 regimes. 
As depicted, the regimes were organized into five \NoverD values in
$[10,20,100,500,1000]$.
\begin{table}[h] \footnotesize
\caption{The (population size, alphabet size) regimes used to generate Zipfian
populations. {\bf This table also indicates the layout of the grid on which
the 2D normalized bias
plots and 3D normalized bias surfaces are laid throughout the paper}.}
\centering
$
\begin{array}{c|ccccc}
& 1B,100M & 1B,50M & 1B,10M & 1B,5M & 1B,1M \\
\multirow{4}{*}{\begin{sideways}\parbox{15mm}\text{$N \rightarrow$}\end{sideways}}& 100M,10M & 100M,5M & 100M,1M & 100M,500K & 100M,100K \\
& 10M,1M & 10M,500K & 10M,100K & 10M,50K & 10M,10K \\
& 1M,100K & 1M,50K & 1M,10K & 1M,5K & 1M,1K \\
\hline
& \multicolumn{5}{c}{\text{ \NoverD $\rightarrow$}}
\end{array}
$
\label{tbl:pop-regimes}\end{table}
 
For each regime, we generated 5 Zipfian populations by varying the Zipfian
skewness parameter \Z through the range $[0,0.5,1,1.5,2]$ that covers most real-world
applications. In this way, we obtain 100 Zipfian
populations.  Note that at high skewness, the number of distinct
classes $D$ in the population is less than the size of the alphabet $A$. 
 
We refer to \NoverD as the \emph{average uniform class size} since it is the
average class size when $\Z=0$ (and up till the time that each alphabet occurs
in the population). We will see
that it is a critical parameter in characterizing estimator error.  

For each of the 100 Zipfian populations, we varied the sampling percentage
through the values $[0.1,1,2,5,10]$.  
For each data-point, we drew 10 random samples without replacement.   We ran our 11 distinct value estimators on each of the 10 samples, thereby generating 10 estimates for each of the 11 estimators. Finally, for each estimator, we computed the average bias of the 10 estimates, as well as the variance across the 10 estimates. 
In this way, we report a total of $100 \times 5 \times 10 \times 11 = 55,000$ experiments. 
We should note that in our study, we experimented with a strictly larger range of parameter values
than what is reported in this paper.  However, to conserve space, we
``compress" to a
subset of our range of parameter values that we feel adequately described
the error patterns.

\section{Results}
\label{sec:results}

Our extensive empirical study generated a large amount of data. Our goal is to
provide a thorough characterization of both, the individual and relative
performance of the estimators, and to organize and understand the error patterns that emerged
from our study.  By mining the generated data, we found that the
parameter \NoverD provides us with this organizing principle: \emph{when the results
of the experiments are arranged in a grid whose X-axis is \NoverD, and
Y-axis is \popsize, we can see regularity in the patterns}.  Accordingly, we
provide  grids of 2D plots of
estimate vs. actual distincts for 
all estimators in a family, as well as grids of 3D surfaces showing normalized bias
$(\est - D)/D$ for each individual estimator.  In both cases varying the
parameters of the underlying population as well as the sampling fraction. For
the 3D surfaces, normalized biases of -1, 0, and 3 are marked on the vertical
axes, and 1 and 2 can be seen in the form of dotted lines.  

We then point out the salient features of
both the relative and the individual behaviors as we vary the parameters of
the population. We observe patterns that arise when we go from left to right 
on each row of the 2D plots. This gives us the variations with the parameter
\NoverD. Likewise, we report variations with \popsize, within each plot with
\Z, and, finally, with \sample. 

To save space, we show only a single 2D
plot for each family: the one at which the maximum ratio error for the most
accurate estimator in the family is at most 5 (Table~\ref{tbl:max-avg:2-5}).
The remaining 2D plots are all included in supplementary material. 

We suggest that when reading the results, the
reader begin with the 3D surfaces for each estimator to understand its individual
performance, followed by inspection of the 2D plots (including those in the
supplementaries) to complete the relative picture.  Note that putting multiple
3D surfaces into a single diagram is not feasible.

We begin with the jackknife family.

\subsection{The Jackknife Estimators}

\begin{figure}[ptb]
\floatpagestyle{empty}
\centering
\subfloat[Grid of 3D bias surfaces for \jkone. ]{
\includegraphics[type=pdf,ext=.pdf,read=.pdf,bb=50 0 862 499,scale=0.55]{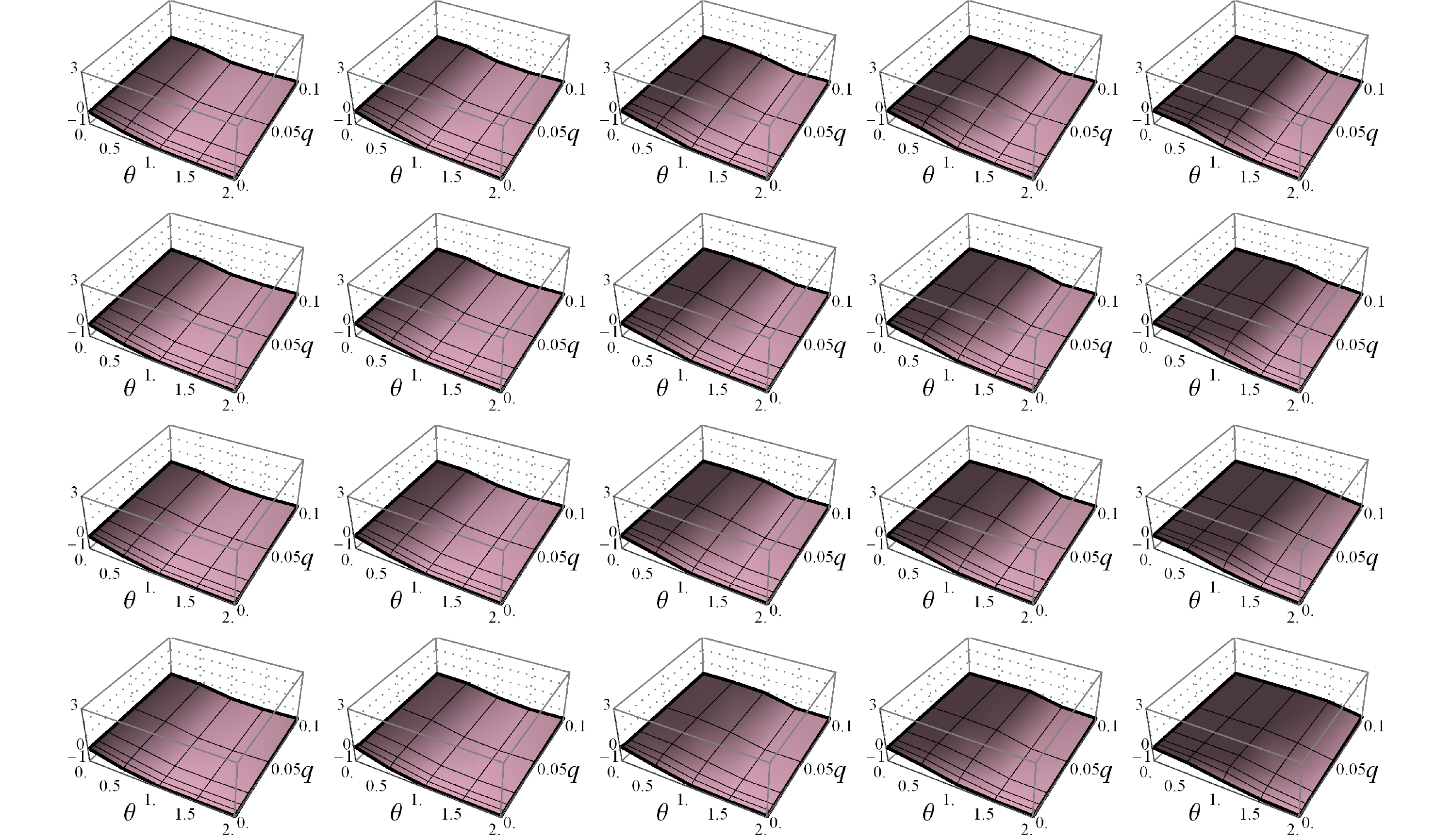}
\label{fig:jk1:grid}}\\
\subfloat[Grid of 3D bias surfaces for \jktwo. ]{
\includegraphics[type=pdf,ext=.pdf,read=.pdf,bb=50 0 862 499,scale=0.55]{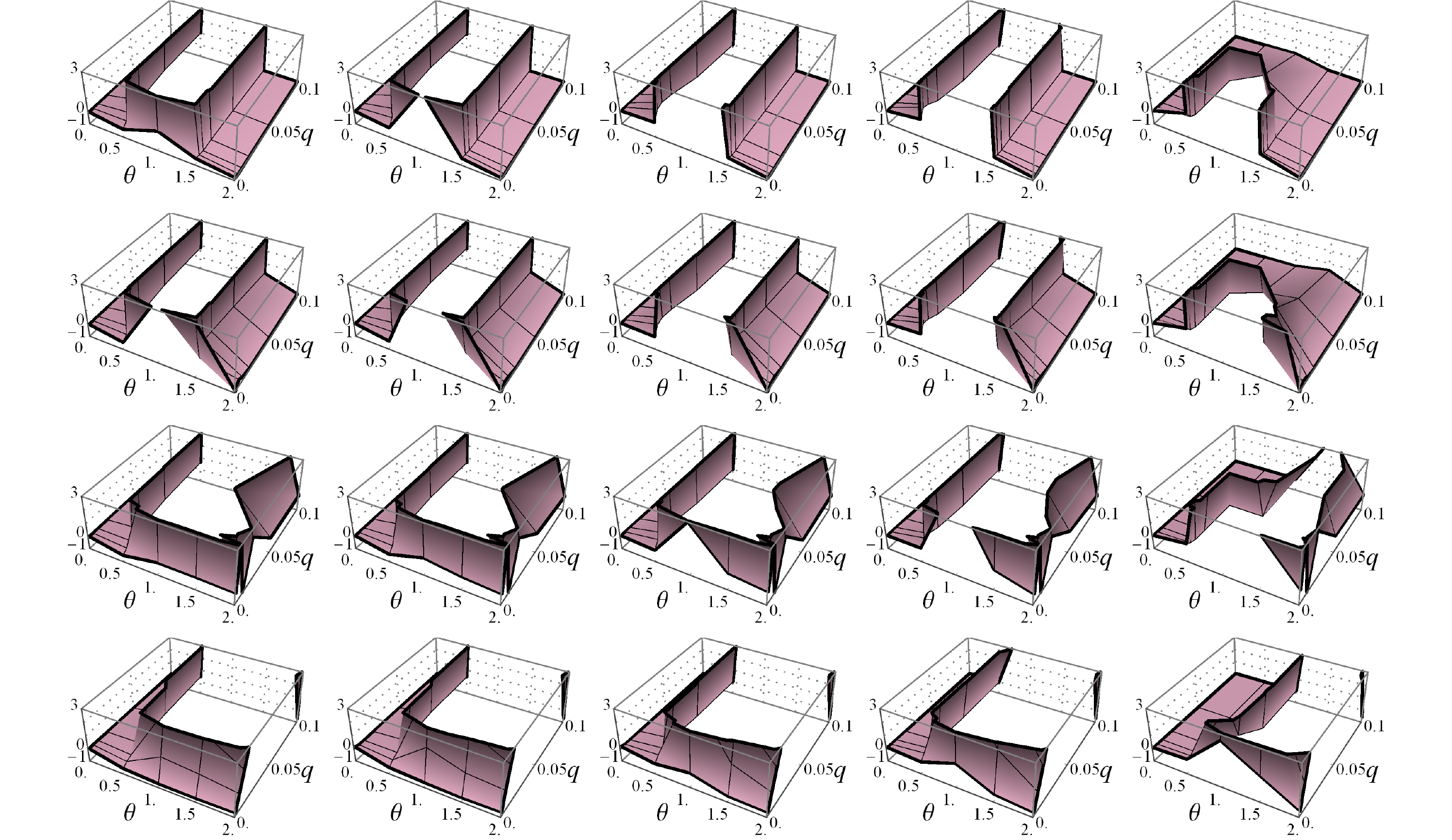}
\label{fig:jk2:grid}}
\caption{Notice that \jkone is a consistent
under-estimator of distinct values. Although second order, \jktwo has
considerably worse and irregular bias behavior as compared to \jkone. Recall
that the layout of the surfaces is on the grid indicated by
Table~\ref{tbl:pop-regimes}.}
\label{fig:jk1:jk2:grids}
\end{figure}

\begin{figure}[htb]
\centering
\subfloat{
\includegraphics[type=pdf,ext=.pdf,read=.pdf,bb=27 2 727
413,scale=0.6]{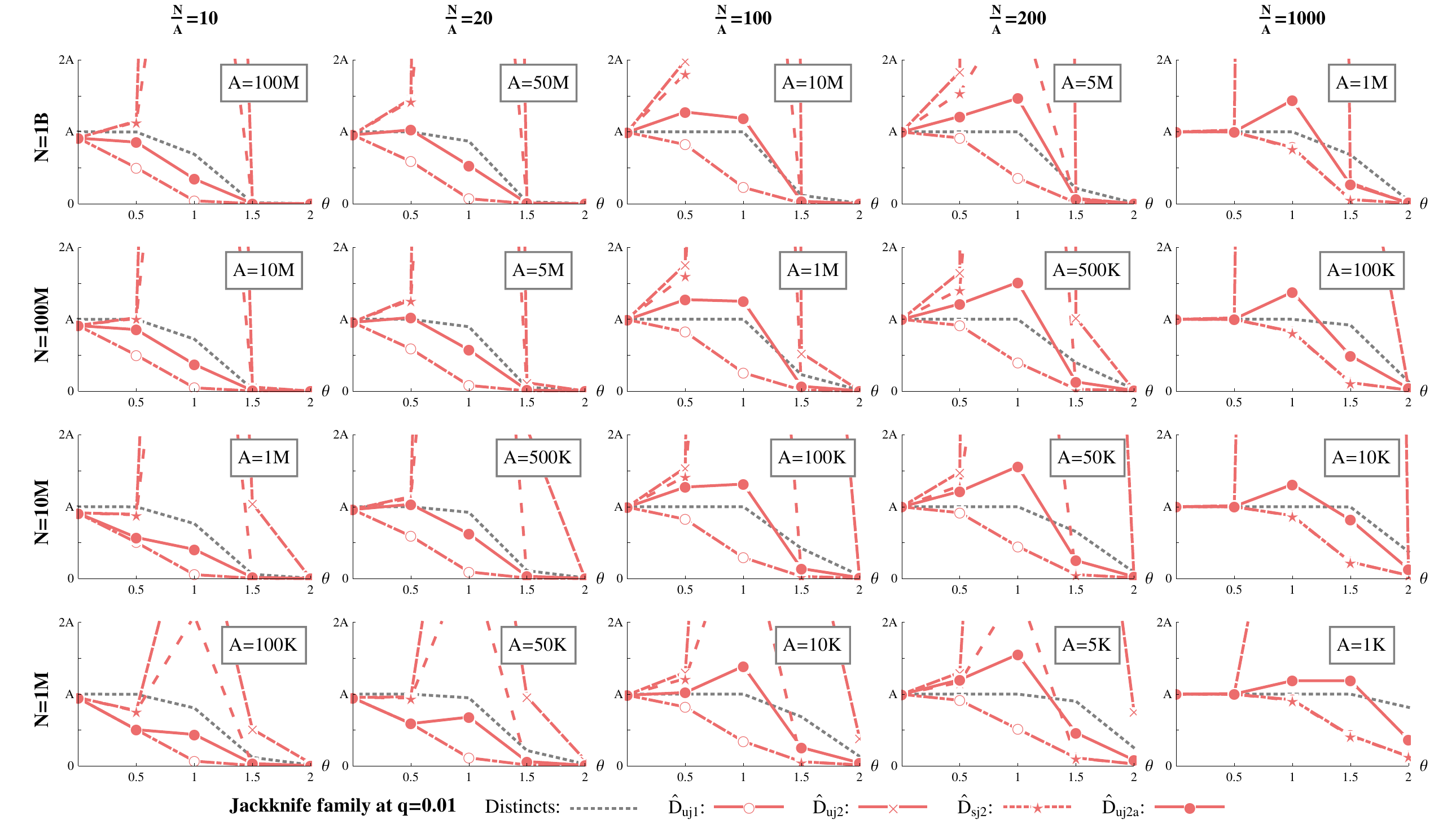}
}\caption{Relative behavior of the jackknife family  at $q=0.01$, which is the
lowest $q$ at which \jktwoa has a maximum error-ratio of 5 (see
Table~\ref{tbl:max-avg:2-5}). Note the
consistent underestimation of \jkone, 
severe mid-skew biases of \jktwo and \jktwos, and effect of stabilization in
\jktwoa.  }
\label{fig:jackknife:1percent}
\end{figure}

\paragraph{Variation with \NoverD:}

\jkone is fairly agnostic to changes in \NoverD, at all values of \sample.  
On the other hand, the positive bias of \jktwo shoots up as \NoverD increases.
The magnitude of the positive bias is highest at mid-skew.

For \jktwos, the bias 3D surface illustrates best the 
severe positive bias at mid-skew. Importantly, the bias magnitude, as also the
region, reduces as \NoverD increases.  In other words, smoothing is more
effective as \NoverD increases. At $\NoverD = 10$, we need considerably
over 10\% sampling to get accurate estimates. 
As \NoverD increases, the
least sampling fraction \sample required to get accurate estimates reduces.

The stabilized estimator \jktwoa is quite accurate overall. 
There is a jump between low and high skew estimates that increases with \NoverD when  $\sample < 0.05$. 
There is a bump at mid-skew for low \sample and high \NoverD; it  
flattens out at lower \NoverD.

\paragraph{Variation with \popsize keeping \NoverD fixed:}

\jkone is fairly agnostic of \popsize. On the other hand, the accuracy of
\jktwo at
high skew varies with \popsize: \jktwo is more accurate at high skew at high
\popsize, and inaccurate at low \popsize.   \jktwos is slightly worse at mid-skew as
\popsize increases.  Finally, \jktwoa is fairly agnostic to \popsize.

\paragraph{Variation with \Z:}

As is known \cite{BF93}, \jkone is a consistent underestimator across all \Z. 
We observe that the  negative bias is the worst at mid-high skew.
The bias profile of \jktwo is best described as ``hat-shaped". 
There is severe positive bias at mid-skew.  There is slight positive bias at
low-skew, and moderate to high positive bias at high-skew up to $N =100M$. 
At $N > 100M$, \jktwo becomes an underestimator at high-skew.
The positive bias at high-skew for low \popsize gets worse as we increase \sample. 
\jktwos shows the same pattern of positive bias as \jktwo, but less prominently owing to the smoothing. 
Smoothing also delays the poor performance at high \NoverD. 

\jktwoa has a crossover in the form of a reflected ``S" shape as we increase
skew. In other words, it overestimates at mid skew and underestimates at high
skew. See also Fig.~\ref{fig:flat:grids}.  
At high \NoverD, the bias goes from positive to negative as skew increases, whereas 
at low \NoverD, it remains largely positive, with some overestimation  at high \sample.

\begin{figure}
\floatpagestyle{empty}
\centering
\subfloat[Grid of 3D bias surfaces for \jktwos]{
\includegraphics[type=pdf,ext=.pdf,read=.pdf,bb=50 0 862 499,scale=0.55]{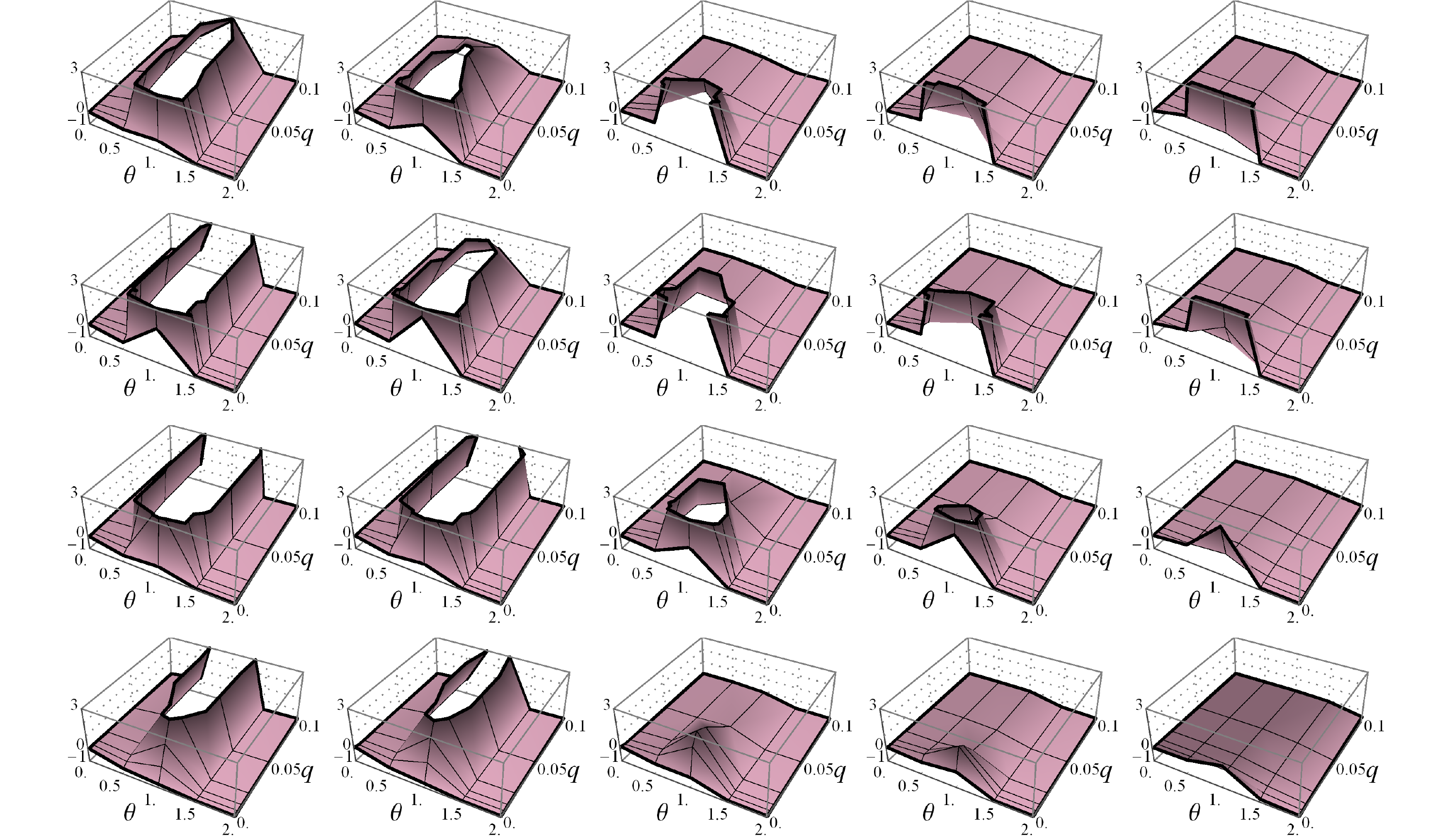}
\label{fig:jk2s:grid}}\\
\subfloat[Grid of 3D bias surfaces for \jktwoa]{
\includegraphics[type=pdf,ext=.pdf,read=.pdf,bb=50 0 862 499,scale=0.55]{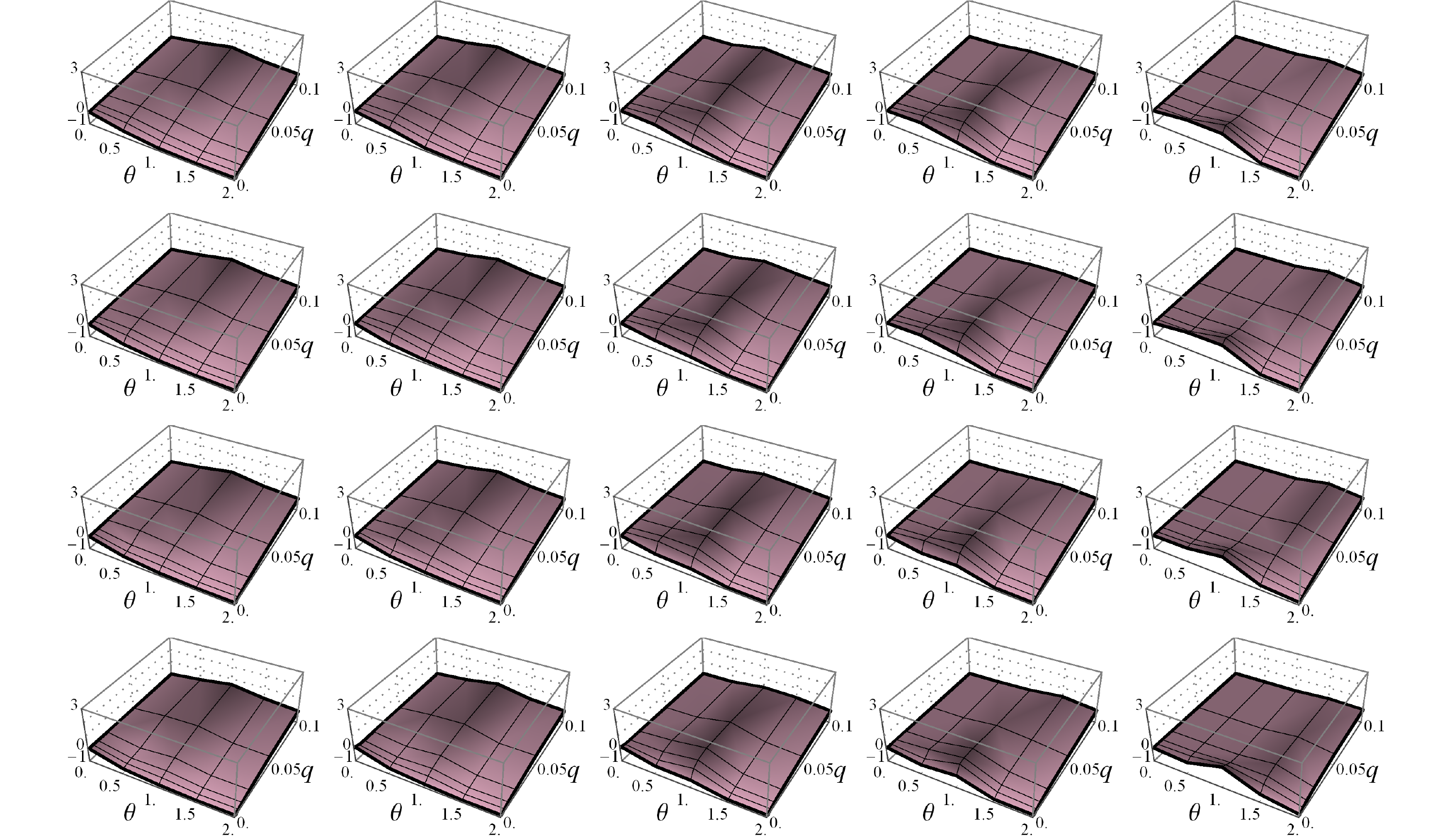}
\label{fig:jk2a:grid}}
\caption{Effect of smoothing versus stabilization can be seen clearly: the
stabilized \jktwoa is one of the most consistently accurate estimators, while
the smoothed \jktwos is suffers severe mid-skew bias at low \NoverD. Note the
effect of \NoverD versus that of \popsize in \jktwos. }
\label{fig:jk2s:jk2a:grids}
\end{figure}

\paragraph{Variation with \sample:}

As we might expect, \jkone reduces its negative bias as \sample increases. 
\jktwo, at high \NoverD, improves dramatically in mid-skew as \sample increases above 5\%. 
Whereas, at low \NoverD, increasing \sample does not help. 

For \jktwos increasing \sample helps more at high \NoverD (100 or greater).
For $\NoverD = 100$, we  need $q > 10\%$, for $\NoverD = 200$ this drops to $q
>  5\%$, and for $\NoverD =1000$, we need only $q > 1\%$ for acceptable 
estimates.  \jktwoa does well at all sampling fractions and improves accuracy
with \sample.  At high
Zipfian skew of $\Z \geq 1.5$, we need only $\sample \geq 0.005$ for accurate
estimates. 

\paragraph{Anomalies:}

The bias of \jktwo is lower at $\sample=0.001$ as compared to higher values of \sample in the vicinity.
The smoothing technique of \jktwos does well at low $\NoverD \sim 10$, and very low sampling (\sample = 0.001), but then exhibits poor performance as either of these parameters increases.

\subsection{The Schlosser Estimators}
\label{sec:results:schlosser}

\begin{figure}
\floatpagestyle{empty}
\centering
\subfloat[Grid of 3D bias surfaces for \sh ]{
\includegraphics[type=pdf,ext=.pdf,read=.pdf,bb=50 0 862 499,scale=0.55]{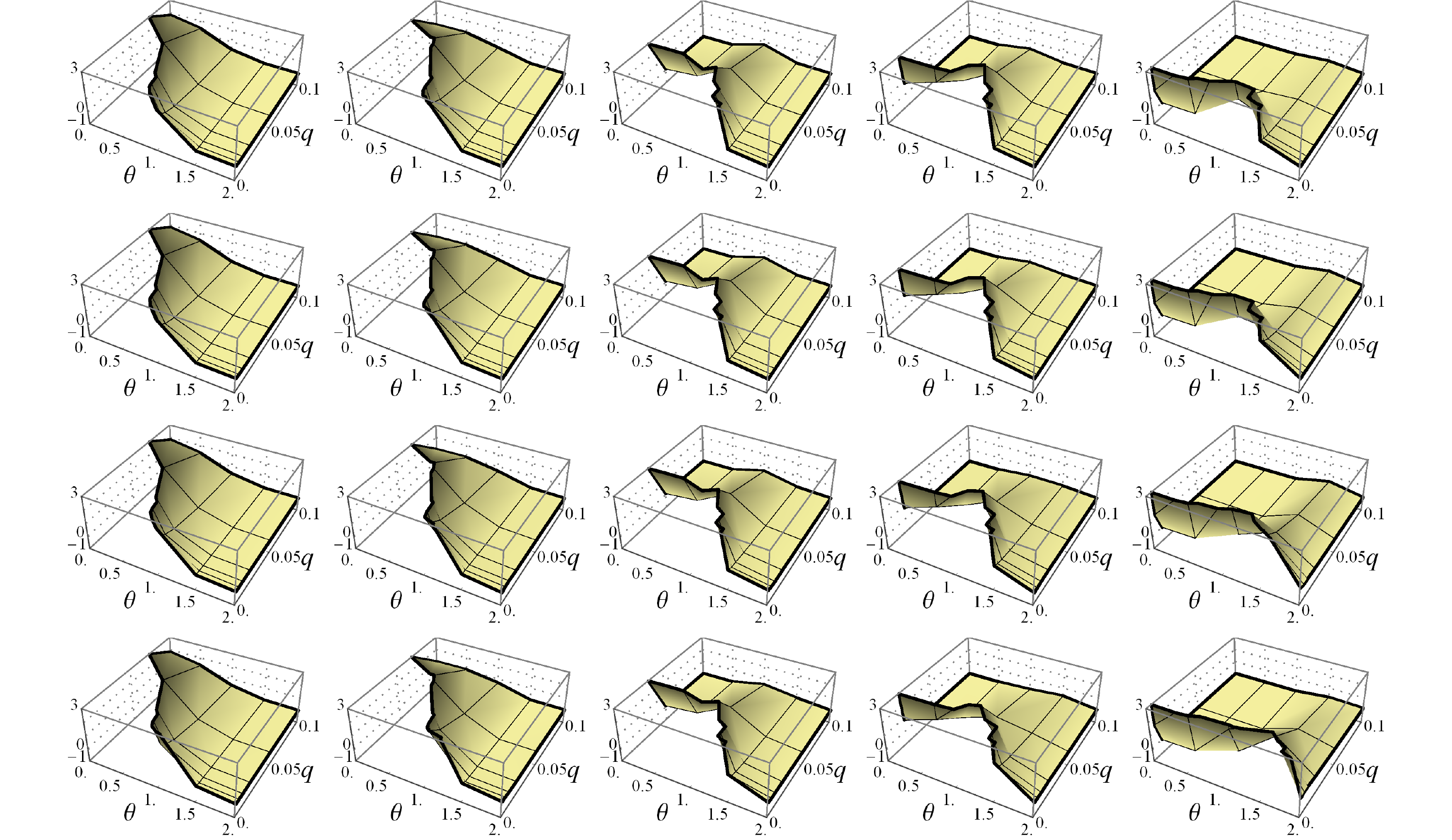}
\label{fig:sh:grid}}\\
\subfloat[Grid of 3D bias surfaces for \shtwo ]{
\includegraphics[type=pdf,ext=.pdf,read=.pdf,bb=50 0 862 499,scale=0.55]{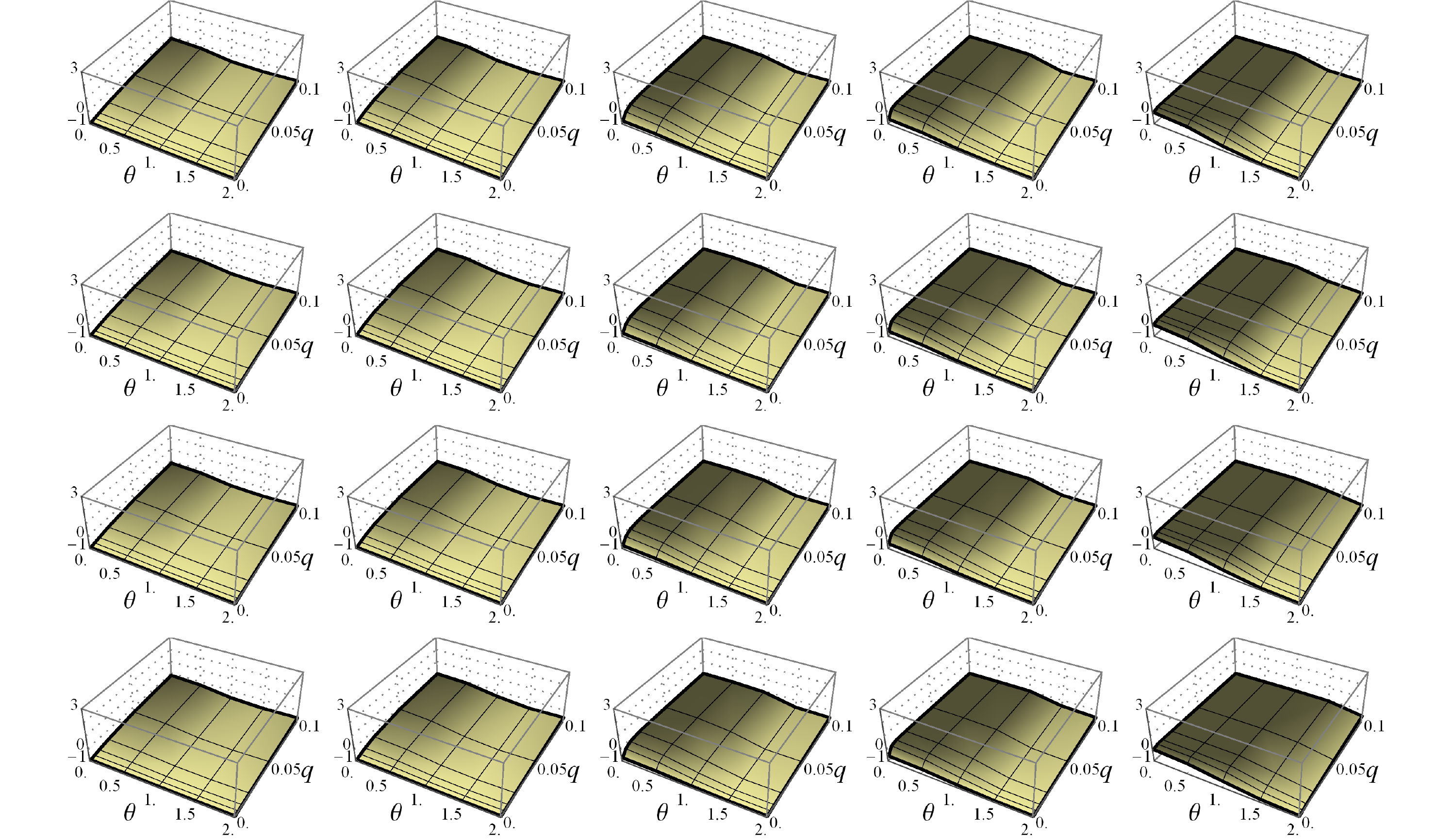}
\label{fig:sh2:grid}}
\caption{\sh shows severe positive bias at lower skew, that is corrected only
with high sampling fractions. \shtwo does not suffer from this problem. Note
the highly regular behavior of both, especially clear with \sh, as a function of \NoverD alone.}
\label{fig:sh:sh2:grids}
\end{figure}

\paragraph{Bias:}
The Schlossers have the general bias profile
$\sh > \shthree > \uec > \shtwo$:  the exception being at low \NoverD and low \sample
where \sh and \shthree underestimate, but only slightly.

\paragraph{Variation with \NoverD:}
For \sh and \shthree, the bias curve is a slope at lower \NoverD and becomes a ``hat" at higher
\NoverD.
The value of \NoverD at which this shape transition happens reduces as \sample increases.
For $\sample = 0.001$, it happens after $\NoverD > 1000$, at $\sample =
0.005$, it happens for \NoverD between 1000 and 200, and for $\sample =0.02$, it
occurs for \NoverD between  200 and 100. 
Increase in \sample improves accuracy as \NoverD increases, with this
improvement manifesting at low skewness.

\shthree is reasonably accurate at $\sample \geq 0.005$. 
\sh requires $\sample \geq 0.01$ for acceptable estimates. 
As \sample increases, the \NoverD at which the accuracy is attained for low
skew reduces, and the range of low-skew where the accuracy is attained also
increases. For low $\NoverD \sim 10$, low skew estimates remain intolerably poor
until we raise the sampling fraction to  $\sample = 0.1$. 

The predominant effect of increasing \NoverD on all three is that the lower skew estimate
becomes reasonably accurate. See also Fig.~\ref{fig:flat:grids} for this
effect in \shtwo.   

\paragraph{Variation with \popsize:}

There is little change in the shape of the bias surfaces as we increase \popsize, keeping \NoverD fixed.

\begin{figure}[htb]
\centering
\includegraphics[type=pdf,ext=.pdf,read=.pdf,bb=50 0 862 499,scale=0.55]{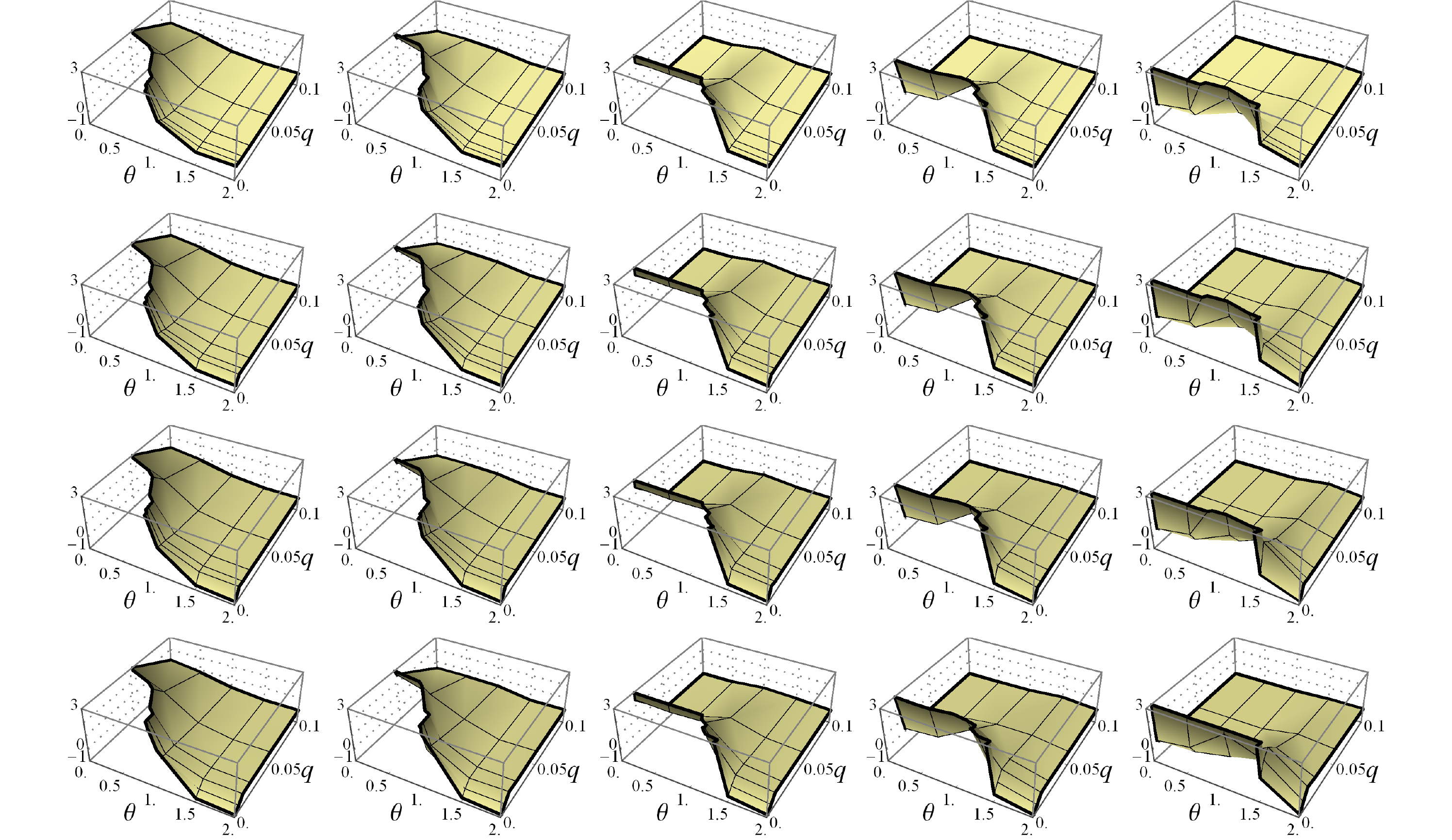}
\caption{Grid of 3D bias surface for \shthree. Like \sh, there is severe positive bias at low skew. Note again
the highly regular behavior of \shthree as a function of \NoverD alone.}
\label{fig:sh3:grid}
\end{figure}

\begin{figure}[htb]
\centering
\subfloat{
\includegraphics[type=pdf,ext=.pdf,read=.pdf,bb=27 2 727
413,scale=0.6]{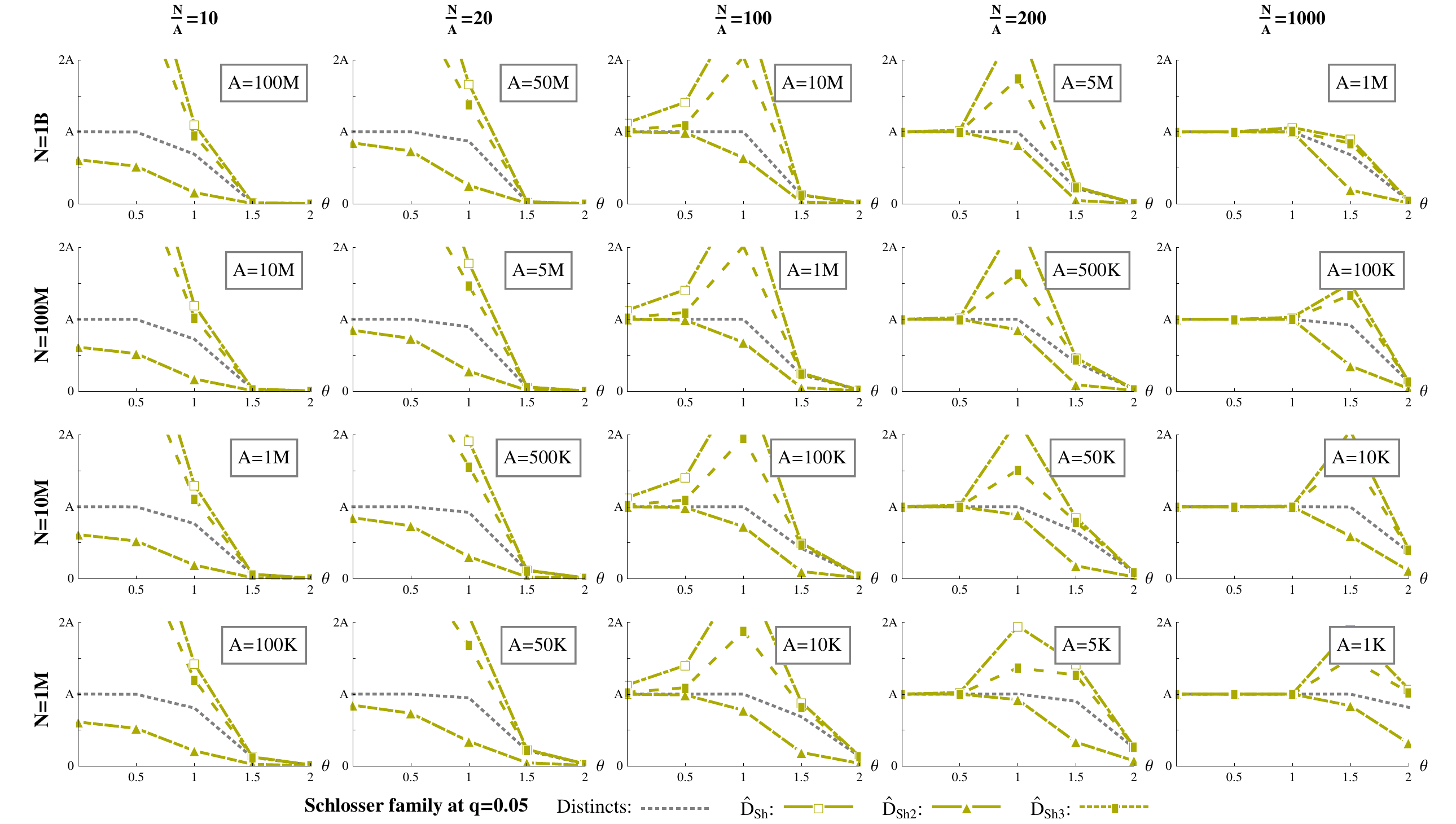}
}\caption{Relative behavior of the Schlosser family at $q=0.05$, which is the
lowest $q$ at which \shtwo has a maximum ratio-error of 5. Even at a
relatively high sampling fraction, \sh and \shthree show severe positive bias
at low-mid skew. In order to use these estimators at low-mid skew, $q \geq
0.1$ is required.}
\label{fig:schlosser:5percent}
\end{figure}

\paragraph{Variation with \Z:}

\sh and \shthree are extremely accurate for high-skew. 
When the bias curve becomes a hat, as described earlier, the estimator is accurate at low skew also (pl. see variation with \NoverD for discussion of when it becomes a hat). 
When this happens, only mid-skew is overestimated. \shtwo has a negative bias
at low-mid skew, but is not as extreme  as \sh and \shthree.   

\paragraph{Variation with \sample:}

This family is very sensitive to \sample, and shows monotonic improvement in
accuracy (which means lesser positive bias) as \sample increases.  
In the range of 5 - 10\% sampling, all three estimators begin giving accurate
estimates for all \NoverD, all skew, and all \popsize, giving a 
maximum ratio error of $\sim 5$. Of these, \shthree is requires the lowest $q$
to provide acceptable estimates across the range of skew (see
Table~\ref{tbl:max-avg:2-5}.

\subsection{\gee and \ade}

\begin{figure}
\floatpagestyle{empty}
\centering
\subfloat[Grid of 3D bias surface for \gee]{
\includegraphics[type=pdf,ext=.pdf,read=.pdf,bb=50 0 862 499,scale=0.55]{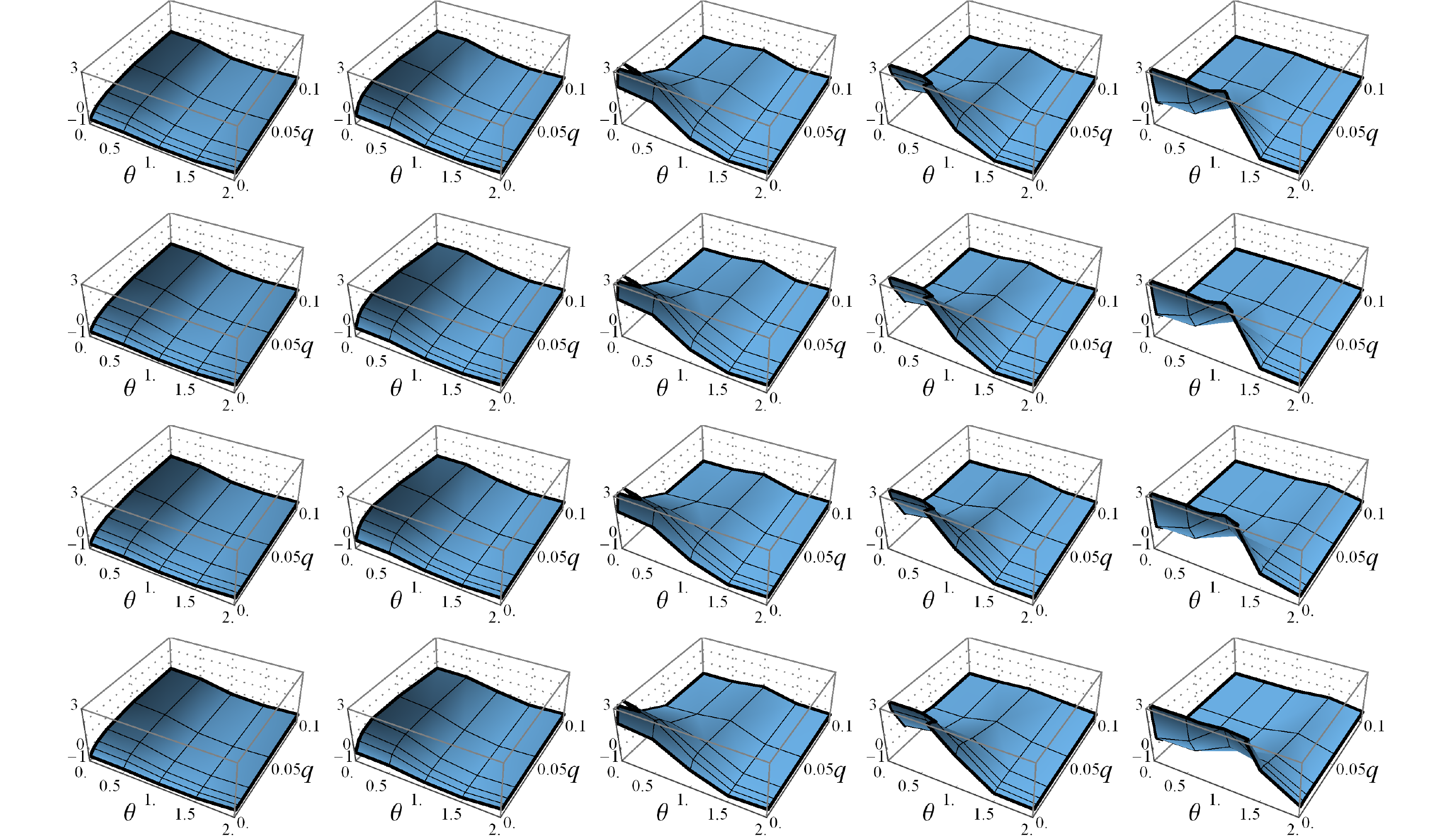}
\label{fig:opt:grid}}\\
\subfloat[Grid of 3D bias surface for \ade]{
\includegraphics[type=pdf,ext=.pdf,read=.pdf,bb=50 0 862 499,scale=0.55]{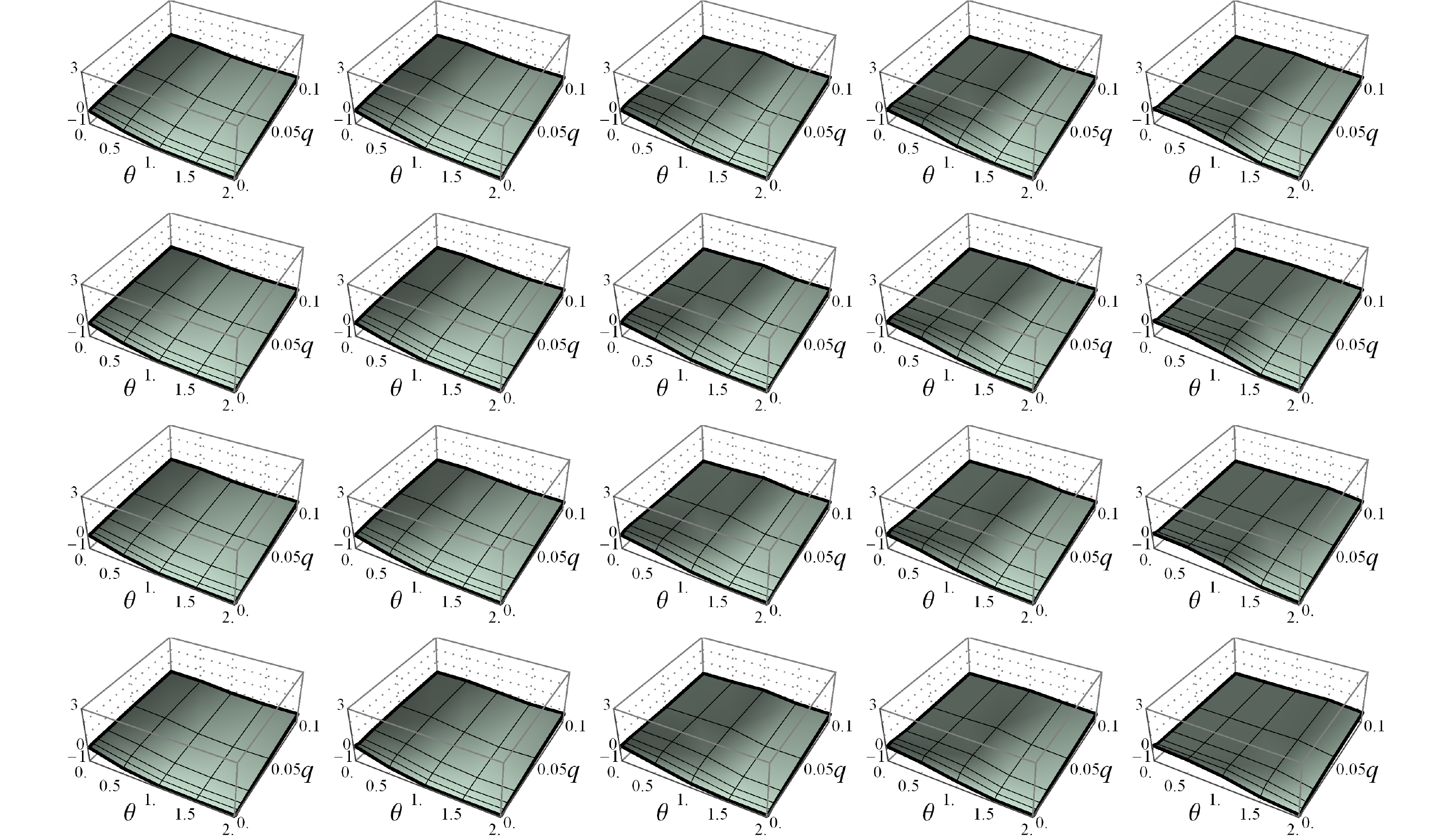}
\label{fig:ae:grid}}
\caption{\gee and \ade are among the most consistent estimators. \gee does
have a region of positive bias at high \NoverD, where it should not be used in
preference to \ade and \jktwoa. Both estimators show highly regular behavior
as a function of \NoverD. }
\label{fig:opt:ae:grids}
\end{figure}

\paragraph{Variation with \NoverD:}

At low $\NoverD \sim 10$, both \gee and \ade underestimate but both are
reasonably accurate. As we increase \NoverD, this picture changes
significantly for \gee, but not for \ade (see Fig.~\ref{fig:opt:ae:grids}. 
\ade shows considerably less change w.r.t. \NoverD. The change of \gee with \NoverD is described next. 
There is a change from a slope to a ``hat", similar to \sh and \shthree, as
\NoverD rises.  However, the degree of overestimation is not comparable the
Schlossers. The ratio errors for the worst case positive bias is less
than 10, compared to over 200 for Schlossers. 
The transition to  ``hat" happens at $\sample = 0.005$ at \NoverD between 1000
and 200; at $\sample = 0.02$ between 200 and 100; and at $\sample = 0.1$
between 100 and 20. 

\paragraph{Variation with \popsize:}

There is little change with \popsize except for some reduction in positive bias
for high-skew for \gee with \popsize.

\paragraph{Variation with \Z:}

\begin{figure}
\floatpagestyle{empty}
\centering
\subfloat[\jkone]{
\includegraphics[type=pdf,ext=.pdf,read=.pdf,bb=100 0 972
191,scale=0.55]{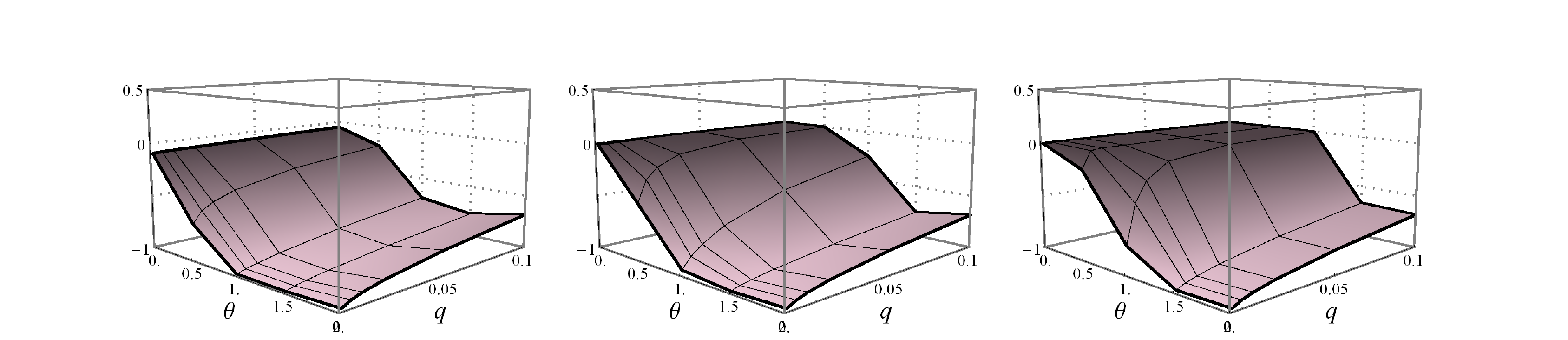}
\label{}}\\
\subfloat[\jktwoa]{
\includegraphics[type=pdf,ext=.pdf,read=.pdf,bb=100 0 972
191,scale=0.55]{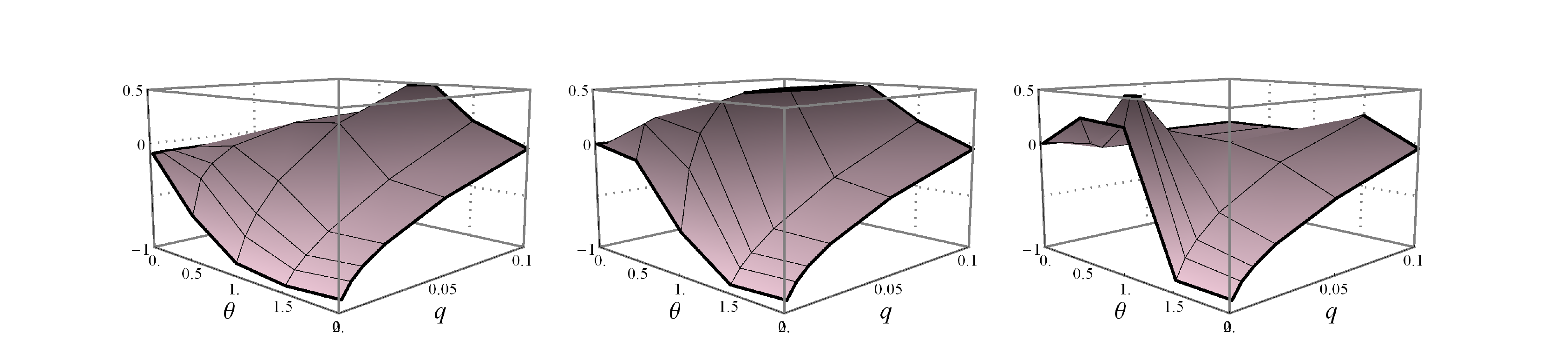}
\label{}}\\
\subfloat[\shtwo]{
\includegraphics[type=pdf,ext=.pdf,read=.pdf,bb=100 0 972
191,scale=0.55]{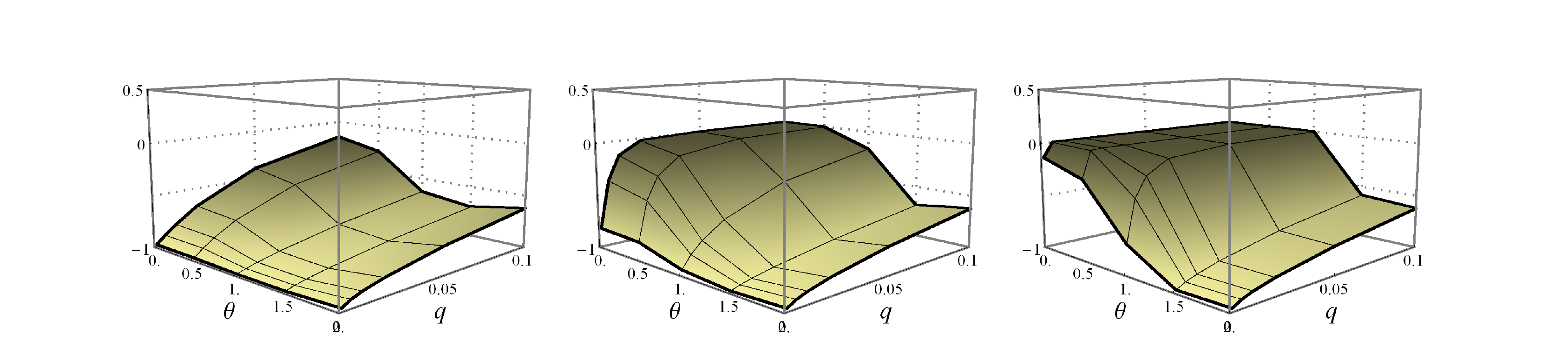}
\label{}}\\
\subfloat[\ade]{
\includegraphics[type=pdf,ext=.pdf,read=.pdf,bb=100 0 972
191,scale=0.55]{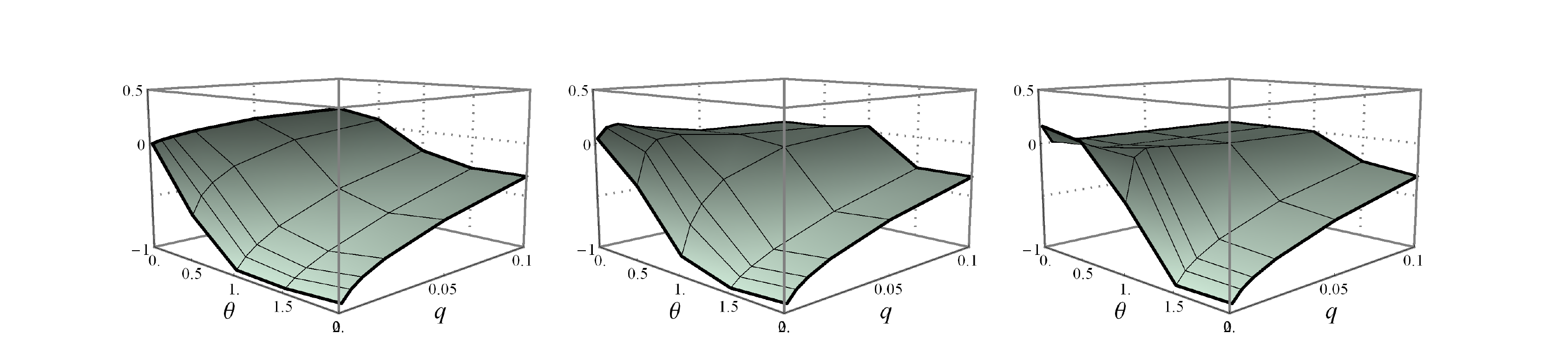}
\label{}}
\caption{The ``flatter" bias surfaces of \jkone, \jktwoa, \shtwo, and
\ade shown with more detail. Only $\popsize = 1B$, and $\NoverD \in [10, 100, 1000]$
shown. Note also that these estimators are relatively agnostic to \popsize,
and show regularity in variation with \NoverD.}
\label{fig:flat:grids}
\end{figure}

\gee is very good at high skew. 
When the transition to ``hat" shape happens,  it becomes good at low skew as
well (see previous discussion of when this happens). \gee is a mid-skew
overestimator (except at very low \sample, where it underestimates). 
\ade is accurate at both high and low skew Zipfian populations, with a
tendency towards negative bias.  See also Fig.~\ref{fig:flat:grids}.  

Since both \gee and \ade appear in the grid of 2D bias plots for the top
three estimators in Sec~\ref{sec:discussion}, we do not show their 2D bias
plots here. Of course, all the 2D bias plots are available in the
supplementaries. 

\paragraph{Variation with \sample:}

Both estimators show monotonic improvement in accuracy (reduction in positive
bias) for increase in \sample. For \gee, at $\sample=0.005$ worst case ratio
error is 5, at $\sample =0.01$, it drops to  4, at $\sample = 0.05$, it is
less than 2.

\paragraph{Anomalies:}
When changing \NoverD from 20 to 100, there is a sudden increase in \gee
positive bias up to $\sample = 0.01$.

\subsection{The Chao-Lee Estimators}

\begin{figure}
\centering
\includegraphics[type=pdf,ext=.pdf,read=.pdf,bb=40 2 862
121,scale=0.5]{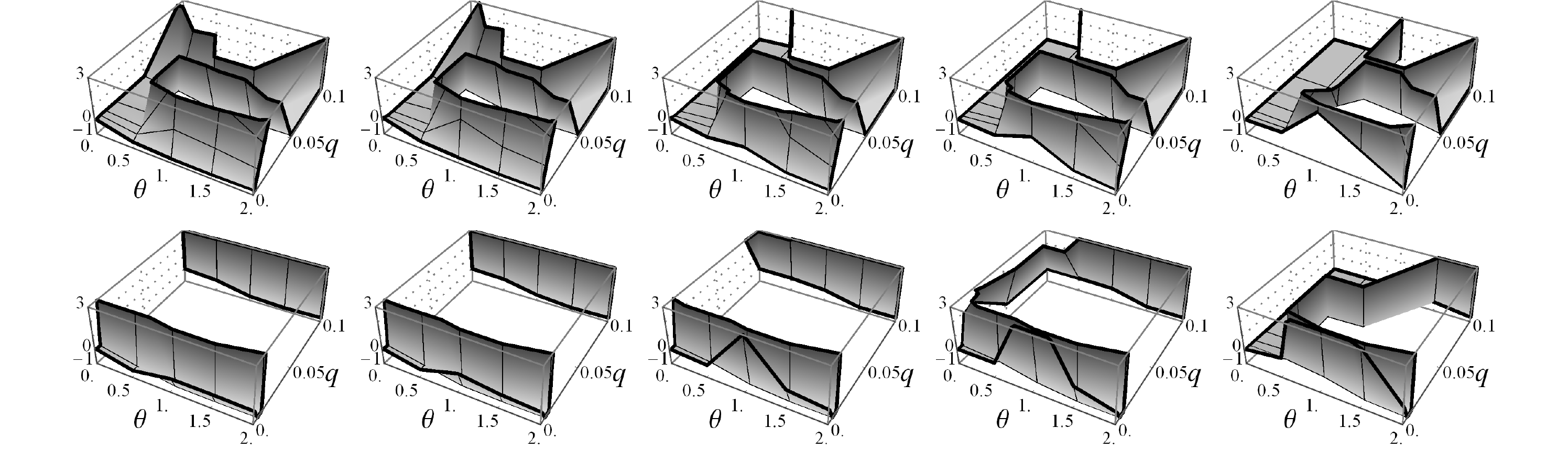}
\caption{The two Chao-Lee estimators show highly irregular behavior that
becomes extreme even at $\popsize = 10M$. Therefore  we show only 1M and 10M
population sizes for \cltwo. The remaining 3D surfaces for \cltwo are available in supplementaries.}
\label{fig:cl1:cl2:1m:grids}
\end{figure}

The Chao-Lee estimators show highly erratic behavior, and perform reasonably
only in following regions. 
\begin{multicols}{2}
\begin{enumerate*}
\item At \sample=0.001, \popsize =1M, 10M
\item At \sample=0.005, \popsize =10M
\item At \sample=0.001, \popsize =100M
\item At \sample=0.5, \popsize =1M and 100M
\item At \sample=0.1, \popsize =10M
\end{enumerate*}
\end{multicols}

The variations described below only pertain to the above regions.\\ 
\noindent{\bf Variation with \NoverD:}
Not much change except at q = 0.001. \\
\noindent{\bf Variation with \popsize:}
 Discontinuity is the predominant effect.\\
\noindent{\bf Variation with \Z and \sample:}
 When $\sample > 0.005$, both underestimate in mid-high skew.

\section{Discussion}
\label{sec:discussion}

The discussion is organized as follows. First we describe the relative
sensitivity of each estimator to the various parameters. Then we address the
question ``which are the best estimators?" in terms of accuracy
over the entire parameter space of our characterization. Next, we identify
regions of the parameter space where certain estimators do well, even though
they may not perform well over the entire parameter space. Finally, we address
the question ``how much sampling do we need?"

\subsection{Sensitivity to Parameters}

\paragraph{Sensitivity to \NoverD:}

The grids of 3D bias surfaces clearly show that every family --- the
Schlossers, jackknives, \gee and \ade, and Chao-Lee --- is sensitive to changes
in \NoverD, and all except the last show regularity in their behaviour as a
function of \NoverD. The parameter \NoverD emerges as the single most
important organizing parameter for estimator behavior overall. The sensitivity
to \NoverD is high in \sh, \shthree,
\jktwo, \jktwos, \gee, \clone, and \cltwo. 
Compared to the above, there was mild sensitivity in \jkone, \jktwoa, \shtwo, and \ade. Therefore,
even within a family, some members are highly sensitive to changes in
\NoverD, while others are not.

\paragraph{Sensitivity to Scale \popsize:}

Among the 11 estimators we tested, \jktwos, and the two Chao-Lee estimators were
highly sensitive to scale. Namely, as we go up vertically along their grids,
their 3D bias surfaces showed
significant changes.  Of these, the change was quite regular and predictable
in \jktwos, but irregular and unpredictable in
the Chao-Lee families. In the case of \jktwos, the bias increases as we
increase the scale, with the increase being most prominent around mid-skew as
can be seen from the 3D bias surfaces for \jktwos (Fig~\ref{fig:jk2s:grid}). 
See supplementaries for complete Chao-Lee estimator grids.

\paragraph{Sensitivity to Sampling Fraction \sample:}

The 2D plots (see supplementaries) are useful to illustrate the effect of
sampling fraction.  
The estimators that improve most as \sample increases are \sh,
\shthree, \gee.  
The estimators that are relatively less sensitive to increases in \sample in our
range, for some values of other parameters, 
are \jktwoa and \ade.  For example, both of them remain relatively accurate
for low skew even at low sampling fractions
(see Tables~\ref{tbl:ratio-error} and \ref{tbl:bias}). 

The estimators that show anomalous or irregular behavior as \sample is
increased are \jktwo, \jktwos, and the Chao-Lee family. In the case of \jktwo
and  \jktwos, we see
anomalous degradation in performance as we go from \sample = 0.001 to 0.005,
especially when $\NoverD < 200$, see 2D plots in supplementaries.

\paragraph{Sensitivity to Zifpian Skew \Z:}

Sensitivity to \Z can be seen more finely in the 2D plots (see
supplementaries). 
The estimators \sh, \shthree, \jktwo, \jktwos, and \gee are highly sensitive to
changes in Zipfian skew.  In particular, \sh and \shthree are quite inaccurate
at low skew. \jktwo and \jktwos perform poorly in mid-skew, while \jktwo also
performs poorly for smaller populations and high skew, even at high sampling
fractions. 

The estimators \shtwo, \jkone, \jktwoa, and \ade are relatively insensitive to
changes in Zipfian skew. Finally, \clone and \cltwo respond irregularly to changes in Zifpian skew.

\subsection{The Best Estimators Overall and their Relative Performance}

From our extensive study, we can conclude that three estimators provide
relatively strong
performance across variations in all underlying parameters. These three are 
\gee, \ade, and \jktwoa (cf. the choice of the provisional estimator in
\cite{BF93}, which is \jktwos). It is perhaps easiest to see this from
Table~\ref{tbl:ratio-error}, and inspect the low and high sampling cases
separately. Of course, which estimator is to be used for an
application depends on the sampling fraction that is available (we return to
this question in $\S.~\ref{sec:sampling-fraction})$, and the skewness of the
population in case it is known. A subtler issue is whether it is the ratio
error that is critical to the application, or the actual value and sign of the bias,
and the role played by \NoverD. 
For example, when the number of distincts is
relatively low,  even large ratio errors do not result in high
absolute value of bias.  Therefore, we
cannot speak meaningfully about ``best estimator" in terms of just ratio error
or bias --- we do need to include the \NoverD factor as well. 
Since in certain database query
optimizers, it may be the value and sign of bias that is the critical factor in change
of a query
plan, while in others, it may be the ratio error, practitioners will find it useful
to have an analysis along both metrics. In both cases, we discuss the low
sampling scenario ($ q < 0.01$) below since that is where differences may be most
manifest.  
 
\begin{figure}[htb]
\centering
\subfloat{
\includegraphics[type=pdf,ext=.pdf,read=.pdf,bb=27 2 727 413,scale=0.6]{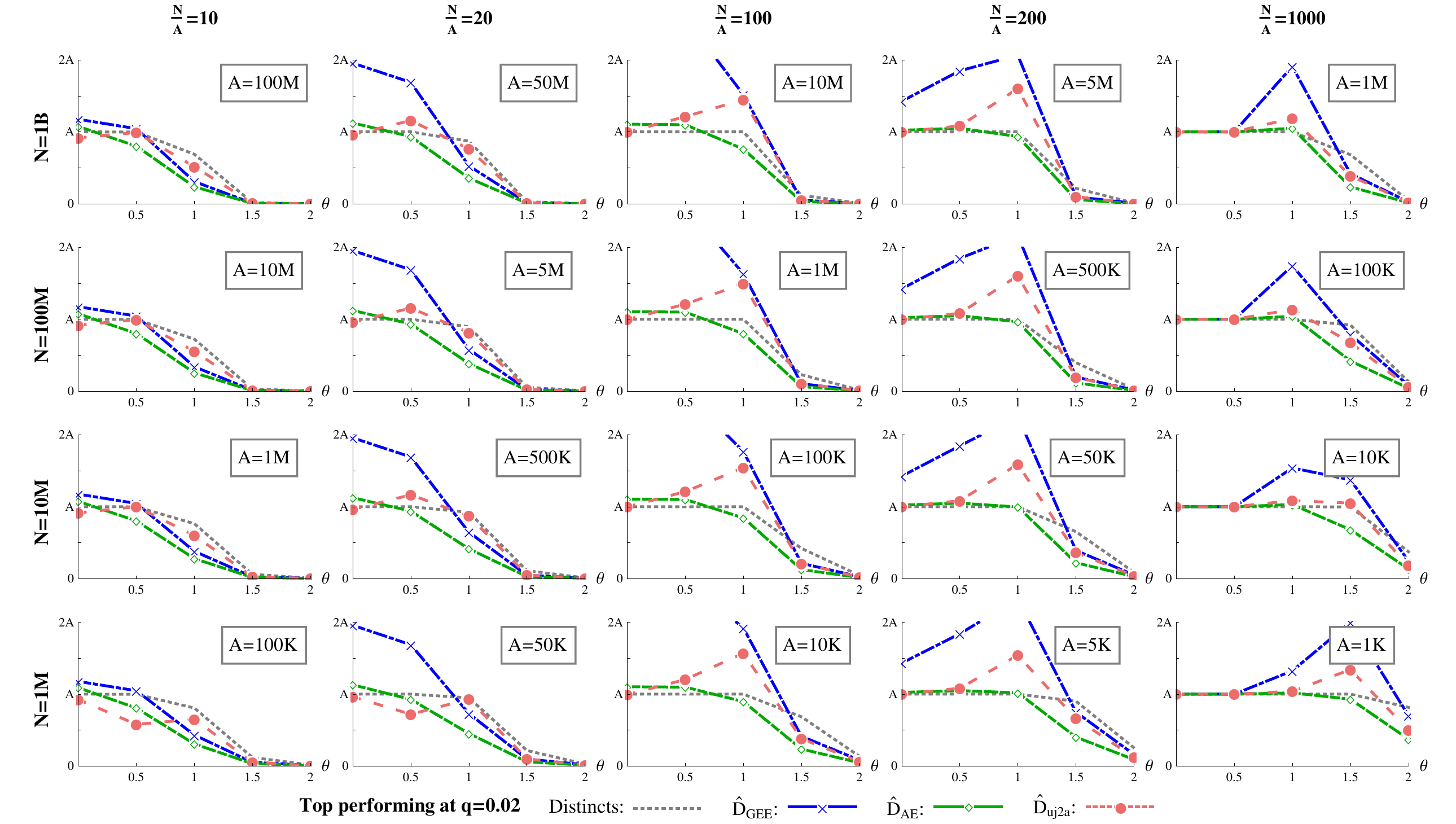}
}\caption{The three most consistent estimators at $q=0.02$, which is the lowest $q$ at
which \ade has maximum ratio error at most 5, see Table~\ref{tbl:max-avg:2-5}
(although \jktwoa offers the same at $q=0.01$). Note that at mid-high 
\NoverD there is a region of significant low-mid skew bias for \gee, described in text.}
\label{fig:top3:2percent}
\end{figure}

\begin{table*}[p]\scriptsize
\caption{Ratio-error vs. \Z and \NoverD, for low
and high sampling fractions. Note that the high and low ranges for \NoverD
overlap on the mid-value of $\NoverD =100$. } \label{} \centering 
\subfloat{
$
\begin{array}{|cc|cccc|cccc|}

\hline%
& & \multicolumn{4}{c|}{\text{$0.001 \leq \sample \leq 0.005$}} &
\multicolumn{4}{c|}{\text{$0.01 \leq \sample \leq 0.1$}} \\
\cline{3-10}
\text{Skew} & \text{\NoverD} & \text{\jkone} &
\text{\jktwo} & \text{\jktwos} &
\text{\jktwoa} & \text{\jkone} &
\text{\jktwo} & \text{\jktwos} & \text{\jktwoa} \bigstrut[t]
\bigstrut[b] \\
\hline%

0\leq \theta \leq 1 & \NoverD\leq 100 & 12.99 & 8.24 & 3.33 & 2.43 &
2.69 & 28.23 & 4.16 & 1.23 \bigstrut[t]\\

1.5\leq \theta \leq 2 & \NoverD\leq 100 & 38.56 & 5.8 & 38.69 & 9.24 & 8.06 & 8.88 & 8.14 & 2.07\\

0\leq \theta \leq 1 & \NoverD \geq 100 & 3.04 & 19.89 & 3.46 & 1.28 & 1.22 & 19.26 & 1.63 & 1.13 \\

1.5\leq \theta \leq 2 & \NoverD \geq 100 & 27.8 & 6.57 & 28.5 &
7.21& 6.23 & 9.46 & 6.27 & 1.83 \bigstrut[b]\\
\hline
\end{array}
$
}

\subfloat{
$
\begin{array}{|cc|ccc|cc|ccc|cc|}

\hline%
& & \multicolumn{5}{c|}{\text{$0.001 \leq \sample \leq 0.005$}} &
\multicolumn{5}{c|}{\text{$0.01 \leq \sample \leq 0.1$}} \\
\cline{3-12} \text{Skew}
& \text{\NoverD} & \text{\sh} & \text{\shtwo} & \text{\shthree}  & \text{\gee}
& \text{\ade} & \text{\sh} &
\text{\shtwo} & \text{\shthree} & \text{\gee} & \text{\ade}
\bigstrut[t] \bigstrut[b] \\
\hline%

0\leq \theta \leq 1 & \NoverD \leq 100 & 19.39 & 25.57 & 16.96 &
2.49 & 4.17 & 5.41 & 3.11 & 4.18 & 1.62 & 1.41 \bigstrut[t] \\
                                                                                                                  
1.5\leq \theta \leq 2 & \NoverD \leq 100 & 1.15 & 36.32 & 2.53 & 2.69 & 14.29& 1.07 & 6.86 & 1.06 & 1.93 & 3.09 \\
                                                                                                                  
0\leq \theta \leq 1 & \NoverD \geq 100 & 64.7 & 3.87 & 45.7    & 4.27 & 1.56 & 4.08 & 1.2 & 2.57  & 1.65 & 1.06 \\
                                                                                                                  
1.5\leq \theta \leq 2 & \NoverD \geq 100 & 1.81 & 26.19 & 2.23  &
2.12 & 10.94& 1.32 & 5.39 & 1.23 & 1.65 & 2.62 \bigstrut[b] \\
\hline

\end{array}
$
}
\label{tbl:ratio-error}
\end{table*}

\begin{table*}[p]\scriptsize
\caption{Percentage bias vs. \Z and \NoverD for low and high
sampling fractions. Shaded cells indicate a change in the sign of the bias in
the individual values within that region. }
\centering
\subfloat{
$
\begin{array}{|cc|cccc|cccc|}
\hline%
& & \multicolumn{4}{c|}{\text{$0.001 \leq \sample \leq 0.005$}} &
\multicolumn{4}{c|}{\text{$0.01 \leq \sample \leq 0.1$}} \\
\cline{3-10}
\text{Skew} & \text{\NoverD} & \text{\jkone} &
\text{\jktwo} & \text{\jktwos} &
\text{\jktwoa} & \text{\jkone} &
\text{\jktwo} & \text{\jktwos} & \text{\jktwoa} \bigstrut[t]
\bigstrut[b] \\
\hline%

 0\leq \theta \leq 1 & \NoverD \leq 100 &  \cellcolor[gray]{0.8}{-50.01} &
\cellcolor[gray]{0.8}{680.77} & \cellcolor[gray]{0.8}{181.94} & \cellcolor[gray]{0.8}{-34.67} & \cellcolor[gray]{0.8}{-33.07}
& \cellcolor[gray]{0.8}{2,719.94} & \cellcolor[gray]{0.8}{310.84} & \cellcolor[gray]{0.8}{6.31} \bigstrut[t]\\
 1.5\leq \theta \leq 2 & \NoverD \leq 100 & -96.19 & \cellcolor[gray]{0.8}{68.14} & -96.22 & -86.06 & -83.27 & \cellcolor[gray]{0.8}{610.59} & -83.36 & \cellcolor[gray]{0.8}{-37.33} \\
 0\leq \theta \leq 1 & \NoverD \geq 100 & \cellcolor[gray]{0.8}{-32.34}
& \cellcolor[gray]{0.8}{1,886.06} & \cellcolor[gray]{0.8}{241.56} &
\cellcolor[gray]{0.8}{1.38} & \cellcolor[gray]{0.8}{-10.24} &
\cellcolor[gray]{0.8}{1,825.79} & \cellcolor[gray]{0.8}{57.9} & \cellcolor[gray]{0.8}{12.77}\\
 1.5\leq \theta \leq 2 & \NoverD \geq 100 & -94.36 &
 \cellcolor[gray]{0.8}{159.33} & -94.41 &
 \cellcolor[gray]{0.8}{-79.67}  & -77.14 &
 \cellcolor[gray]{0.8}{661.2} & -77.24 &
 \cellcolor[gray]{0.8}{-28.31} \bigstrut[b]\\
\hline
\end{array}
$
}

\subfloat{
$
\begin{array}{|cc|ccc|cc|ccc|cc|}
\hline%
& & \multicolumn{5}{c|}{\text{$0.001 \leq \sample \leq 0.005$}} &
\multicolumn{5}{c|}{\text{$0.01 \leq \sample \leq 0.1$}} \\
\cline{3-12} \text{Skew}
& \text{\NoverD} & \text{\sh} & \text{\shtwo} & \text{\shthree}  & \text{\gee}
& \text{\ade} & \text{\sh} &
\text{\shtwo} & \text{\shthree} & \text{\gee} & \text{\ade}
\bigstrut[t] \bigstrut[b] \\
\hline%
 0\leq \theta \leq 1 & \NoverD \leq 100 & \text{1,839.04} & -87.59 & \text{1,596.3}             & \cellcolor[gray]{0.8}{13.26} & \cellcolor[gray]{0.8}{-38.34} & 441.38        & \cellcolor[gray]{0.8}{-46.03} & \cellcolor[gray]{0.8}{318.15} & \cellcolor[gray]{0.8}{35.78} & \cellcolor[gray]{0.8}{-10.28} \bigstrut[t] \\
 1.5\leq \theta \leq 2 & \NoverD \leq 100 & \cellcolor[gray]{0.8}{10.14} & -95.93 & \cellcolor[gray]{0.8}{-31.19}           & -58.5 & -90.66 & \cellcolor[gray]{0.8}{5.12}                  & -79.94 & \cellcolor[gray]{0.8}{3.59}            & -46.19 & -58.73                      \\
0\leq \theta \leq 1 & \NoverD \geq 100 & \text{6,370.46} & \cellcolor[gray]{0.8}{-49.79} & \text{4,470.08}  & \cellcolor[gray]{0.8}{326.4} & \cellcolor[gray]{0.8}{-13.12} & \cellcolor[gray]{0.8}{307.72} & \cellcolor[gray]{0.8}{-9.85} & \cellcolor[gray]{0.8}{156.69}  & \cellcolor[gray]{0.8}{64.86} & \cellcolor[gray]{0.8}{0.89}       \\
1.5\leq \theta \leq 2 & \NoverD \geq 100 & \cellcolor[gray]{0.8}{76.9} & -93.98 & \cellcolor[gray]{0.8}{-3.22} & \cellcolor[gray]{0.8}{-38.63} & -86.62     & \cellcolor[gray]{0.8}{30.47}   & -73.45 & \cellcolor[gray]{0.8}{21.91}           & \cellcolor[gray]{0.8}{-29.78} & \cellcolor[gray]{0.8}{-51.53}
  \bigstrut[b] \\
\hline
\end{array}
$
}
 \label{tbl:bias}
\end{table*}

 \begin{table*}[p]\scriptsize
\caption{Percentage RMSE vs. \cvar and \NoverD for low
and high sampling fractions}
\centering
\subfloat{
$
\begin{array}{|cc|cccc|cccc|}
\hline%
& & \multicolumn{4}{c|}{\text{$0.001 \leq \sample \leq 0.005$}} &
\multicolumn{4}{c|}{\text{$0.01 \leq \sample \leq 0.1$}} \\
\cline{3-10} 
\text{\cvar} & \text{\NoverD} & \text{\jkone} &
\text{\jktwo} & \text{\jktwos} &
\text{\jktwoa} & \text{\jkone} &
\text{\jktwo} & \text{\jktwos} & \text{\jktwoa} \bigstrut[t]
\bigstrut[b] \\
\hline%
 0\leq \cvar<1 & \NoverD \leq 100 & 63.68 & \text{2,293.07} & 575.77
 & 48.91 & 45.78 & \text{5,816.36} & 705.15 & 26.77 \bigstrut[t]\\
 1\leq \cvar\leq 50 & \NoverD \leq 100 & 96.21 & 228.61 & 96.24 & 86.3       & 83.7 & \text{1,571.14} & 83.79 & 47.92 \\
 \cvar>50 & \NoverD \leq 100 & 95.63 & \text{2,300.52} & 580.57 & 80.51      & 80.49      & \text{5,956.07} & 708.07 & 45.69 \\
 0\leq \cvar<1 & \NoverD \geq 100 & 46.63 & \text{4,998.23} & 627.08 & 28.2  & 21.49 & \text{4,528.98} & 230.89 & 23.17 \\
 1\leq \cvar\leq 50 & \NoverD \geq 100 & 94.48 & 467.77 & 94.52 & 81.3       & 78.61 & \text{1,577.23} & 78.7 & 43.14         \\
 \cvar>50 & \NoverD \geq 100 & 89.48 & \text{5,083.31} & 640.39 &
 70.98      & 68.54 & \text{4,775.1} & 242.68 & 42.17
 \bigstrut[b] \\
\hline 
\end{array}
$
}\\
\subfloat{
$
\begin{array}{|cc|ccc|cc|ccc|cc|}
\hline%
& & \multicolumn{5}{c|}{\text{$0.001 \leq \sample \leq 0.005$}} &
\multicolumn{5}{c|}{\text{$0.01 \leq \sample \leq 0.1$}} \\
\cline{3-12} \text{\cvar}
& \text{\NoverD} & \text{\sh} & \text{\shtwo} & \text{\shthree}  & \text{\gee}
& \text{\ade} & \text{\sh} &
\text{\shtwo} & \text{\shthree} & \text{\gee} & \text{\ade}
\bigstrut[t] \bigstrut[b] \\
\hline%
 0\leq \cvar<1 & \NoverD \leq 100 & \text{2,856.32} & 88.96 &
 \text{2,563.94} & 115.98 & 55.95& 653.14 & 55.16 & 478.76 & 79.1 &
 28.25  \bigstrut[t] \\
 1\leq \cvar\leq 50 & \NoverD \leq 100 & 23.05 & 95.96 & 47.43                & 59.92 & 90.77 & 11.04 & 80.63 & 9.46    & 47.21 & 61.86 \\
 \cvar>50 & \NoverD \leq 100 & 403.1 & 95.5 & 218.4                           & 60.89 & 88.46 & 131.95 & 77.48 & 80.6   & 46.66 & 56.92 \\
 0\leq \cvar<1 & \NoverD \geq 100 & \text{9,061.75} & 58.72 & \text{6,462.05} & 433.9 & 32.17 & 610.17 & 20.07 & 339.7  & 106.91 & 8.92 \\
 1\leq \cvar\leq 50 & \NoverD \geq 100 & 185.16 & 94.12 & 62.61               & 57.05 & 87.24 & 64.65 & 75.33 & 45.18   & 41.94 & 56.27 \\
 \cvar>50 & \NoverD \geq 100 & \text{1,434.17} & 89.1 & 636.45
 & 130.03 & 78.32& 224.9 & 65.37 & 124.14  & 61.5  & 47.23
 \bigstrut[b] \\
\hline
\end{array}
$ 
}
\label{tbl:rmse}
\end{table*}

\subsubsection{By Ratio Error}
See Table~\ref{tbl:ratio-error}. 
For low sampling fractions, for low \NoverD, \jktwoa and \gee are 
the best estimators when $ 0 \leq \Z \leq 1$. For higher
skew, only \gee continues to perform well. At high \NoverD, and low-mid skew,
\jktwoa and \ade are the best, which \gee again does very well as \NoverD increases.

Caveat: If \NoverD is mid-high, and the data has low-mid skew, then \gee should \emph{not} be
used unless the sampling fraction is greater than 0.01.  

\subsubsection{By Bias}
See Table~\ref{tbl:bias}. 
For low sampling fractions and for low \NoverD, all three --- \jktwoa, \gee,
\ade --- do well measured by bias. 
For high \NoverD, for low-mid skew, \jktwoa and \ade are the best estimators. As we
raise skew, \gee is more accurate than both; however, all three are good.

Note that by either metric --- bias or ratio error --- the provisional choice
\cltwo estimator given in \cite{BF93} is poor.  

\subsection{Regions of Good Performance}

The three estimators discussed above do reasonably well in all regions.
However, there are other estimators that actually do better than these, but in
small sub-domains. However, these sub-domains are clearly delineated, and
therefore we can potentially use these estimators when our data lies in the
corresponding domains.

\subsubsection{Regions Defined by Zipfian Skew }

The best example of such estimators are \sh and \shthree. Whether sampling
fractions are low or high, these estimators absolutely shine in the high skew
region of $ \Z > 1.5 $ (see Table~\ref{tbl:ratio-error}). Indeed, their ratio
errors are  an order of
magnitude lower than other estimators in this region.  Interestingly, \shtwo,
which is, on the average, a far better estimator than \sh and \shthree due to
its reasonable performance at low-mid \Z, does not offer as good of a
accuracy gain in this high-skew region.  We also note that for high
$\NoverD \sim 1000$, the bias profile of \sh and \shthree turns from a ``slope" to
a ``hat", and then they also offer reasonably accurate estimates at
low-skew (see $\S.~\ref{sec:results:schlosser}$ for a discussion of this
phenomenon).

\subsubsection{Regions Defined by Coefficient of Class Variation}

Earlier studies \cite{HNSS95, HS98} have reported some trends by coefficient
of class variation. We validate some of these at higher scale and
dimensionality of parameter space. On the other hand, other reported trends no
longer continue to hold in our large-scale study. Note that we vary the
sampling percentage through a wider range of values than  
previous studies.   

\paragraph{$ 0 \leq \cvar \leq 1 $:}

For low sampling fractions ($ \leq 0.005$), \jktwoa is the best estimator in the
region $ 0 \leq \cvar \leq 1$; however, \jkone, \shtwo, and \ade are comparable (cf. 
\cite{HS98}, where \jktwo was declared the best estimator in this region). 

For high sampling fraction ($ > 0.005$), the picture remains the same,
except for high \NoverD ($> 100$), where \ade emerges as the best estimator.  
For sizes below 1B, \shtwo is the best estimator in this region.

\paragraph{$ 1 \leq \cvar \leq 50 $:}

For low and high sampling fractions, \jktwoa is the best among the jackknives, but
comparable to \jkone and \jktwos. However, it is the Schlossers \sh and \shthree
that are the best, by a comfortable margin in this region. Finally \shtwo, \gee and
\ade are comparable to \jktwoa. Again, this shows that the optimal estimator for
this region, which was \jktwoa in the study of \cite{HS98}, is no longer optimal as we increase the scale
and dimensionality of the underlying characterization (this includes reducing
the average $q$). 

We also note that at our scale and dimensionality, \shtwo shows similar
accuracy to 
the jackknife families, but with the exception of \jktwo, for low to medium
\cvar (cf. \cite{HS98}, who do not exclude \jktwo). 

\paragraph{$ \cvar > 50 $:}

For both low and high sampling fractions, the best estimators are the same as the three best
estimators overall, namely \jktwoa, \gee, and \ade, with \shtwo being
comparable. Among these, for low \NoverD,
\gee is the best, while for high \NoverD, \jktwoa is the best.
The reasoning that \sh is a good estimator when \cvar is large
since its derivation does not depend on a Taylor-series expansion in \cvar
\cite{HNSS95} does not find evidence.  Indeed, in the very high \cvar regions
of mid-Zipfian skew, all the Schlosser estimators do very poorly.  

\shtwo is the most accurate among the Schlossers  in this region, by a
considerable margin, as opposed to \shthree which was declared
the best estimator in this range in \cite{HS98}. 

\subsection{Smoothing versus Stabilization in Jackknives}

Our study also validates, at a higher scale, some observations made in
\cite{HS98}. Namely, stabilization works far more effectively than smoothing.
In the mid-skew regions, the second order
jackknives exhibit poor performance, while the stabilized \jktwoa retains
acceptable performance.

\subsection{Sampling Percentages Required}
\label{sec:sampling-fraction}

In today's large commercial databases, hundreds of millions  of rows are
standard, and billions of rows are frequently encountered. Therefore, the cost of
sampling is significant, and is a major design consideration. In our
experience, it is the second question asked by designers behind the
choice of estimator.  The ``default" value of sampling fraction in industrial
databases is
0.02, but there is increasing pressure to reduce this as database sizes
increase. 

From our experience of working closely with query optimizer designers, we
observed a wide gulf between the accuracy that distinct value estimators can
provide, and the accuracy that query optimizer designers expect. It is
important to understand that without essentially scanning the entire relation,
we cannot hope to achieve the accuracies that are expected for arbitrary
datasets. We feel that this is a communication gap that should be addressed. The published literature that deals with required accuracies \cite{ASW87} is now fairly old, and was suitable to the small tables encountered then. Today's query optimizers should be designed with the understanding that obtaining ratio errors of less than 10 consistently, with the sampling fractions that are feasible for such large tables, is itself a non-trivial problem.    
For instance, the ratio error bound on the GEE --- the only estimator to have error bounds --- at even 10\% sampling, is $\sqrt{10} \sim 3.2 $.  At the more feasible sampling rate of 2\%, this bound is $\sqrt{50} \sim 7.1$ and at the sampling rate of 1\%, it is 10. 
Note that even a ratio error of 3.2 is enough at large database sizes to cause
the query optimizer to formulate highly inefficient plans. 

In Table~\ref{tbl:max-avg:2-5}, we provide the best estimator as a function of both maximum and average
ratio error, for the ratio error values of two and five.
Table~\ref{tbl:max-avg:2-5} indicates that if $q = 0.01$, then \gee or \jktwoa may
be the better choice over \ade. However, we should note that \gee has a
region of relatively poor performance at very low sampling fractions, and so
we should not use \gee for low-mid skew data if the sampling fraction is to be dropped below 0.005.  

\begin{table} %\scriptsize 
\caption{Sampling percentage required for maximum/average ratio error of 2 and
5}
\label{tbl:max-avg:2-5} \centering 
$
\begin{array}{cccccccccccc}
\toprule
\text{Error} &  \jkone & \jktwo & \jktwos & \jktwoa & \sh & \shtwo & \shthree & \gee & \ade &
\clone & \cltwo \\ 
\toprule
\text{Max 5} &  0.1 & na & na & 0.01 & 0.1 & 0.1 & 0.05 & 0.01 & 0.02 & na & na \\
\text{Max 2} &  na & na & na & 0.05 & na & na & na & 0.1 & 0.1 & na & na \\
\text{Avg 5} &  0.05 & na & na & 0.005 & 0.05 & 0.05 & 0.02 & 0.005 & 0.01 & na & na \\
\text{Avg 2} &  na & na & na & 0.02 & 0.1 & na & 0.1 & 0.05 & 0.05 & na & na \\
\end{array}
$
\end{table}

\subsection{Ease of Implementation}

There is no significant difference in the ease of implementation among the
estimators (besides our simplification for \shtwo on
p.~\pageref{eq:shtwo}). While some of the estimators require storing values of $f_i, i >
2$, the \gee requires only the storage of $f_1$. However, this is not a factor
in today's systems. Likewise, the iterations required for the
Newton-Raphson method in the \ade are far too few to be a design factor. In
all our experiments, Newton-Raphson converged in less than ten iterations. In
summary, the choice of estimator depends only on accuracy, and not on
implementation considerations.

\section{Conclusion and Future Work}
\label{sec:conclusion} 
\label{sec:futurework}

\paragraph{Conclusions.}

There are two kinds of principles at play in statistics --- the theoretical,
and the empirical. The literature in this area has mostly postulated the
former. This extensive study aimed to uncover the latter. 

Our high-level conclusion is that there exist stable patterns of relative behavior
of distinct value estimators over populations of real-world size and
frequently occurring distributions. This provides us with the best
characterization yet of the answer to ``which estimators do well on
which datasets?" and therefore also sheds light on the question
of ``what properties of datasets allow certain estimators to do well on
them?" We have proposed a systematic methodology, which integrates 
visualization, for characterization of errors of distinct value
estimators.   

Some of the conclusions that arise from our study are below.
\begin{enumerate*}
\item The parameter \NoverD is critical in characterizing datasets.
\item Scale effects cannot be ignored: conclusions drawn through studies on
small datasets can lead to erroneous choices for large real world datasets. 
\item  Three distinct value estimators ---
\gee, \ade, \jktwoa --- are the best estimators across a wide range of
parameters, and at large scales. Each represents a different approach to
estimation. Moreover, the choice of estimator should be informed by
finer-grained behavior w.r.t. parameters.  
\item Estimators obtained through ``second order" methods that require
estimation of \cvar are highly inaccurate, especially as \cvar
increases. 
\item A sampling fraction of 2\% may be considered optimum
in the sense that it is at the high end of what may be considered
feasible for today's large commercial databases, and at the low end of
obtaining acceptable ratio errors provided good estimators are chosen.  
\item Visualization of error patterns is a powerful methodology to gain
insight into the behavior of distinct value estimators.  
\end{enumerate*}

This paper was written for both the practitioner and the researcher. 
The practitioner seeks to answer questions such as sampling percentage, choice of estimator for his dataset, etc. 
The researcher will find the relative performance of estimators a source of
challenging problems: why do certain estimators behave in certain ways
relative to one another for certain population parameters.

\paragraph{Future Work.}

It has been remarked before \cite{HNSS95} that perhaps the reason that there is
relatively less literature on the distinct values problem in the database
community is that the problem is hard, and our understanding of it is limited.
We hope that with the characterization we provide in this work, there will be
more clarity on the accuracies we can expect for various datasets of commercial
importance. An important question is: how well can we estimate \NoverD for a
dataset? Can we then use the resulting better understanding of errors incurred to
make the query optimizer more robust?  

We also hope that the  understanding of the relative performances of estimator
families that emerges from this study should lead to better hybrid estimators.

Finally, the empirical characterization of this work could be used
to improve the theory and methods for existing families of estimators. For
example, why is \NoverD such a critical parameter for error patterns?
Can we design estimators that operate
very well for specified ranges of \NoverD? 
Can a modified form of stabilization be used on other estimators in
light of its effectiveness in \jktwoa? Can we understand the ``slope" to
``hat" transitions that happen in multiple estimators?  

%\section*{Acknowledgements}
%
%We thank Lakshminarayan Choudur for introducing us to the subject of distinct
%value estimation in 2007 in the context of
%hybrid estimators through the reference \cite{HS98}, and Umeshwar
%Dayal for setting the business context in that project. Though the present
%study was carried out in 2011, its motivations owe to the project done in
%2007.  We thank Alistair Veitch for a timely suggestion to build a
%special-purpose testbed in C that could handle the scales required for this
%study.  

%\section{Supplementary Material}
%
%A large amount of data was generated in our study. Due to space
%considerations, the following supplementary material is not included in the
%main body of the paper, but is provided in the supplementary material, and
%will be made available online. 
%\begin{enumerate*}
%\item Grids for 2D bias plots for the following families: jackknife, Schlosser, \gee
%and \ade at each sampling fraction $q \in [0.001, 0.005, 0.01, 0.02, 0.1]$
%excluding those that appears in the main body of the paper.  
%\item Grids for the
%top three estimators --- \jktwoa, \gee, \ade --- for  each sampling fraction
%$q \in [0.001, 0.005, 0.01, 0.1]$.
%\item Grid of 3D bias surfaces, and 2D plots at $q=0.1$, for \cltwo.
%\end{enumerate*}

\section*{Dedication}
This study was carried out in 2011: the $40^{\textup{th}}$ year of the genocide of
two million Hindus in 1971. This work is dedicated to  their sacred memory, and especially to the women
violated during that genocide; and also to Flt. Lt.
Vijay Vasant Tambay. 

\appendix

\section*{SUPPLEMENTARY MATERIAL}
A large amount of data was generated in our study.  The following supplementary material is not included in the
main body of the paper, but is provided in this section. 
\begin{enumerate*}
\item Grids of 2D bias plots for the following families: Jackknife, Schlosser, \gee
and \ade at each sampling fraction $q \in [0.001, 0.005, 0.01, 0.02, 0.1]$,
excluding those that appear in the paper. 
\item Grids of 2D bias plots for the
top three estimators --- \jktwoa, \gee, \ade --- for  each sampling fraction
$q \in [0.001, 0.005, 0.01, 0.1]$.
\item Grid of 3D bias surfaces, and 2D plots at $q=0.1$, for \cltwo.  
\end{enumerate*}
For (1) and (2), each grid is labelled in-figure, and therefore not captioned.  

\newpage

\noindent\includegraphics[type=pdf,ext=.pdf,read=.pdf,bb=27 2 727 413,scale=0.6]{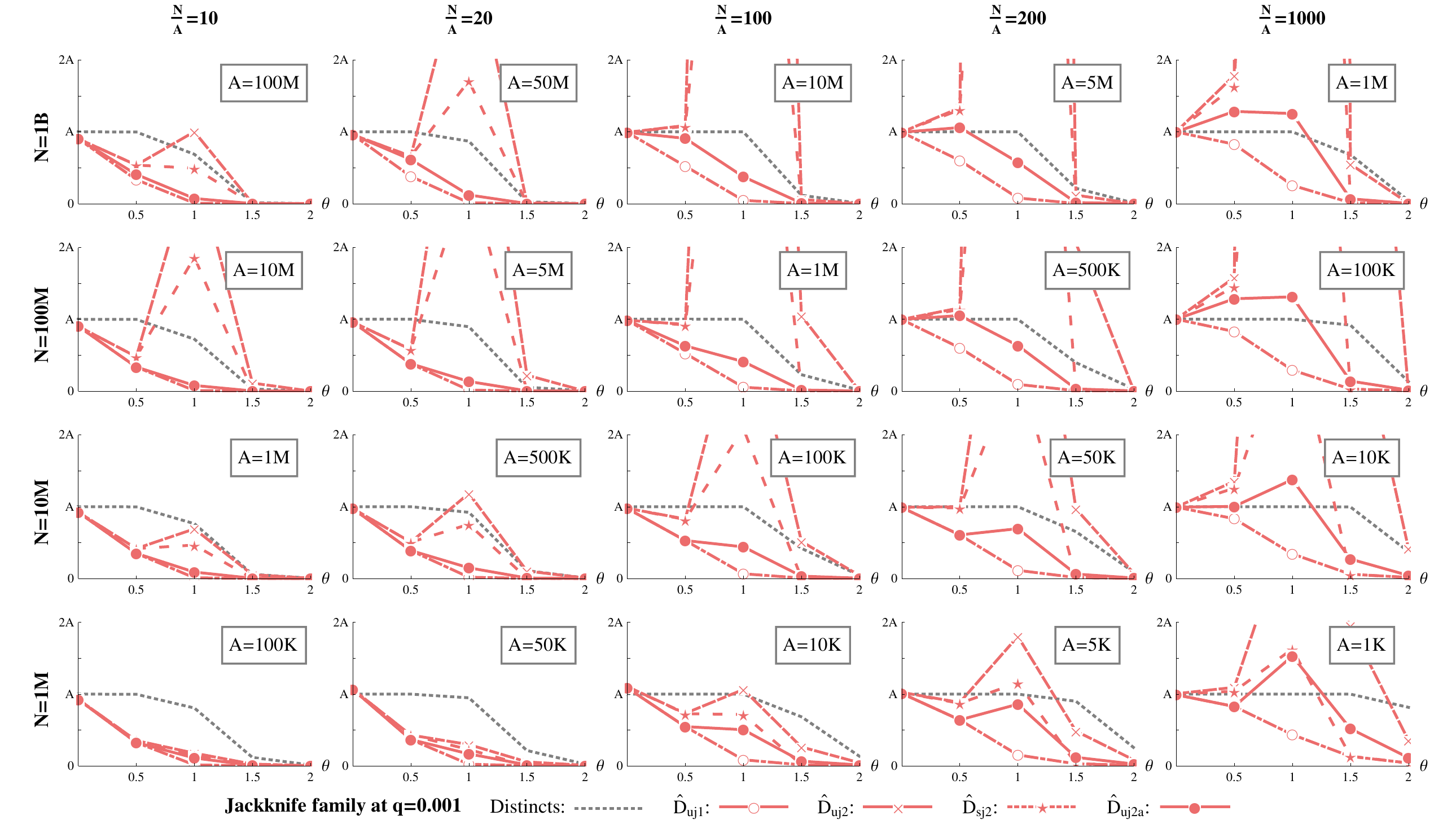}\vspace{1in}	
\includegraphics[type=pdf,ext=.pdf,read=.pdf,bb=27 2 727 413,scale=0.6]{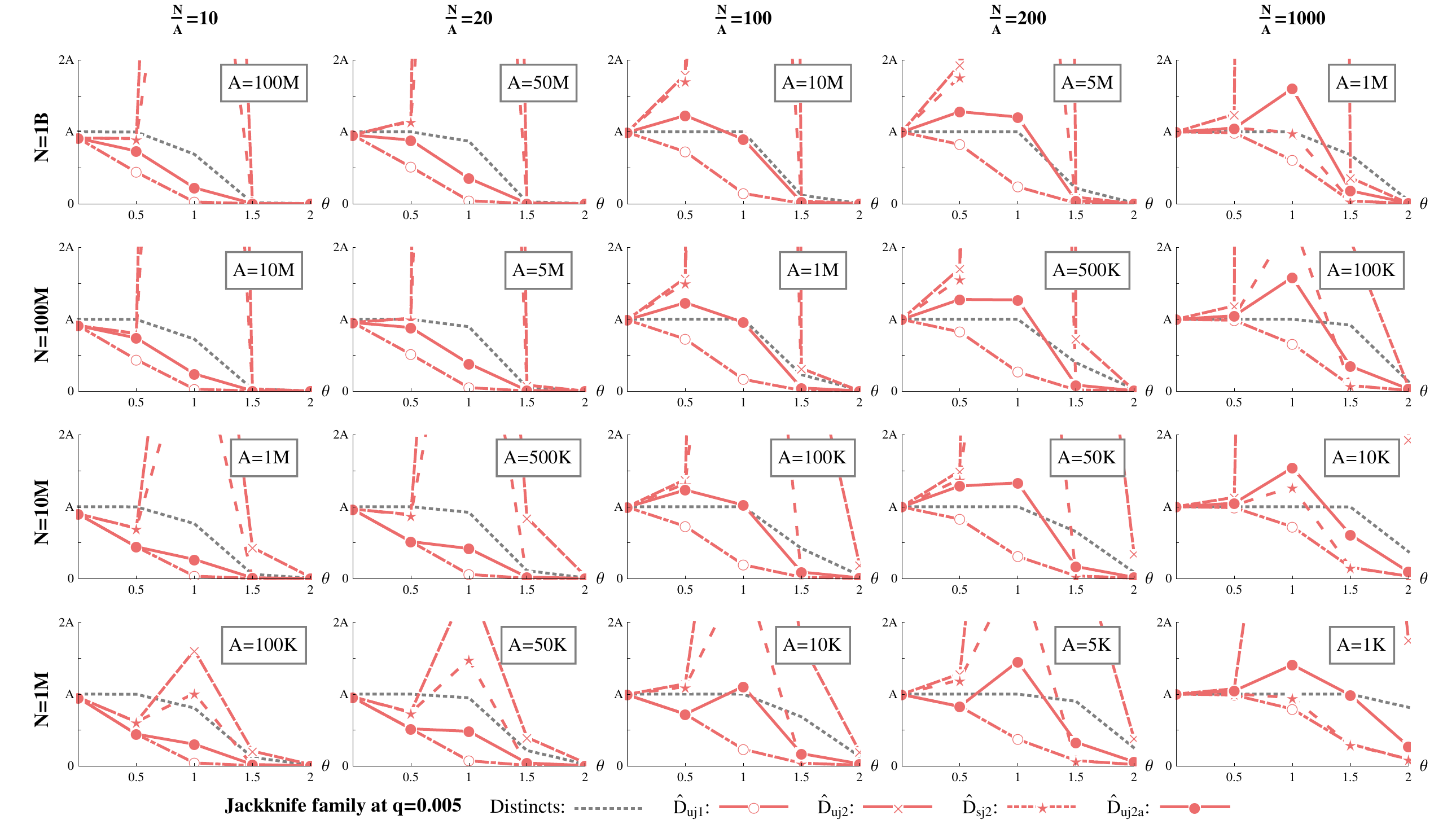}
\includegraphics[type=pdf,ext=.pdf,read=.pdf,bb=27 2 727 413,scale=0.6]{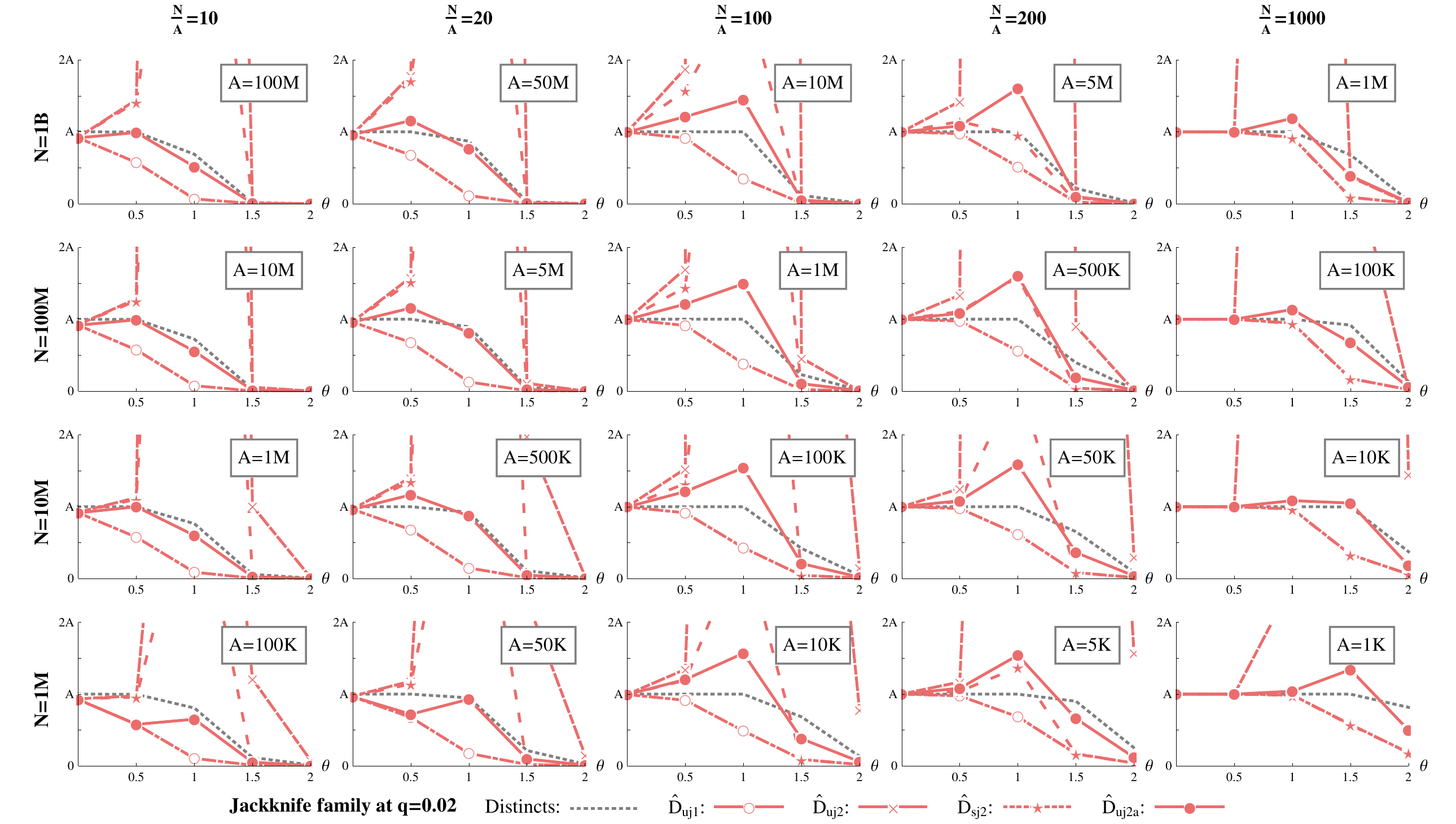}\vspace{1in}
\includegraphics[type=pdf,ext=.pdf,read=.pdf,bb=27 2 727 413,scale=0.6]{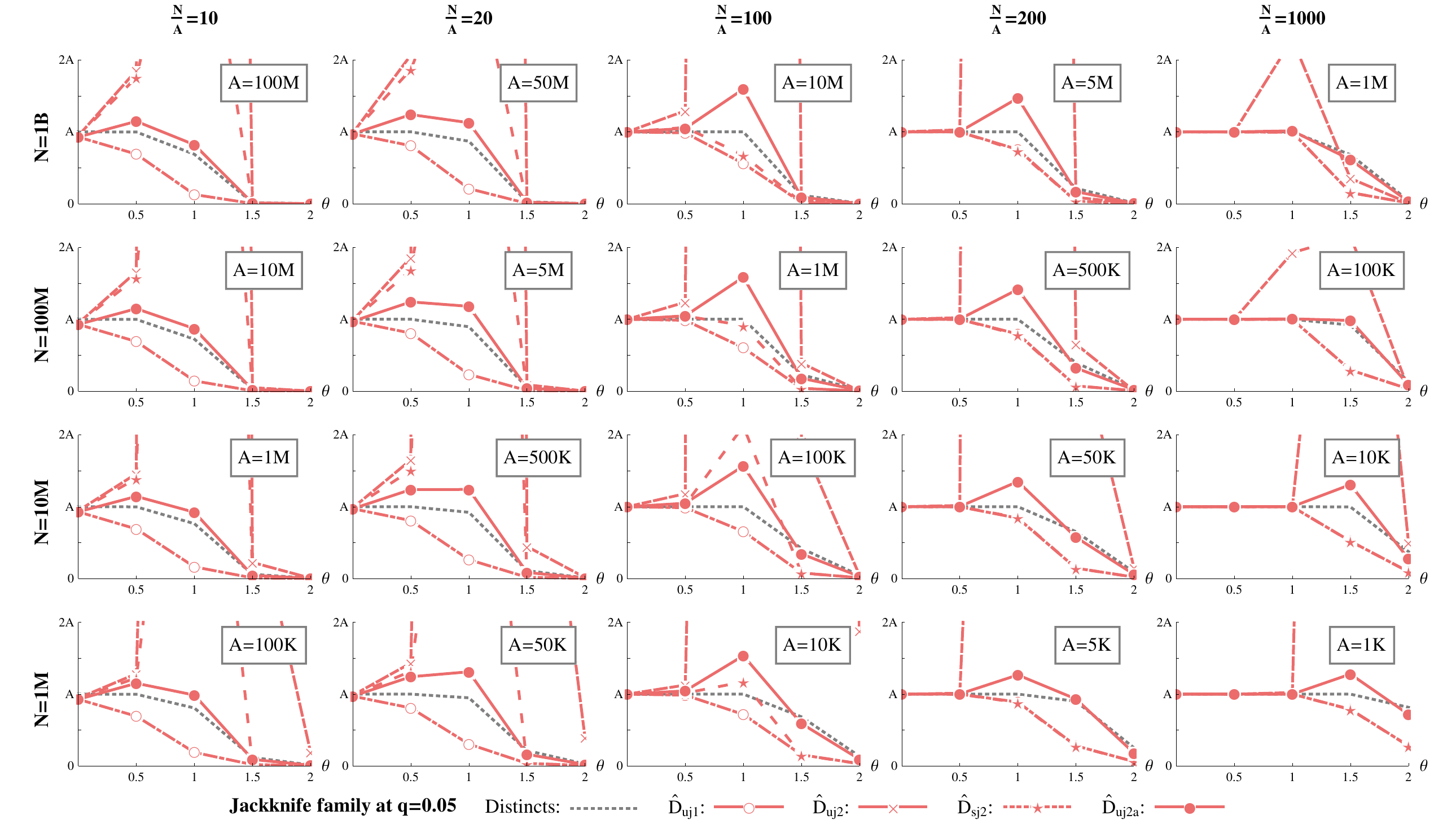}
\includegraphics[type=pdf,ext=.pdf,read=.pdf,bb=27 2 727
413,scale=0.6]{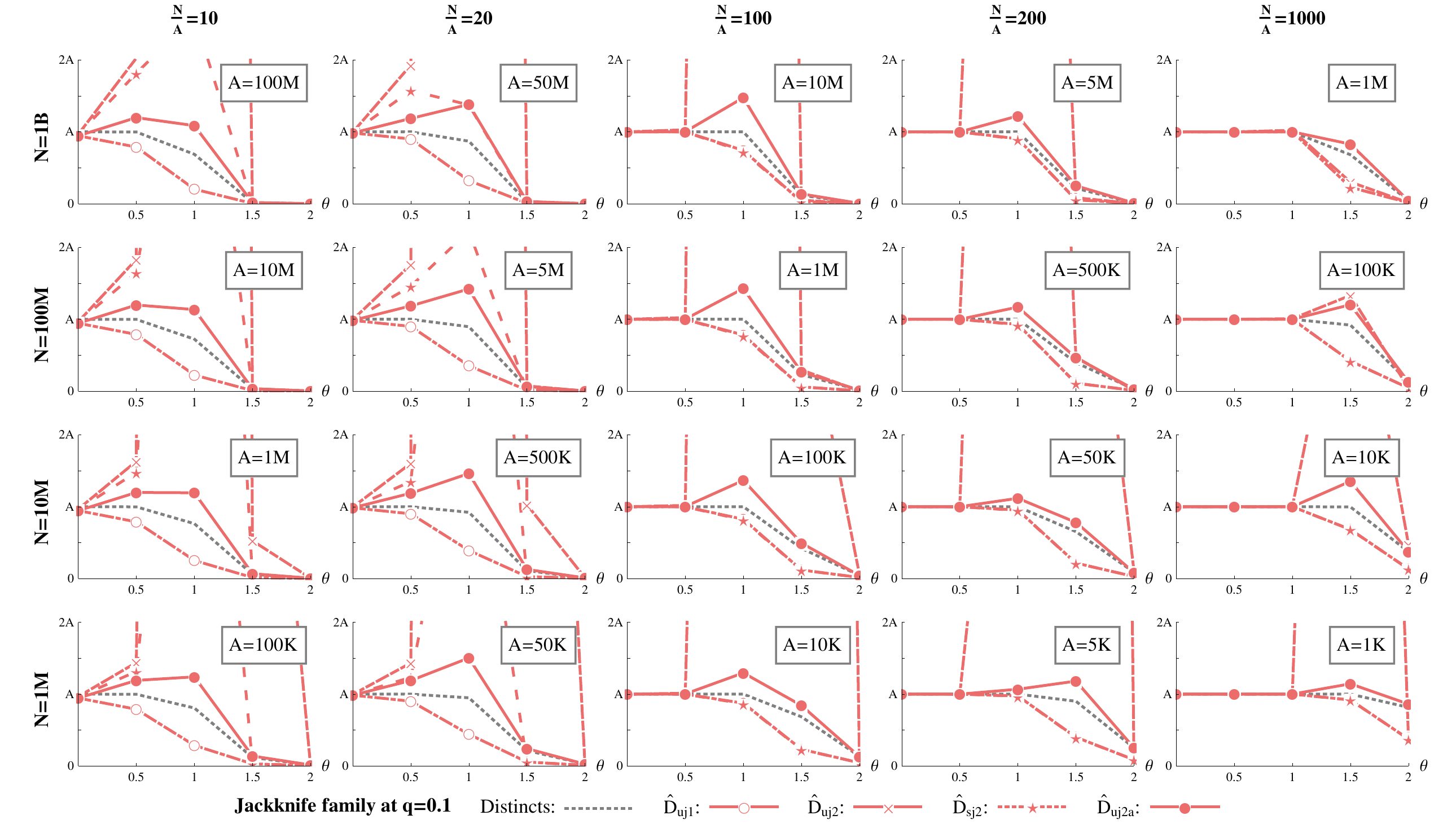}\vspace{1in}
\includegraphics[type=pdf,ext=.pdf,read=.pdf,bb=27 2 727 413,scale=0.6]{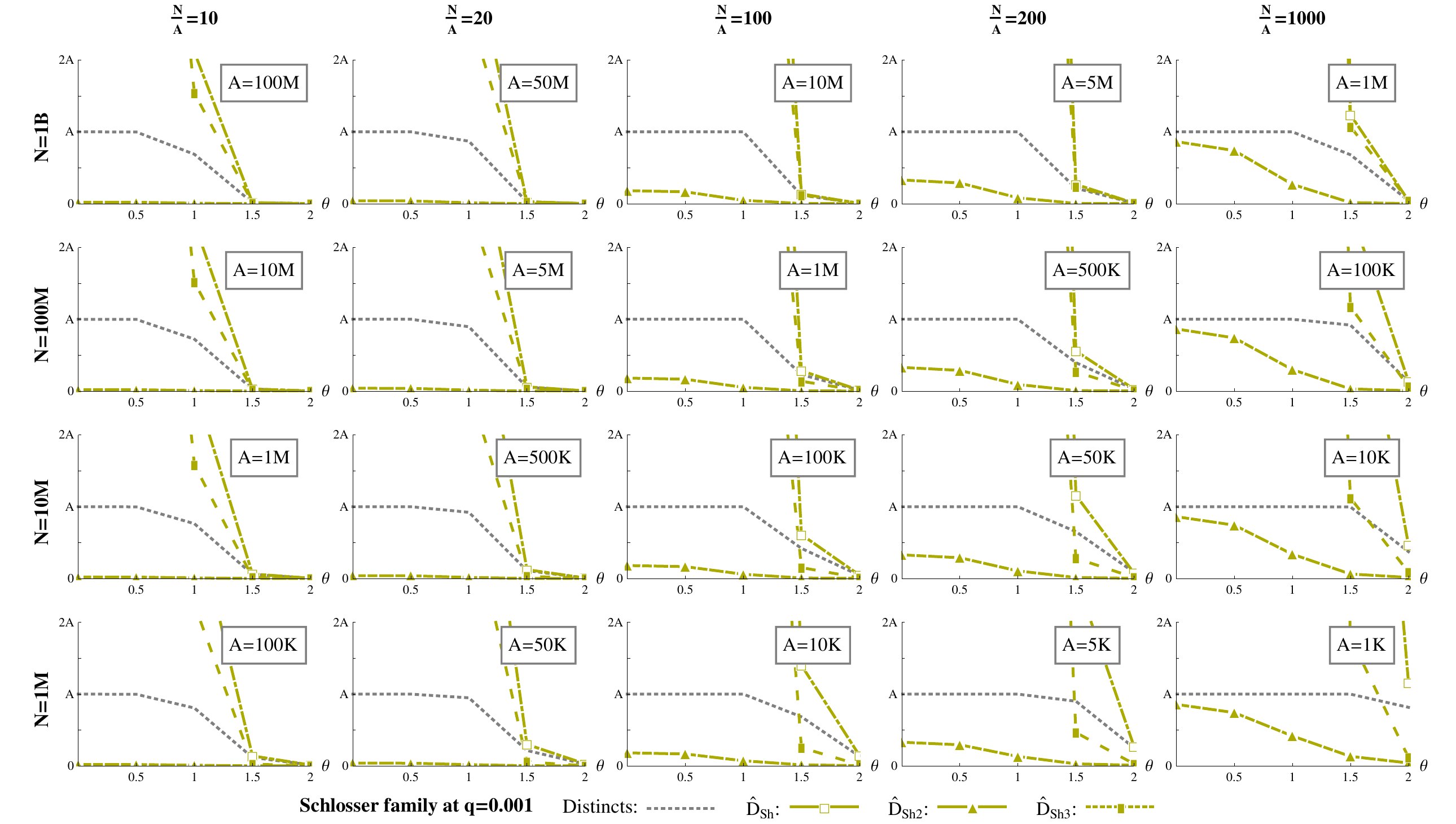}
\includegraphics[type=pdf,ext=.pdf,read=.pdf,bb=27 2 727
413,scale=0.6]{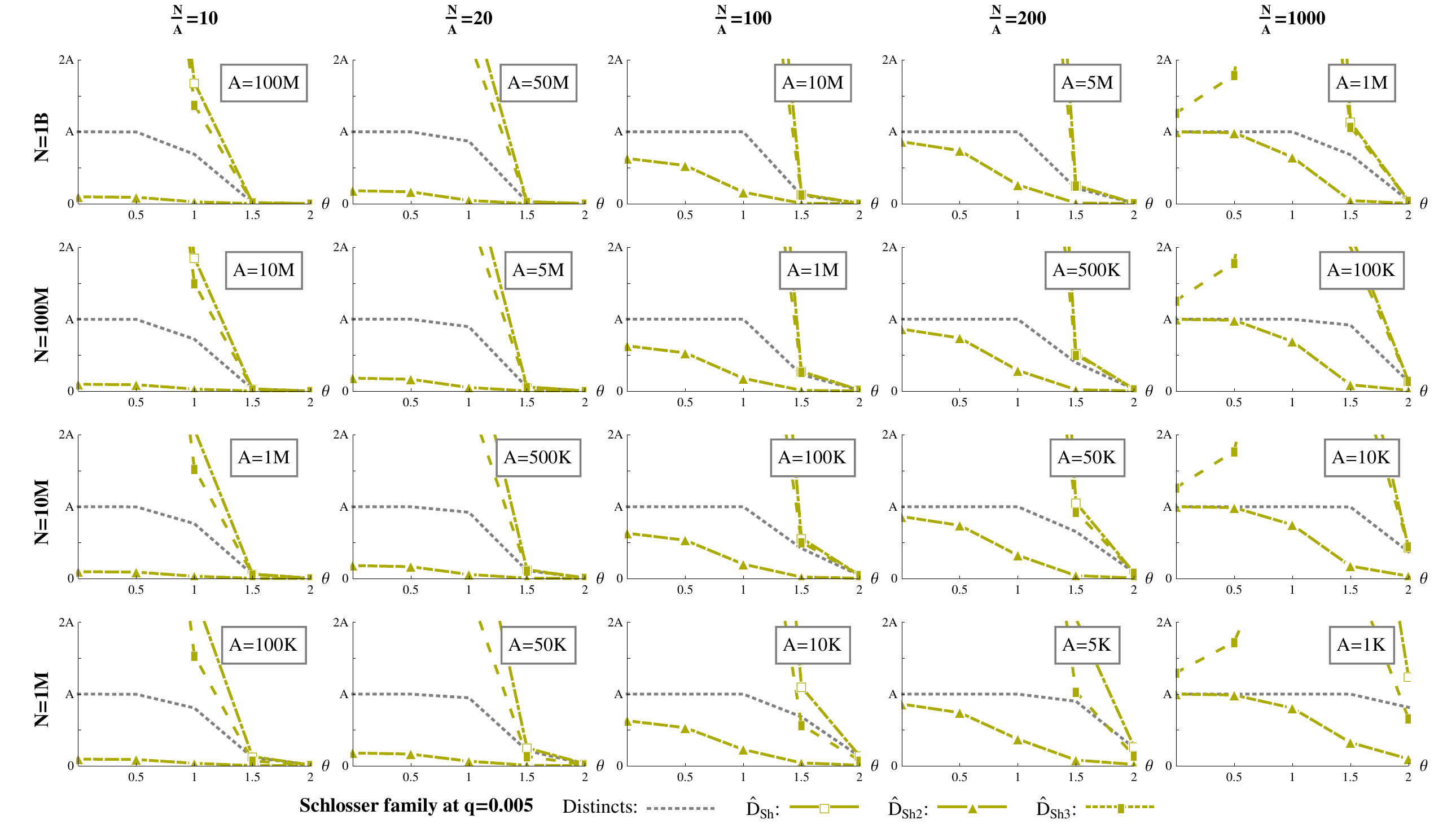}\vspace{1in}
\includegraphics[type=pdf,ext=.pdf,read=.pdf,bb=27 2 727 413,scale=0.6]{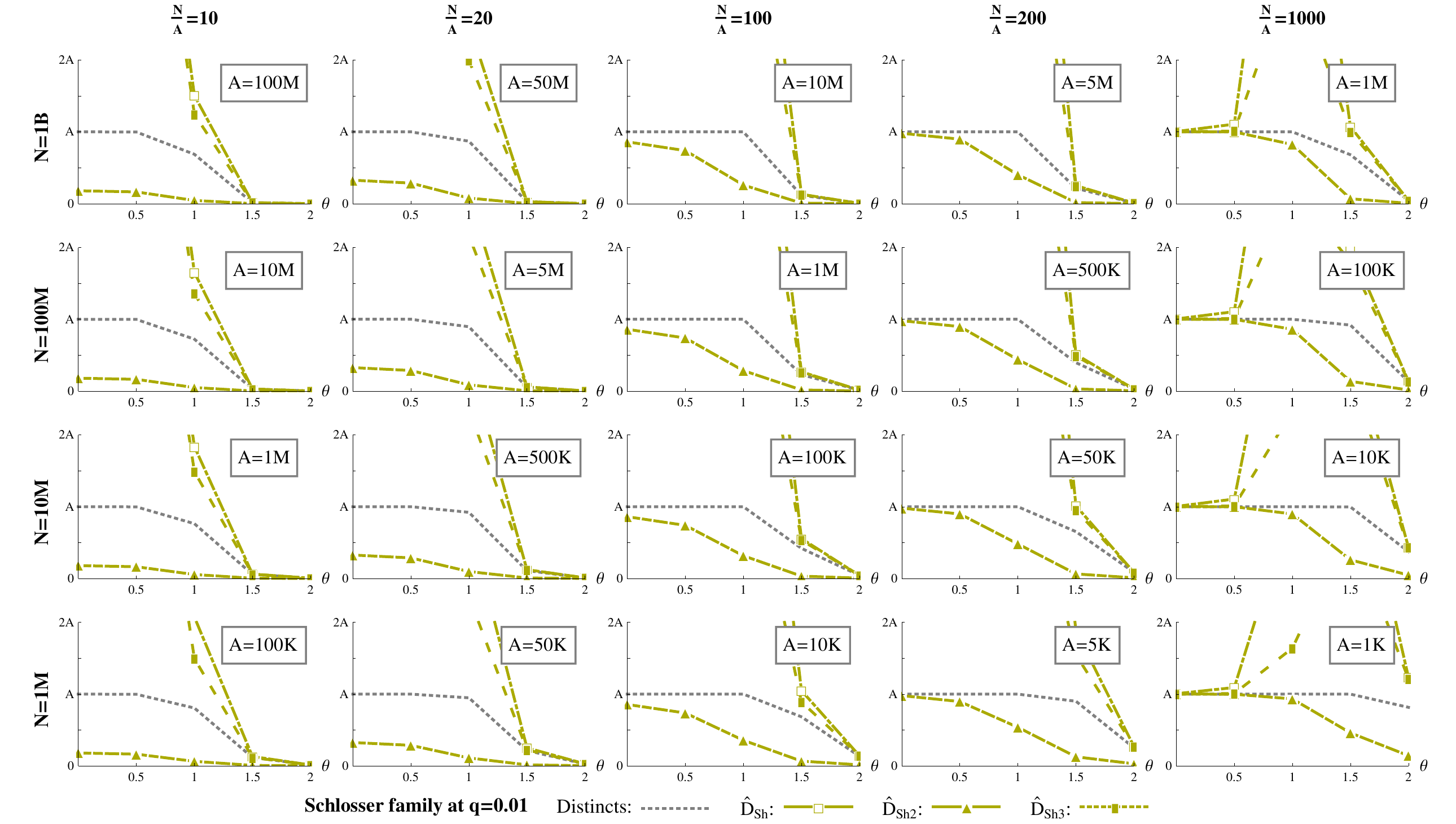}
\includegraphics[type=pdf,ext=.pdf,read=.pdf,bb=27 2 727 413,scale=0.6]{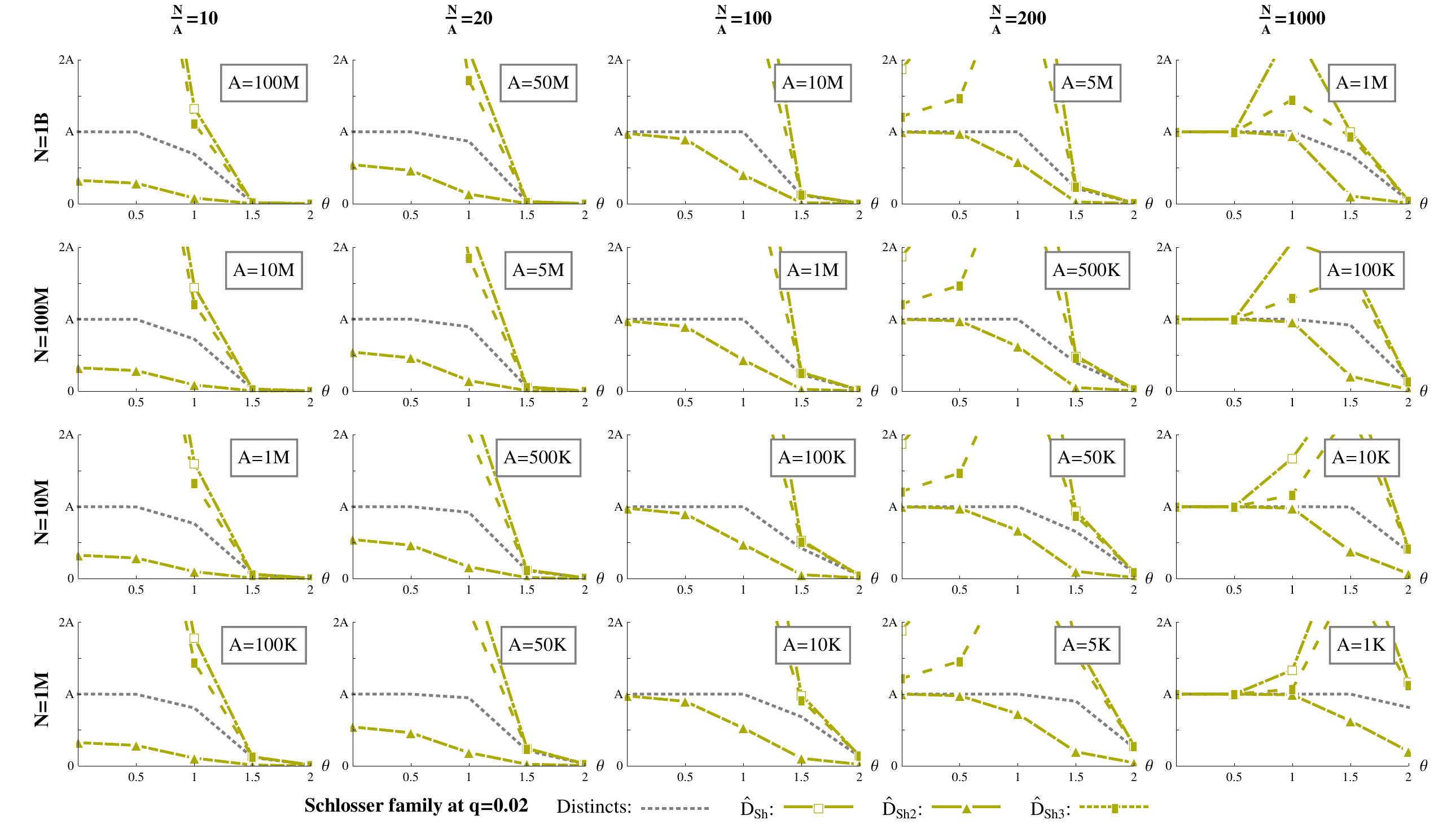}\vspace{1in}
\includegraphics[type=pdf,ext=.pdf,read=.pdf,bb=27 2 727 413,scale=0.6]{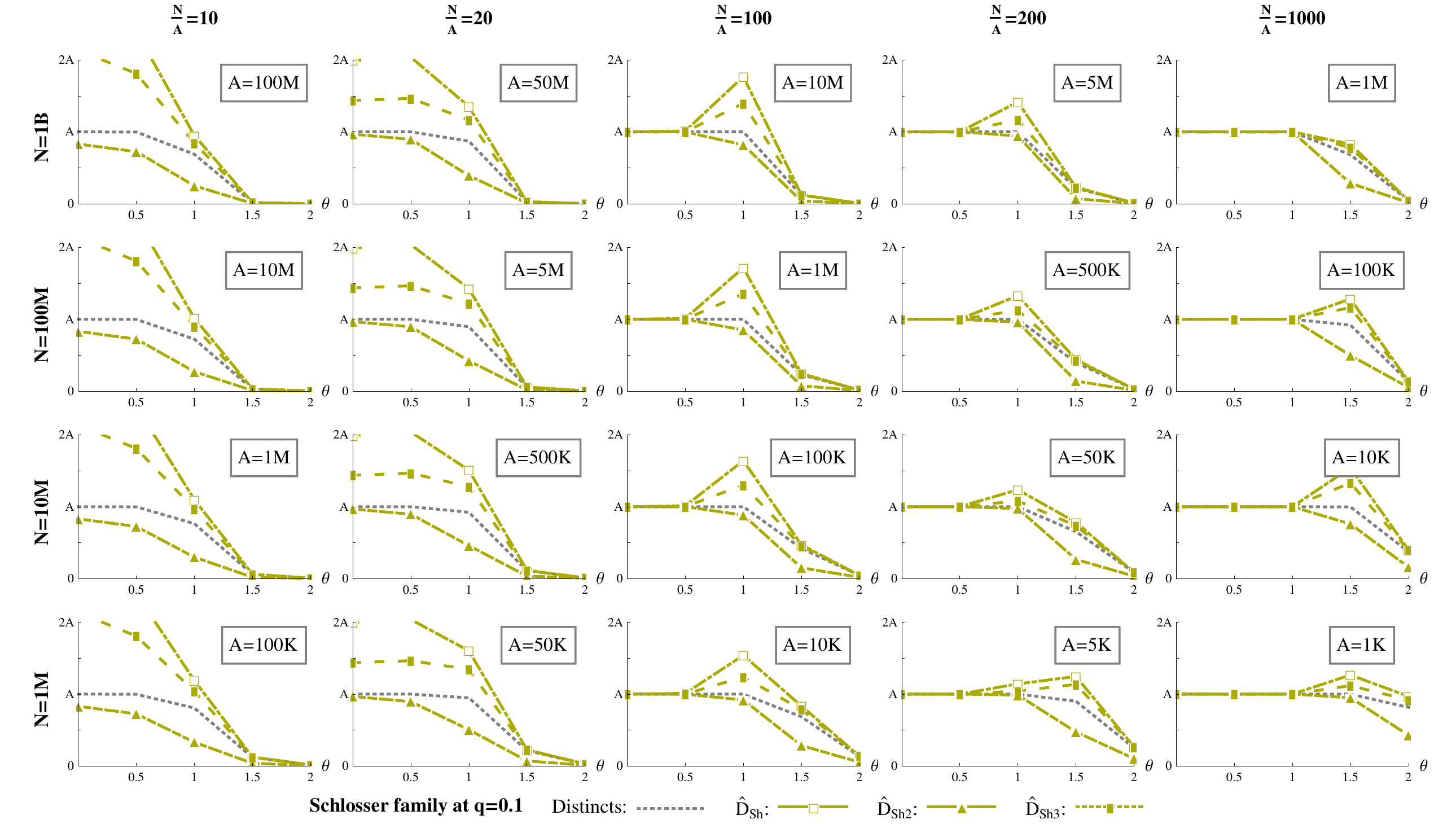}
\includegraphics[type=pdf,ext=.pdf,read=.pdf,bb=27 2 727 413,scale=0.6]{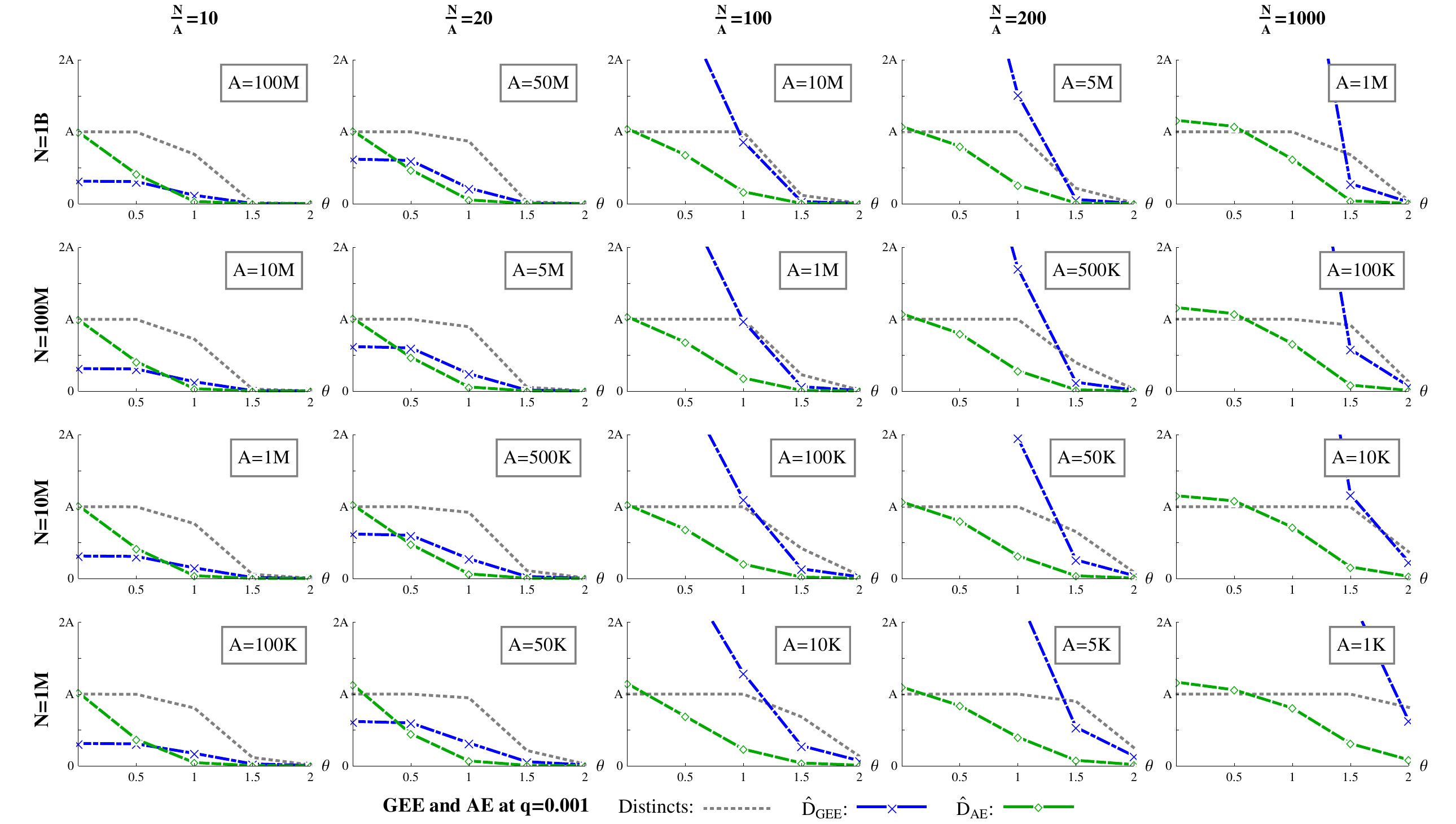}\vspace{1in}
\includegraphics[type=pdf,ext=.pdf,read=.pdf,bb=27 2 727 413,scale=0.6]{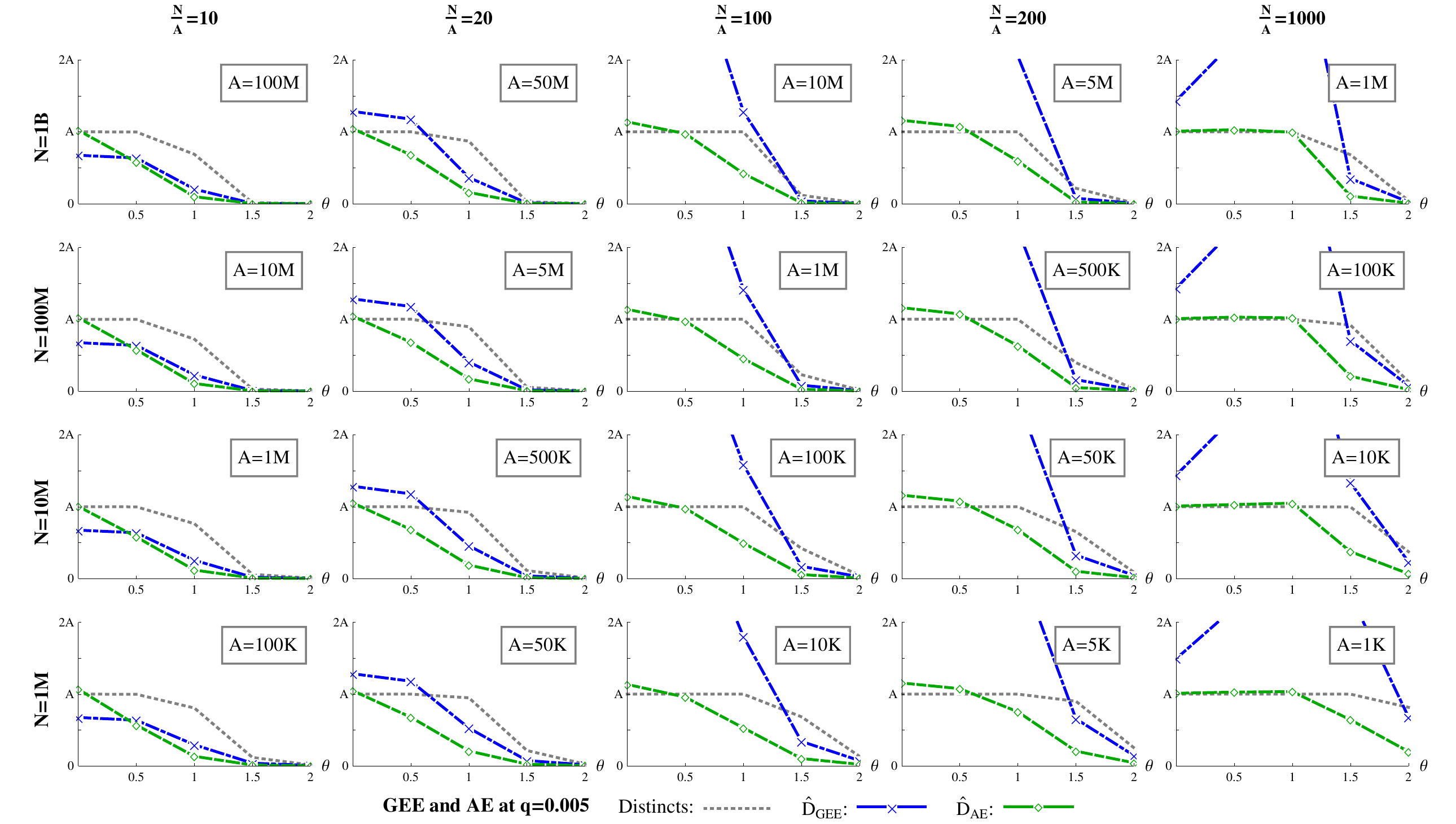}
\includegraphics[type=pdf,ext=.pdf,read=.pdf,bb=27 2 727 413,scale=0.6]{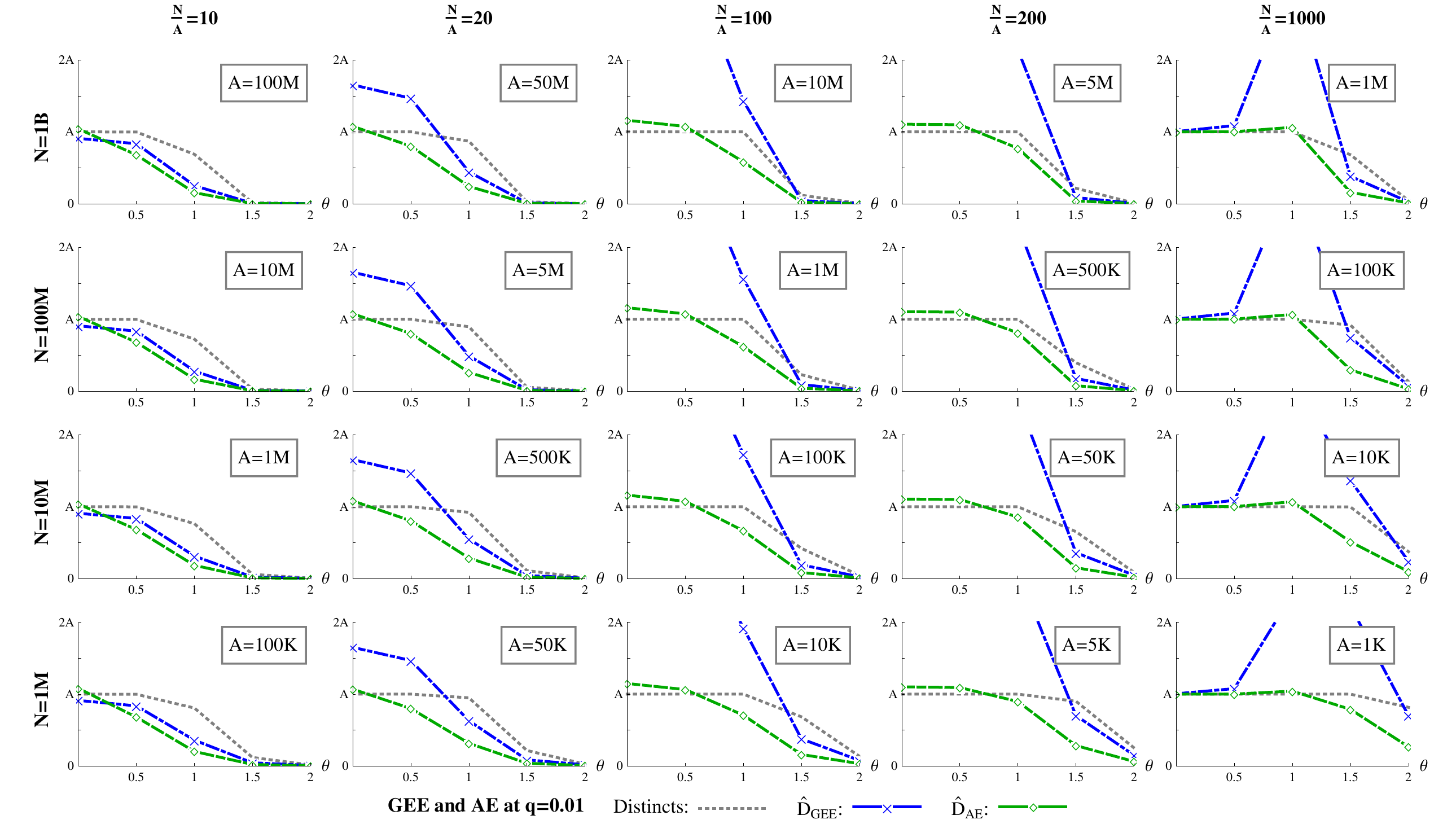}\vspace{1in}
\includegraphics[type=pdf,ext=.pdf,read=.pdf,bb=27 2 727 413,scale=0.6]{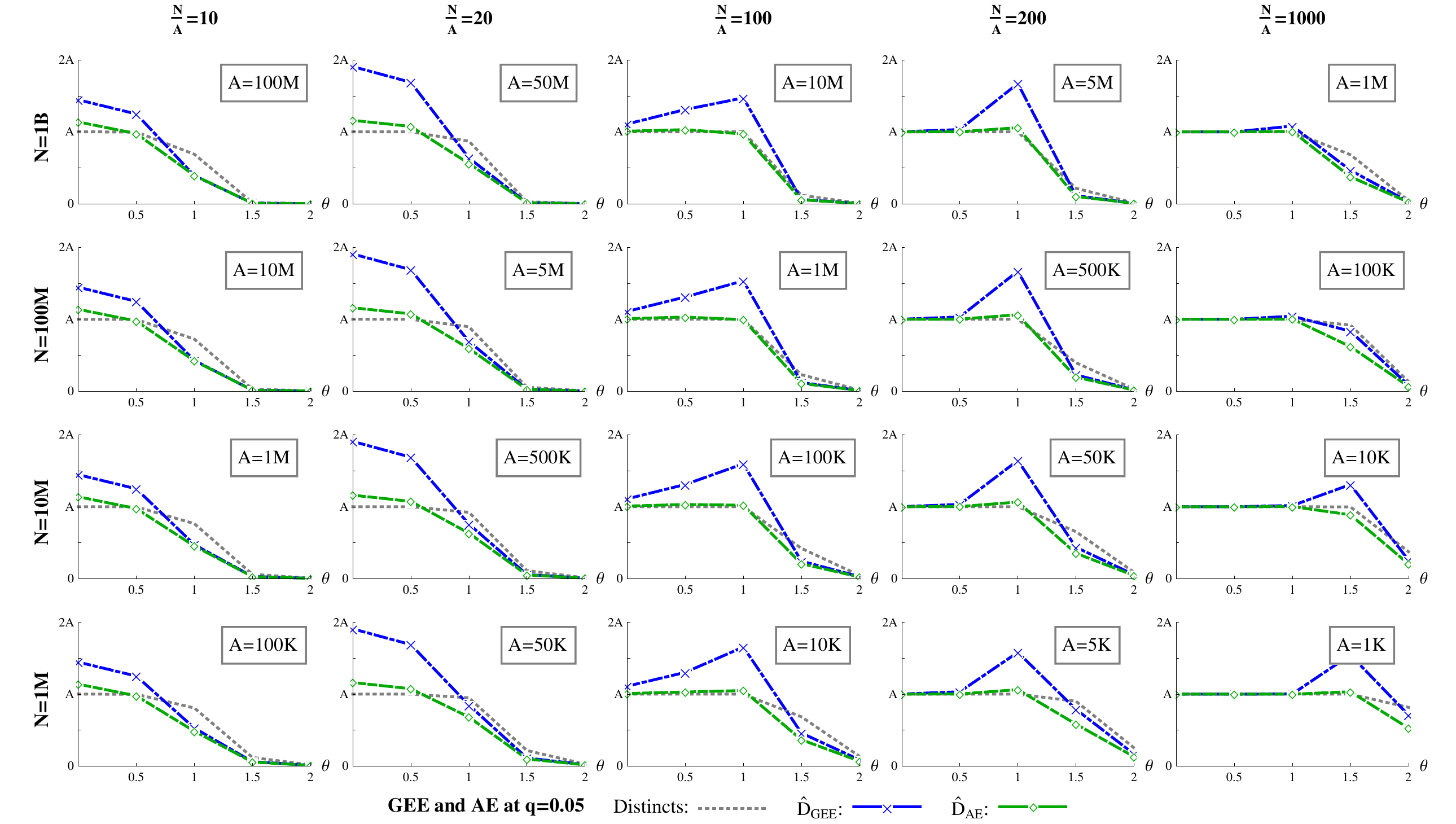}
\includegraphics[type=pdf,ext=.pdf,read=.pdf,bb=27 2 727 413,scale=0.6]{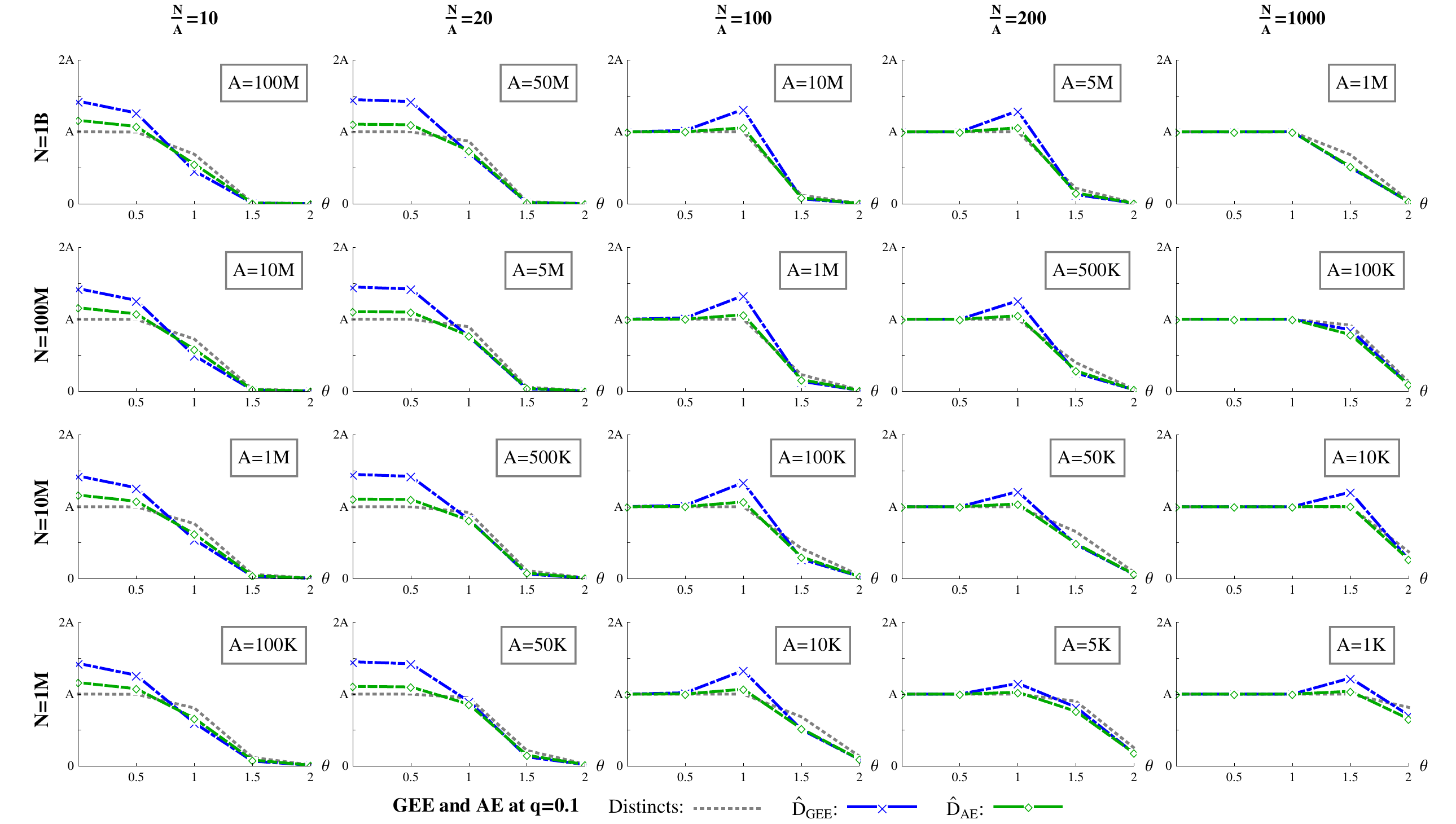}\vspace{1in}
\includegraphics[type=pdf,ext=.pdf,read=.pdf,bb=27 2 727 413,scale=0.6]{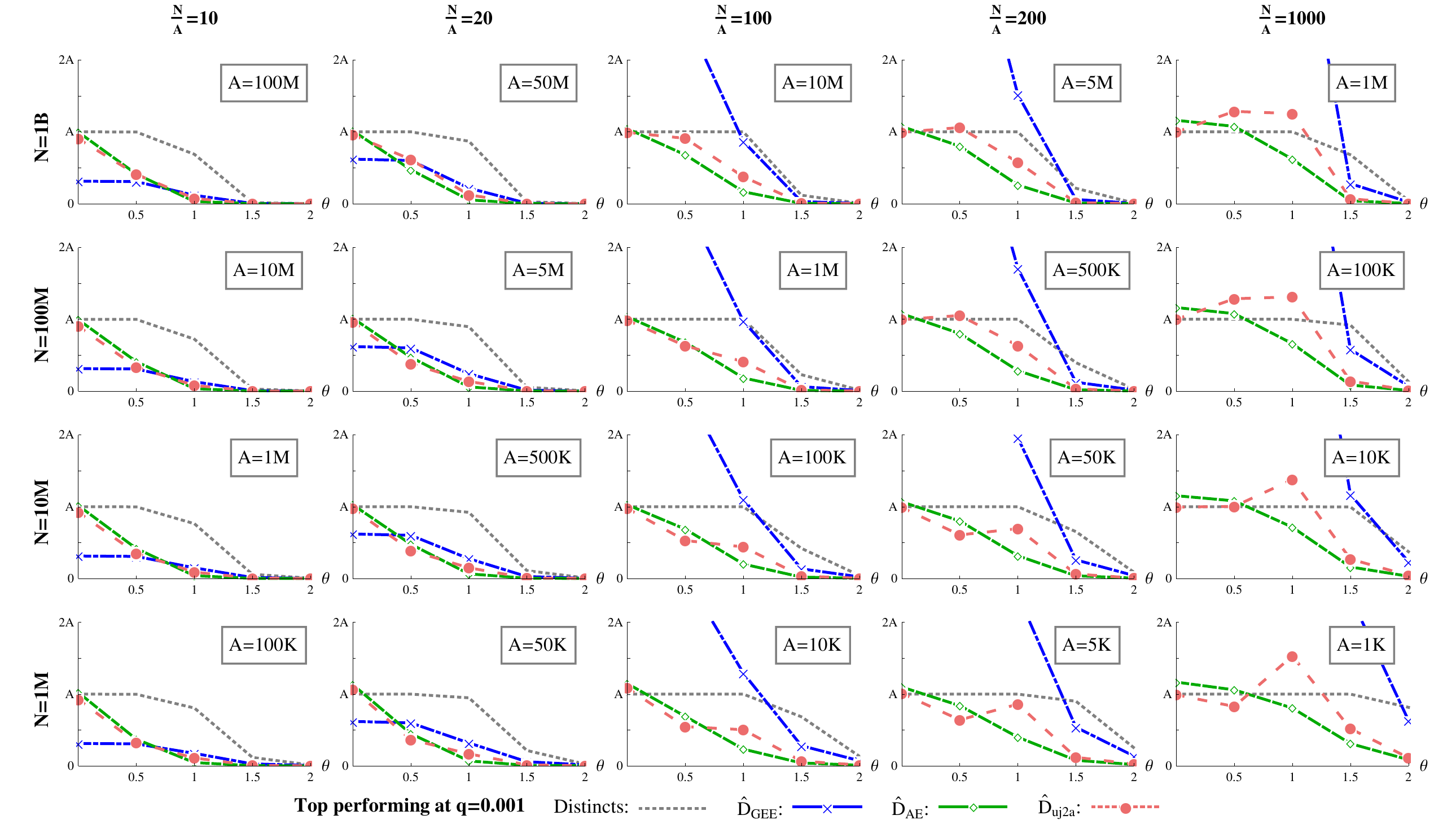}
\includegraphics[type=pdf,ext=.pdf,read=.pdf,bb=27 2 727 413,scale=0.6]{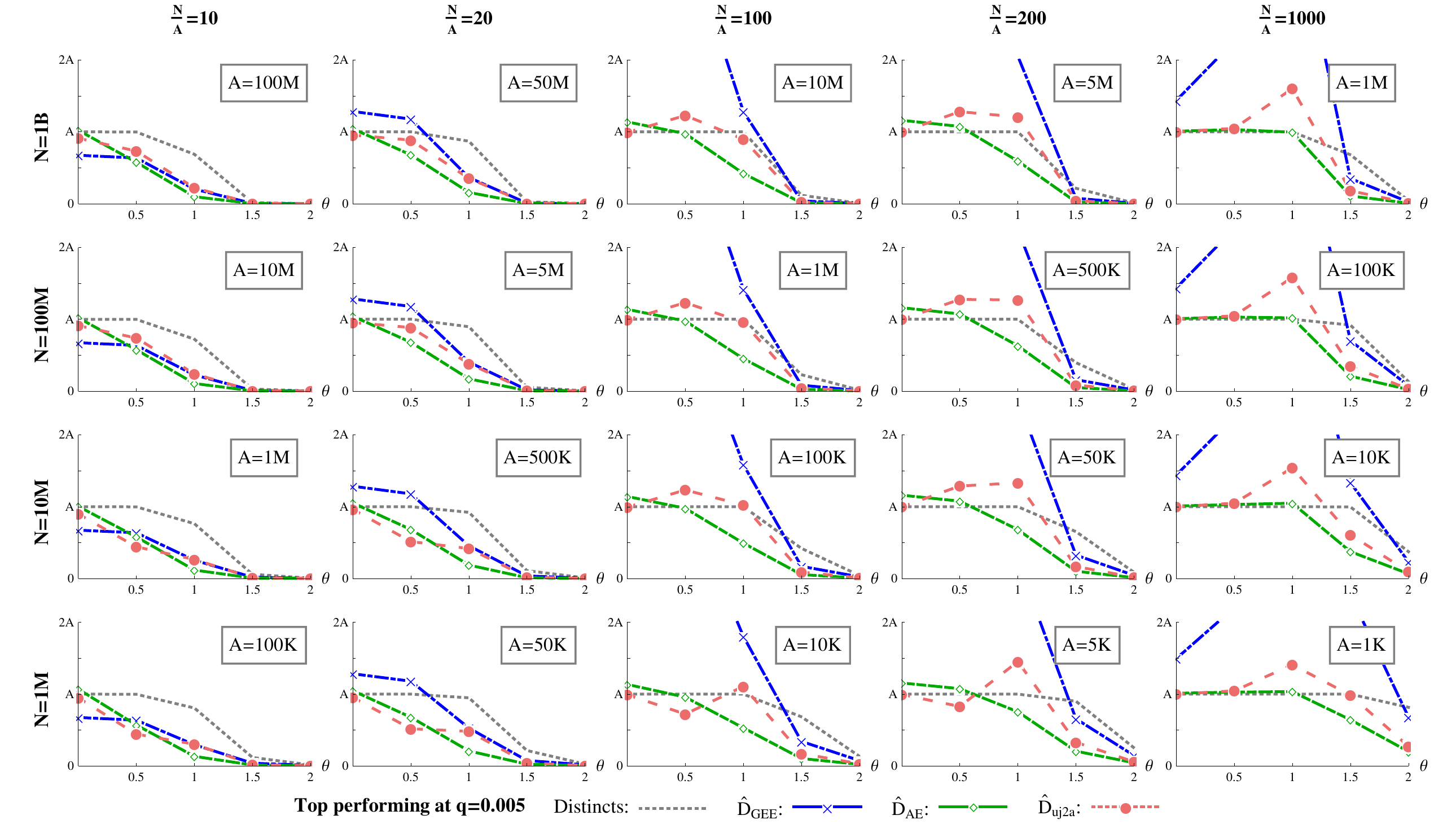}\vspace{1in}
\includegraphics[type=pdf,ext=.pdf,read=.pdf,bb=27 2 727 413,scale=0.6]{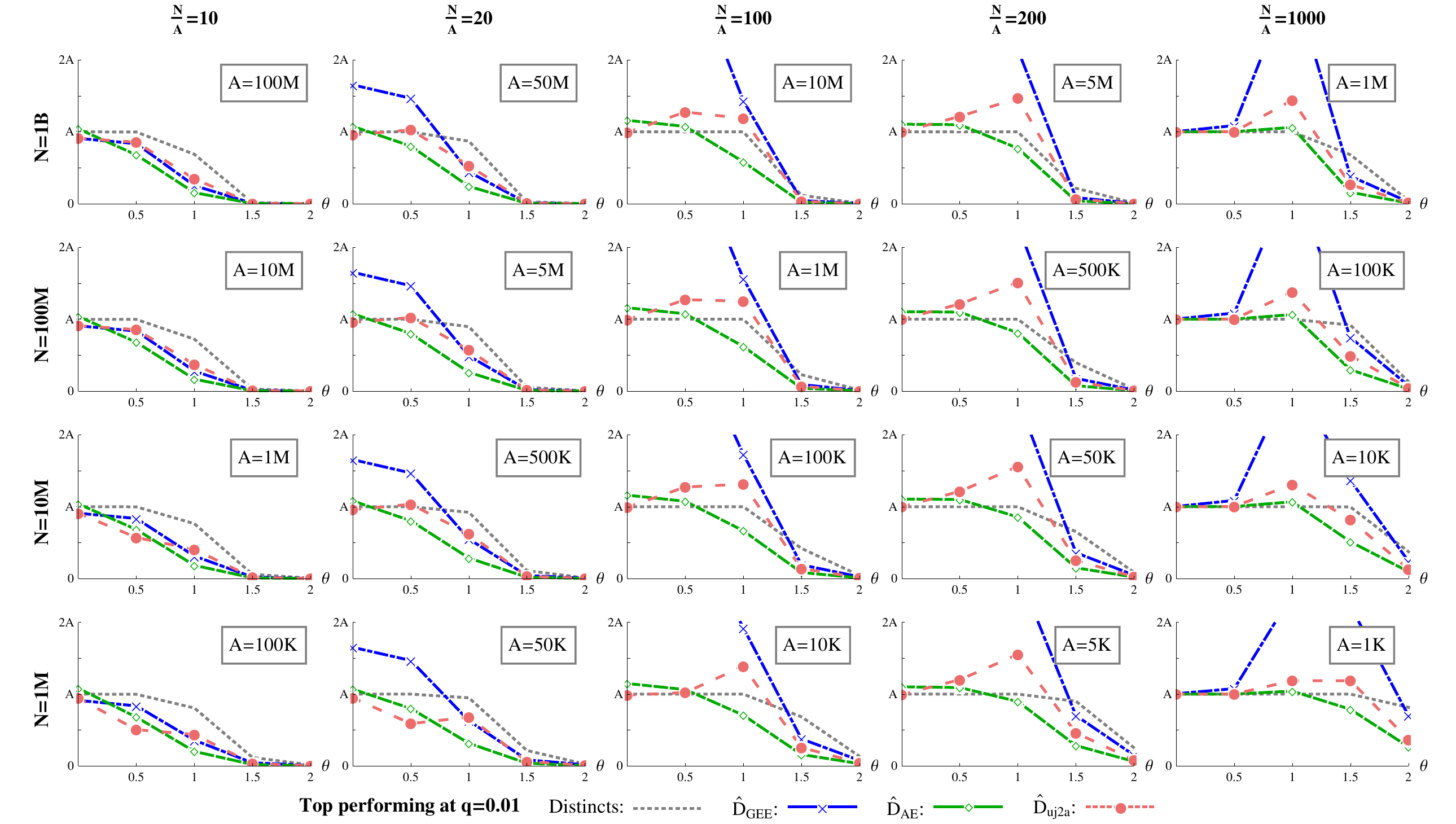}
\includegraphics[type=pdf,ext=.pdf,read=.pdf,bb=27 2 727 413,scale=0.6]{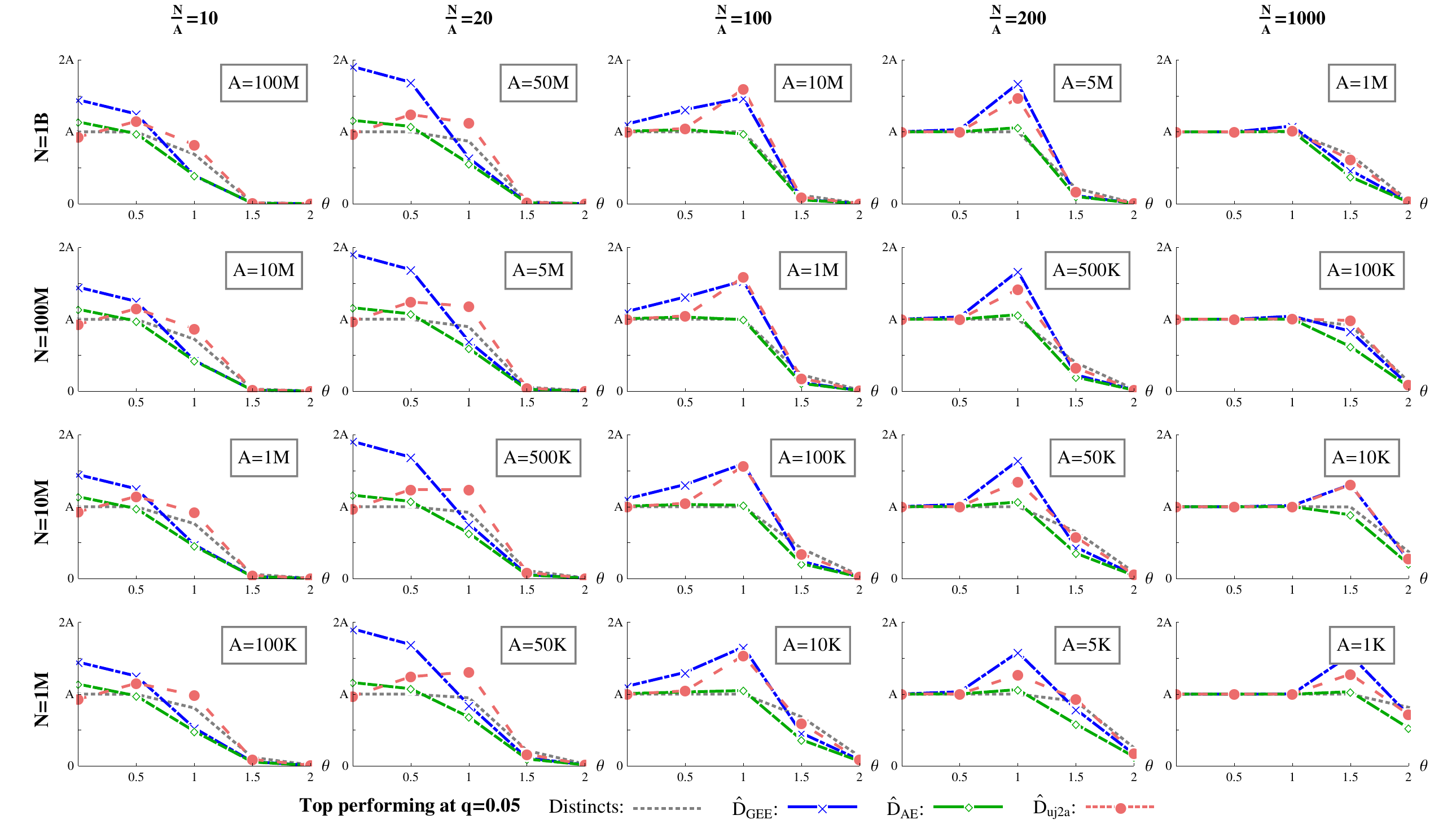}\vspace{1in}
\includegraphics[type=pdf,ext=.pdf,read=.pdf,bb=27 2 727 413,scale=0.6]{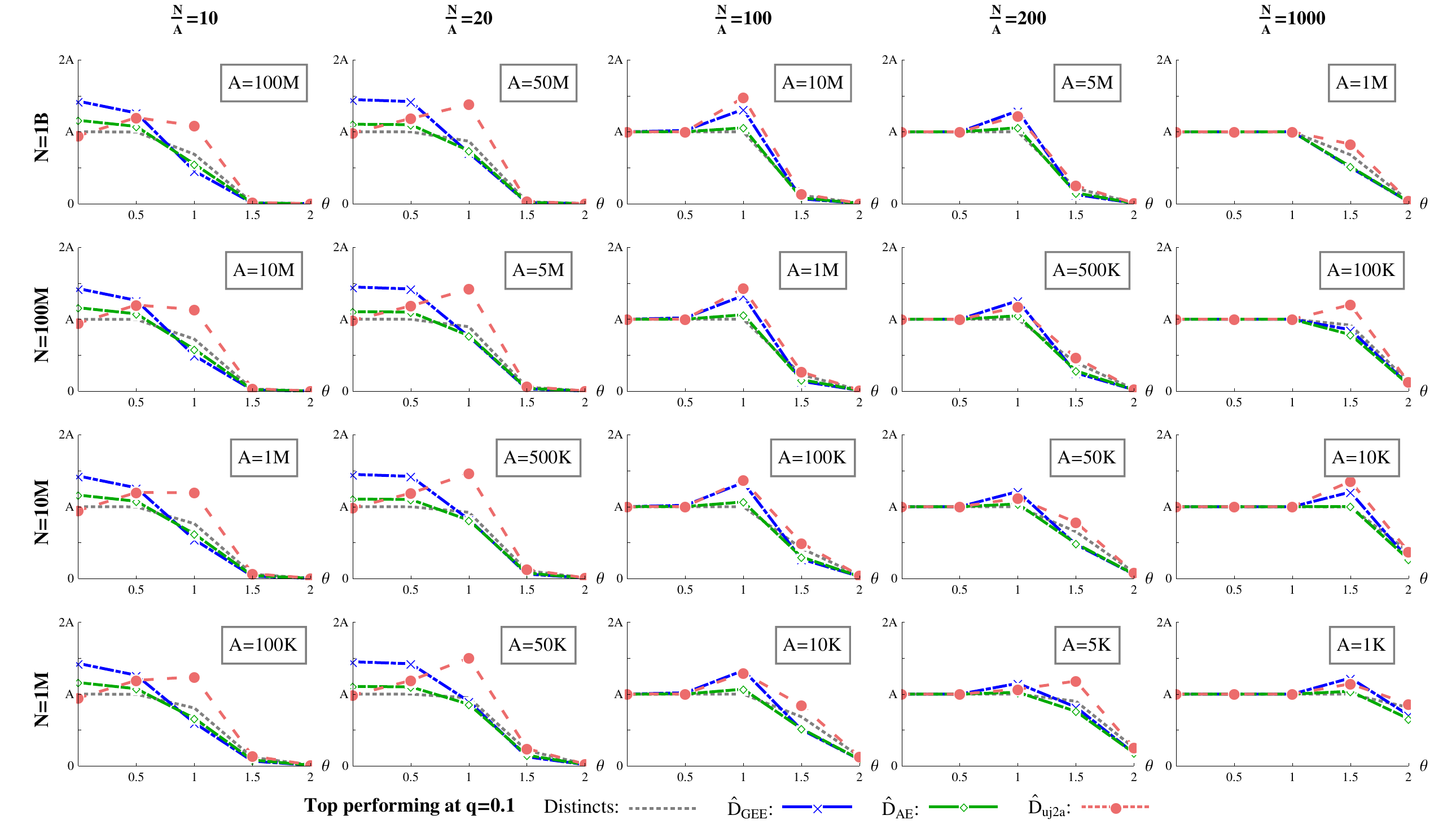}
\clearpage
\begin{figure}[h]
\centering
\floatpagestyle{empty}
\subfloat[3D bias surfaces for \cltwo. The almost vertical surfaces indicate
onset of severe positive bias.]{
\includegraphics[type=pdf,ext=.pdf,read=.pdf,bb=50 0 862
499,scale=0.55]{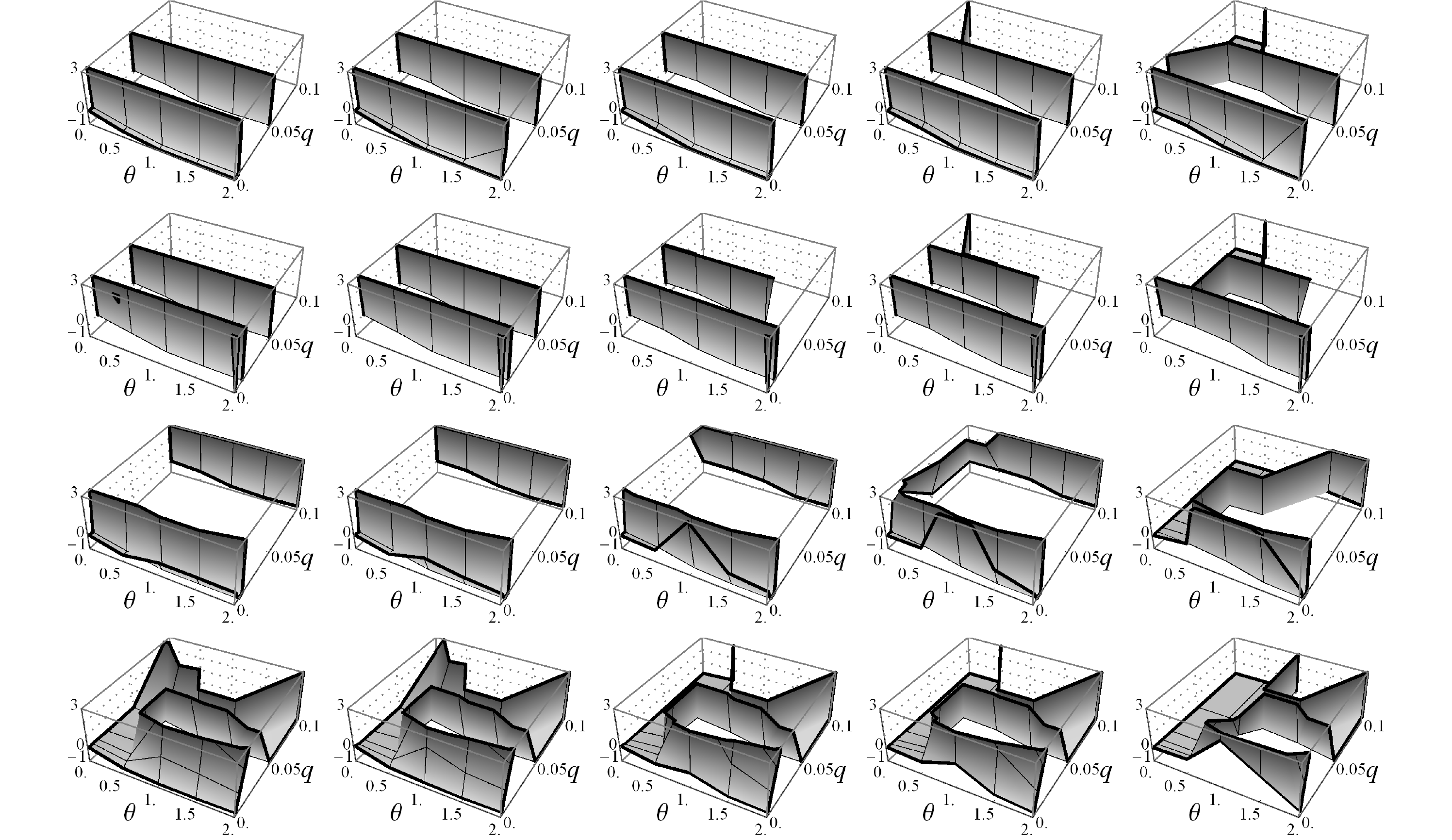}
\label{fig:cl1:grid}}\\
\subfloat[\cltwo at $q=0.1$. Even at this high sampling fraction, \cltwo is accurate
only at 10M
population size. Surprisingly, it is inaccurate at the lower population size
of 1M, as well as at higher sizes.]{
\includegraphics[type=pdf,ext=.pdf,read=.pdf,bb=27 2 727
413,scale=0.6]{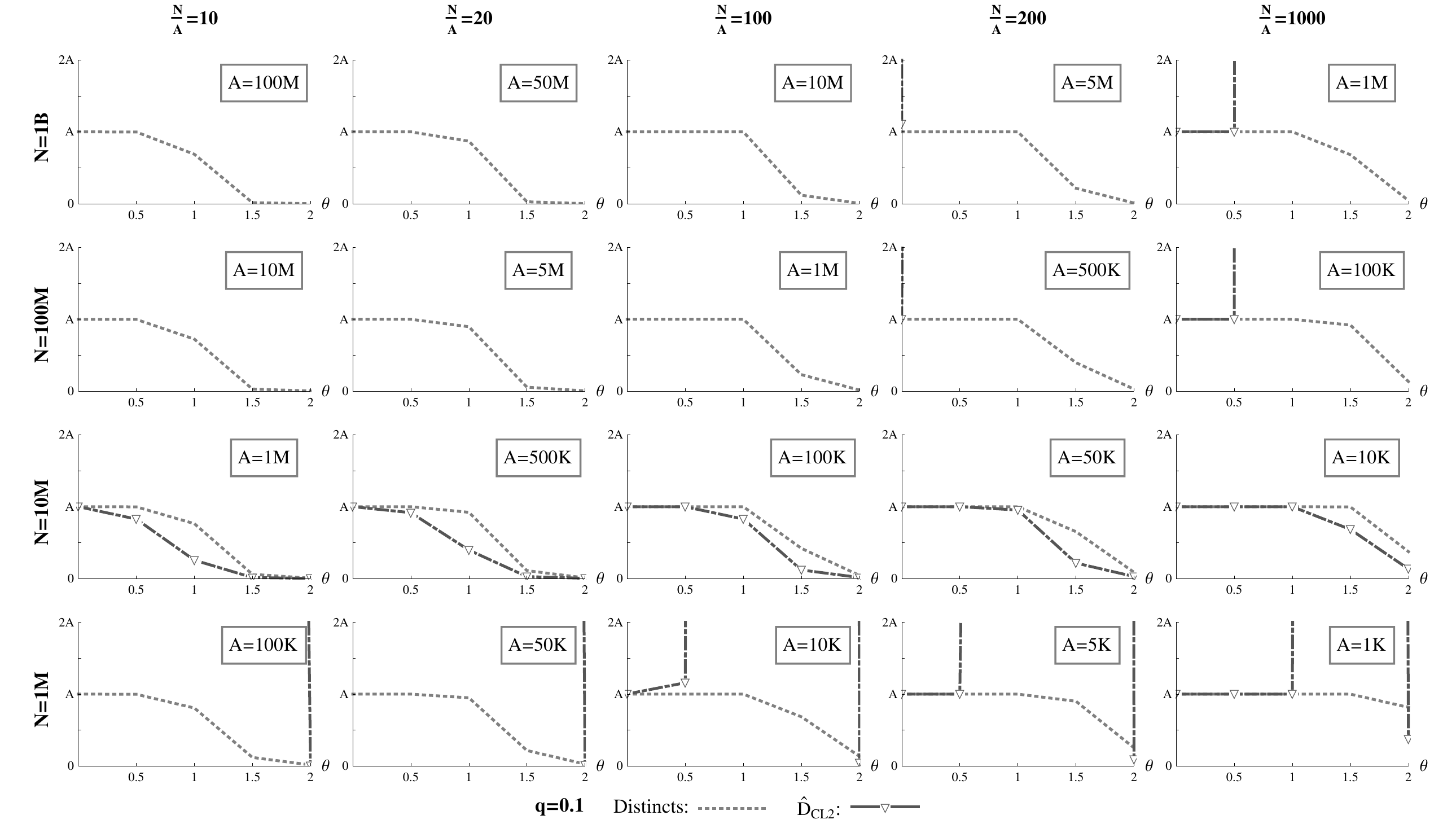}
\label{fig:cl2:grid}}
\caption{The highly irregular bias behavior of the Chao-Lee estimators. The
difference between \clone and \cltwo is insignificant, hence only \cltwo shown.}
\label{fig:cl1:cl2:grids}
\end{figure}

\begin{spacing}{0.9}
{\small	
	\bibliographystyle{plain}
	\bibliography{uecbib}
}
\end{spacing}
\end{document}